\documentclass[manuscript,nonacm]{acmart}

\usepackage{url}            
\usepackage{booktabs}       
\usepackage{color}          
\usepackage{gensymb}        

\usepackage{amsmath, amsthm}       
\usepackage{nicefrac}       
\usepackage{gensymb}        

\usepackage{algorithm}
\usepackage[noend]{algpseudocode}

\usepackage{graphicx}
\graphicspath{ {./images/} }
\usepackage{caption}
\usepackage{subcaption}

\newcommand{\LCP}{LCP}

\newcommand{\OBD}{OBD}
\newcommand{\OMP}{OMP}
\newcommand{\PRP}{PRP}
\newcommand{\SDP}{SDP}
\newcommand{\DLE}{DLE}
\newcommand{\Detect}{Detect}
\newcommand{\Communicate}{Redirect}
\newcommand{\vin}{\vec{v_{in}}}
\newcommand{\vout}{\vec{v_{out}}}
\newcommand{\vrot}{\vec{v_{rot}}}

\def\Collect{\textbf{Collect}}

\newtheorem{theorem}{Theorem}
\newtheorem{claim}[theorem]{Claim}

\newtheorem*{remark*}{Remark}
\newtheorem*{claim*}{Claim}
\newtheorem{observation}[theorem]{Observation}
\newtheorem*{observation*}{Observation}
\newtheorem*{definition*}{Definition}
\newtheorem*{proposition*}{Proposition}

\begin{document}

\title{Efficient Deterministic Leader Election for Programmable Matter}

\author{Fabien Dufoulon}
\email{dfabien@campus.technion.ac.il}
\affiliation{%
  \institution{Technion - Israel Institute of Technology}
    \city{Haifa} 
  \country{Israel} 
}
\orcid{0000-0003-2977-4109}

\author{Shay Kutten}
\email{kutten@technion.ac.il}
\affiliation{%
  \institution{Technion - Israel Institute of Technology}
    \city{Haifa} 
  \country{Israel} 
}
\orcid{0000-0003-2062-6855}

\author{William K. Moses Jr.}
\email{wkmjr3@gmail.com}
\affiliation{%
  \institution{University of Houston}
  \city{Houston} 
  \country{USA} 
}
\orcid{0000-0002-4533-7593}

\begin{abstract}
It was suggested that a programmable matter system (composed of multiple computationally weak mobile particles) should remain connected at all times since otherwise, reconnection is difficult and may be impossible. At the same time, it was not clear that allowing the system to disconnect carried a significant advantage in terms of time complexity. We demonstrate for a fundamental task, that of leader election, an algorithm where the system disconnects and then reconnects automatically in a non-trivial way (particles can move far away from their former neighbors and later reconnect to others). Moreover, the runtime of the temporarily disconnecting deterministic leader election algorithm is linear in the diameter. Hence, the disconnecting -- reconnecting algorithm is as fast as previous randomized algorithms. When comparing to previous deterministic algorithms, we note that some of the previous work assumed weaker schedulers. Still, the runtime of all the previous deterministic algorithms that did not assume special shapes of the particle system (shapes with no holes) was at least quadratic in $n$, where $n$ is the number of particles in the system. (Moreover, the new algorithm is even faster in some parameters than the deterministic algorithms that did assume special initial shapes.)

Since leader election is an important module in algorithms for various other tasks, the presented algorithm can be useful for speeding up other algorithms under the assumption of a strong scheduler. This leaves open the question: ``can a deterministic algorithm be as fast as the randomized ones also under weaker schedulers?'' 
\end{abstract} 

\maketitle


\section{Introduction}

The study of the interplay between movement and computation has long been an area of interest spanning multiple research fields~\cite{FPS19}. Programmable matter, introduced by Toffoli and Margolus~\cite{TM91}, is an attempt to use mobile computational agents to model matter that can change its physical properties, e.g., change shape. While ``regular'' matter is composed of ``dumb'' particles, each of the ``smart'' particles that compose programmable matter is equipped with computational power and the ability to move. Like ``dumb'' particles, smart particles need to be small. Thus, they are also weak in their
memory size, computational power, communication ability, movement ability, etc. To accomplish significant tasks, they must cooperate.

Among the multiple concrete realizations of programmable matter, the amoebot model, first proposed by Derakhshandeh et al.~\cite{DDGRSS14} (also explained in detail by Daymude et al.~\cite{DHRS19}), has gained much traction in recent years. In this model, each particle can move from one grid point of a triangular grid to a neighboring grid point in a way that resembles an amoeba. Many problems of interest were addressed, including coating of materials~\cite{DGRSST14,DGRSS17b,DDGPRSS18}, bridge building~\cite{ACDRR18}, energy distribution~\cite{DRW21}, shape formation~\cite{DGRSS15,CDRR16,DGRSS16,DFSVY20shapeformation,DGHKSR20,DFSVY20mobileram}, and shape recovery~\cite{DFPSV18}. Towards solving these problems, a dominant strategy has been to elect a unique leader among the particles, which then coordinates all the movements.

Early deterministic leader election algorithms did not use the movement ability as an aid to the computation and the communication. This resulted in algorithms that were based on assumptions on the initial shape of the particle system (intuitively, that it had no ``holes'')~\cite{GAMT18,DFSVY20shapeformation} or elected up to 6 leaders instead of one in some cases~\cite{BB19}. Recently, it was shown that the particles' movements could be leveraged in order to elect a unique leader (or up to 3 leaders when actions are not atomic) without imposing a restriction on the initial shape \cite{EKLM19,DDDNP20ieee}. Unfortunately, the above improvements came at the cost of a considerably higher runtime, compared to the algorithms that assumed no holes. 

In the current paper, we present a linear time algorithm to elect a unique leader deterministically when assuming a strong scheduler, regardless of the initial shape. (We make the rather common assumption that the particles have the same chirality; see Section~\ref{subsec:system} and~\cite{DGRSS16,DDGPRSS18,PR18,DGSBRS15,DGRSS17a,BB19,GAMT18}.) 
The runtime of all the previous deterministic algorithms that did not assume special shapes of the particle system (shapes with no holes) was at least quadratic in $n$, where $n$ is the number of particles in the system.
One may then ask: can a deterministic algorithm be as fast as the randomized ones even when holes are present (at least when assuming a strong scheduler)? 
Since leader election is an important module in algorithms for various other tasks, the presented algorithm can be useful for speeding up other algorithms. 

Previous literature assumed that the algorithm ensures that the system is connected at all times, although (see~\cite{DHRS19}) the amoebot model does not require that the system remains connected.\footnote{There are some partial exceptions. In \cite{DFPSV18}, faults kill some of the particles, possibly disconnecting the non-faulty ones. In \cite{DFSVY20shapeformation}, the model does not allow a particle to contract to the tail; they simulate such a contraction by the particle moving to a neighboring node (possibly disconnecting) and immediately moving back (and reconnecting).} Allowing the system another operation (i.e., to disconnect) obviously gives algorithm designers another potential tool. Still, it was not clear that such a tool was, in fact, useful. On the contrary, in \cite{DHRS19}, the opposite is implied.
According to them, ``If a particle system disconnects, there is little hope the resulting components could ever reconnect. Since each particle can only see and communicate with its immediate neighbors and does not have a global compass, disconnected components have no way of knowing their relative positions and thus cannot intentionally move toward one another to reconnect.''

In this paper, we demonstrate that disconnecting and reconnecting again can actually be very useful. In fact, the significant improvement in runtime achieved by the algorithm presented here is obtained thanks to the disconnection and reconnection.

\subsection{Related Work}

The type of scheduler used affects the leader election results (which are compared with our result in Table~\ref{table:results-comparison}). Typically in the literature~\cite{DGSBRS15,DGRSS17a,GAMT18,EKLM19}, the ``strong'' scheduler activates particles atomically. 
In ``weaker'' schedulers~\cite{BB19,DFSVY20shapeformation,DDDNP20ieee}, where activations are non-atomic, it becomes impossible to elect a unique leader deterministically in some cases. (Recall that without particle movements, it may be the case that it is impossible to always elect a unique leader even under a strong scheduler \cite{BB19}.)

Leader election in the amoebot model was initially studied by Derakhshandeh et al.~\cite{DGSBRS15}. They assumed that particles have common chirality and proposed a randomized algorithm to solve leader election (of exactly one leader) in $O(L_{max})$ rounds in expectation, where $L_{max}$ is the length of the largest boundary in the shape. 
Daymude et al.~\cite{DGRSS17a,DGRSS17aarxiv} assumed common chirality too. 
They improved upon the previous result by presenting a randomized algorithm that elected a unique leader (with probability 1) in $O(L_{out} + D)$ rounds w.h.p.\footnote{With high probability, that is, with probability of success at least $1-n^{-c}$ where $c$ is any constant equal or greater than 1.}, where $L_{out}$ is the length of the outer boundary of the shape and $D$ is the diameter of the shape.

The first deterministic leader election algorithm was presented by Di Luna et al.~\cite{DFSVY20shapeformation} for the natural special case that the shape did not contain holes; multiple  (up to three) leaders could be elected in some cases. The runtime was $O(n)$ where $n$ is the number of particles. The paper used the elected leader(s) to perform shape transformation. Gastineau et al.~\cite{GAMT18} assumed common chirality and an initial shape with no holes.They presented a $O(r + mtree)$ round deterministic leader election algorithm (of exactly one leader), where $r$ and $mtree$ are terms specific to their paper. It can be shown that $r + mtree = \Omega(D)$. In addition to election, they also assigned local identifiers to particles. (A subsequent paper by Gastineau et al.~\cite{GAMT20} extended these results to a proposed three dimensional variant of the amoebot model and took $O(n)$ time to elect a leader deterministically subject to constraints specific to their model.) The deterministic algorithm of Bazzi and Briones~\cite{BB19} also assumed common chirality. They could elect (up to six) leaders deterministically even when holes were present in the shape. However, their runtime was $O(n^2)$. Emek et al.~\cite{EKLM19} addressed the use of movement for electing a leader in the ameobot model. They showed that a unique leader can be elected if movement is allowed assuming a strong scheduler even if there are holes in the shape and even if common chirality is not assumed. The runtime of their algorithm was $O(L_{out} n^2)$ rounds. 
 
D'Angelo et al.~\cite{DDDNP20ieee} proposed a variant of the amoebot model, called SILBOT, where particles could not communicate the content of their variables but could ``view'' the position of other particles within 2-hops from them. Moreover, they could see whether another such particle was in the ``process of moving'' to a different grid point. In this altered model, they first presented a deterministic leader election algorithm (of up to three leaders) in $O(n)$ rounds when the initial shape had no holes. Subsequently, they presented a deterministic leader election algorithm (of up to three leaders) on any initial shape when particles were empowered with yet another sense -- the ability to determine whether an empty grid point within the 2-hop view was a part of a hole or the outer face of the shape. This assumption is stronger than the assumption that particles know initially which boundary is the outside boundary, since the extra sense could be used at any point during the execution.  

\subsection{Technical Challenges}
To improve the runtime, we had to address the issues of particle movements, those of particles messaging each other, and the combination thereof.
Previous deterministic election algorithms used an erosion process that did not require particle movements. The idea was that a particle $p$, once it knew it was on the outer boundary, removed itself from candidacy provided $p$ could be sure the system remained connected (even when not counting $p$)~\cite{DFSVY20shapeformation,GAMT18,DDDNP20ieee}. To ensure such connectivity, each particle first verified some local conditions. Intuitively, it seems that the conditions in~\cite{DFSVY20shapeformation,DDDNP20ieee} were restrictive to the point that they limited parallelism and thus slowed the erosion process down. The conditions in the process of~\cite{GAMT18} seem much less restrictive and indeed allowed the authors there to prove a faster process. However, the conditions may have been too ``liberal'' in the sense it may have allowed the adversarial scheduler too much freedom in choosing an order of erosion that would slow the process down. The conditions we chose for a particle to erode itself allowed us to prove an existentially tighter upper bound, using some basic tools from metric graph theory \cite{BC08,CDV02}: we prove $O(D_A)$ for the erosion, where $D_A$ is the diameter of the whole sub-grid surrounded by particles (including the holes). Note that this can be smaller than $D$.

Initially, we assume that at the beginning of the execution, particles residing on a boundary know if that boundary borders a hole or the outer face. We remove this assumption later at the cost of $O(L_{out} +D)$ rounds (Note that this assumption, nevertheless, is weaker than that of \cite{DDDNP20ieee}.) When this assumption is used, we manage to improve the runtime of the algorithm, even for shapes with holes, to $O(D_A)$. 
This requires particle movement (as opposed to the use of erosion only, when no holes exist). Intuitively, with no erosion, any algorithm seems to be doomed to $\Omega(D)$ runtime. Moreover, without particle movement, it seems that no algorithm can guarantee the choice of a unique leader (this was shown in \cite{BB19}, at least for the case of a weak scheduler). 
 However, movement can complicate the algorithm. In particular, algorithms that use particle movements can face the problem of different groups of particles standing in the way of each other. Such a phenomenon could cause the algorithm of \cite{EKLM19} to reset many times, increasing its runtime significantly. On the other hand, careful coordination of the movement is difficult, given that no leader has been selected yet. We avoid these resets since particles move ``inwards'' in our algorithm. 

A similar type of inward movement can be seen in D'Angelo et al.~\cite{DDDNP20ieee}, where they remove holes from the shape while running the erosion process. However, their process still takes $O(n^2)$ rounds. We are able to speed things up by allowing the particle system to disconnect. Intuitively, there is less space the further one goes inwards. Hence, not all the particles could move inwards simultaneously.  
Arranging them in queues and moving the queues of particles inwards is, possibly, what increases the runtime to $O(n^2)$. (This, at least, was the bottleneck when we designed an algorithm without disconnection, an algorithm not presented here.)

An additional challenge is a result of the shape disconnection we utilize. Due to the particles' limited memories, it is non-trivial to reconnect a shape once particles are disconnected. Not only do we need to ensure that all particles (the total number of which we cannot count) are collected, but we must do this quickly. We overcome this issue by disconnecting the shape in a careful manner; informally, we disconnect particles such that we leave ``breadcrumbs'', which a leader can follow to collect everyone.

Finally, the outer boundary detection primitive required quadratic time in \cite{EKLM19}. It used a subroutine that elected up to 6 leaders on each boundary. One runtime bottleneck of that subroutine was a process of comparison between one set of particles to another. The comparison was performed in a sequential manner -- comparing certain inputs of two particles at a time, because the memory size of a particle is constant and could not contain many inputs values. We managed to pipelines these comparisons carefully. One reason that care is needed is that each of the compared sets can be changed during the comparison. (In previous algorithms, the sets were, in essence, ``frozen'' for the duration of the comparison.)

\begin{table}[ht]
	\caption{This table compares the results on leader election to those of previous papers. In column three, D \& R correspond to deterministic solution and randomized solution, respectively, while S \& W correspond to strong scheduler and weaker scheduler, respectively. The assumptions \& relaxations used in papers include (i) assuming particles have common chirality, (ii) assuming the initial shape has no holes, (iii) assuming each particle can detect whether an empty grid point within its visibility range is part of a hole or the outer face (boundary detection), either \emph{initially} or \emph{throughout} the execution of the algorithm, and (iv) relaxing the solution of leader election to allow multiple leaders. The length of the largest boundary in the initial configuration is $L_{max}$. The length of the outer boundary in the initial configuration is $L_{out}$. The number of particles in the configuration is denoted by $n$. The terms $r$ and $mtree$ are specific to the algorithm of~\cite{GAMT18}.}
	\centering \vspace{1em}
		\resizebox{1.0\columnwidth}{!}{
	\begin{tabular}{c c c c}
		\toprule
		Paper & Running Time & Randomness/ & Assumptions/Relaxations \\
		& (rounds) & Scheduler & \\
		\midrule
		\cite{DGSBRS15} & $O(L_{max})$ on expectation  & R/S  & Chirality \\
		\cite{DGRSS17a,DGRSS17aarxiv} & $O(L_{out} + D)$ w.h.p.  & R/S & Chirality  \\
		\cite{BB19} & $O(n^2)$   & D/W & Chirality; Multiple leaders \\
		\cite{DFSVY20shapeformation} & $O(n)$  & D/W & Multiple leaders;  No holes \\
		\cite{GAMT18}* & $O(r + mtree)$   & D/S & Chirality; No holes \\
		\cite{DDDNP20ieee}** & $O(n^2)$ & D/W & Boundary detection throughout \\
		\cite{EKLM19} & $O(L_{out} n^2)$  & D/S & -\\
		Current Paper & $O(D_A)$ & D/S & Boundary detection initially; Chirality \\
		Current Paper & $O(L_{out} + D)$ & D/S & Chirality \\
		\midrule
		\multicolumn{2}{l}{*Note that $r + mtree$ = $\Omega(D)$.
		} &
        \multicolumn{2}{l}{**Their model is a bit different from the amoebot model.}\\
        \bottomrule
	\end{tabular}
		}
	\label{table:results-comparison}
\end{table}

\subsection{Results and Paper Organization}

We present the first linear time deterministic algorithm for (unique) leader election that can handle holes, under a strong  scheduler. Specifically, when we assume that particles recognize the outer boundary initially, the runtime is linear in $D_A$ -- that may be smaller than $D$. (The new algorithm is even faster in some parameters than deterministic algorithms that did assume special initial shapes.) Note that some of the previous algorithms did not assume common chirality. When we do not assume that particle recognize the outer boundary initially, the runtime increases to $O(L_{out}+D)$. The presented deterministic algorithms are at least as fast as the current best randomized algorithms. These algorithms demonstrate the power of using disconnection (at least so far, it is not known how to obtain the same results without using the power of disconnection). A comparison of the results with previous ones can be found in Table~\ref{table:results-comparison}.

The algorithm is composed of two parts, the algorithm for leader election where the system may disconnect is presented in Section \ref{subsec:DisconnectingLeaderElection} and is analysed in Section \ref{subsec:AnalysisDLE}. The reconnection procedure and its analysis appear in Section \ref{subsec:ReconnectingLeaderElection}. 
The model description and most of the definitions appear in Section \ref{sec:prelim}. This includes the definition of chirality as well as the assumption that the particles initially agrees on it (similar to the assumptions, e.g., in \cite{DGRSS16,DDGPRSS18,PR18,DGSBRS15,DGRSS17a,BB19,GAMT18}).

\section{Preliminaries}
\label{sec:prelim}

The particles of a particle system occupy points of a triangular grid as detailed later in this section. For short, we usually refer to grid points just as ``points''. The grid is assumed to be embedded in the plane, though the embedding is not known to the particles. In particular, we may talk about $\{W,NW,NE,E,SE,SW\}$, referring, e.g., to the West side of the embedding, or to the Northwest, but the particles do not know which direction is which. Given a subset of the (grid) points, the graph they induce includes these points as nodes. It also includes an edge between two of points iff these two points are neighbors on the grid.  

\subsection{Shapes}
\label{subsec:shape}

A \emph{shape} is a finite set of points from the triangular grid $G$ -- see Figure \ref{fig:ShapesExample}. 
By an abuse of notation, the (finite) subgraph of the triangular grid induced by such a shape is also called a shape. 
In what follows, let $S$ be an arbitrary connected shape. One may neglect mentioning $S$ when the considered shape $S$ is evident.

\paragraph*{Boundaries and Holes in Shapes.} The clockwise (and counter-clockwise) successor and predecessor edges are defined for every point's incident edges in the natural way. Clockwise (and counter-clockwise) cyclic intervals are defined similarly.
A shape partitions the plane into \emph{faces} (regions bounded by its points and edges) including exactly one unbounded face, the outer one. The set of points that lie on the outer face is called the \emph{outer boundary}. An inner face containing at least one point (that is not in the shape) is said to be a \emph{hole} in the shape. The set of points bounding a hole is called an \emph{inner boundary}. A shape without any holes is said to be \emph{simply-connected}. The \emph{length} of a boundary is the number of points of that boundary. 
We denote by $L_{out}(S)$ the length of the outer boundary of a shape $S$ and by $L_{max}(S)$ its maximum boundary length.
Points on boundaries are referred to as \emph{boundary points}, other points in the shape as \emph{interior points}, and points in the shape's holes as \emph{hole points}. 
See Figure \ref{fig:ShapesExample}.

\paragraph*{Area of a Shape.} 

The \emph{area} of shape $S$ consists of $S$ and all of $S$'s holes points -- see Figure \ref{fig:ShapesExample}.
For any pair of points in $S$, their \emph{distance} with respect to (w.r.t.) some shape $S^* \supseteq S$ -- denoted by $dist_{S^*}$ -- is the length of the shortest path (within $S^*$) between these two points. Then, the \emph{eccentricity} of point $v \in S$ w.r.t. $S^*$ -- denoted by $\epsilon_{S^*}(v)$ -- is the greatest distance (w.r.t. $dist_{S^*}$) from $v$ to any other point in $S$. The \emph{diameter} of $S$ w.r.t. $S^*$ is the greatest eccentricity (w.r.t. $S^*$) of any point in $S$. When $S^* = S$, these are simply said to be the distance, eccentricity and diameter of $S$.

\begin{observation}
\label{obs:parameterObservations}
The following holds:
\begin{enumerate}
\item The diameter of $S$ is greater than or equal to the diameter of $S$ w.r.t. the area of $S$. 
\item If $S$ is a simply-connected shape with $n_S$ points and diameter $D_S$, then $n_S = O(D_S^2)$. 
\item If $S$ is simply-connected, then the length of the outer boundary of $S$ is greater than or equal to $S$'s diameter.
\end{enumerate}
\end{observation}

\begin{proposition}
\label{prop:holesNodesAreOnShortestPath}
For any hole point $h$ of $S$, there exists $v_1,v_2 \in S$ such that $h$ is on a shortest path between $v_1$ and $v_2$.
\end{proposition}

\begin{proof}
Take any two opposite incident edges $e_1,e_2$ of $h$ (i.e., separated by two edges clockwise, in the cyclic interval of $h$'s incident edges). Consider the two straight paths (on the hole points of $S$ as well as the points of $S$) starting at $h$ and going along the directions of $e_1$ and $e_2$, respectively. Intuitively, since $S$ is a finite boundary encircling $h$, these two paths intersect $S$ in two grid points. In other words, these paths reach points $v_1,v_2 \in S$, respectively, and $v$ is on the shortest path between $v_1$ and $v_2$.
\end{proof}

\paragraph*{Local Boundaries in Shapes and Convex Vertices.}
Let $v$ be a boundary point of $S$ (and $|S| \geq 2$). The \emph{local boundary} $B$ (w.r.t. $S$) of $v$ is a clockwise cyclic interval of incident edges leading to points not in $S$. Denote by $|B|$ the size (the number of edges) of the interval. Note that $v$ may have up to 3 local boundaries, and these local boundaries may be a part of the same (inner or outer) global boundary of that shape. The clockwise \emph{successor} (respectively, \emph{predecessor}) \emph{point} of $v$ with respect to $B$ is defined as the point reachable by the clockwise successor of $B$'s final edge (resp., predecessor of $B$'s first edge). The counter-clockwise successor and predecessor points of $v$ are defined similarly.
The \emph{boundary count} of $v$ w.r.t. $B$ is defined as $c(v,B)=|B|-2$, which by definition (of a local boundary) is necessarily in $\{-1,\ldots,3\}$ -- see Figure \ref{fig:exampleCounts}.\footnote{Note that when a shape consists of a single point -- which is not considered here -- that point has boundary count $4$. } 
If $c(v,B) > 0$, then $v$ is said to be \emph{(strictly) convex} w.r.t. $B$. For simplicity, when $v$ has a single local boundary $B^*$ w.r.t. $S$, we say that $v$ has a boundary count of $c(v,B^*)$ w.r.t. $S$ and if $c(v,B^*) > 0$, that $v$ is (strictly) convex w.r.t. $S$.

\paragraph*{Virtual Nodes and (Oriented) Rings on Global Boundaries}
Each global boundary can be transformed into an (oriented) virtual ring. 
These rings are used in the proof of Proposition \ref{lem:AllConvexOuterBoundaryNodesErodable} and in Section \ref{sec:boundaryDetection}. 
Some definitions are given first. A boundary point $v \in S$ is subdivided into one, two, or three \emph{virtual nodes} (v-nodes), each corresponding to one of $v$'s local boundaries. The v-node associated with $v$'s local boundary $B$ is denoted by $v(B)$ and has \emph{boundary count} $c(v(B)) = c(v,B)$. The \emph{clockwise} successor (resp., \emph{predecessor}) \emph{v-node} of $v(B)$ is defined as the v-node $v'(B')$ satisfying (1) $v'$ is the clockwise successor (resp., predecessor) point of $v$ w.r.t. $B$ and (2) there exist two unique edges $e \in B, e' \in B'$ with a common unoccupied endpoint $u$. Point $u$ is said to be the \emph{common point} of $v(B)$ and $v'(B')$. (Note that $v$ and $v'$ have exactly two common adjacent points and $u$ is one of them.)
Any v-node has a successor (resp., predecessor) v-node and can compute their common point, by Observation \ref{obs:existenceOfSuccessorAndPredecessor}.
Furthermore, note that $B$ and $B'$ correspond to the same global boundary $\tilde{B}$ -- the one that borders the face containing the common point of $v(B)$ and $v'(B')$. The counter-clockwise successor and predecessor v-nodes of $v(B)$ are defined similarly.

Let $\tilde{B}$ be a global boundary. Then, the first corresponding (oriented) virtual ring consists of all v-nodes $v(B)$ such that $B$ is part of $\tilde{B}$, and its edges are directed from a v-node to its clockwise successor. 
This ring is oriented clockwise if $\tilde{B}$ is the global outer boundary, and counter-clockwise otherwise -- see Figure \ref{fig:exampleBoundaryTraversal}.
The second ring is defined similarly except that its edges go from a v-node to its counter-clockwise successor (and thus has the inverse orientation). 
 
 \begin{observation}
\label{obs:existenceOfSuccessorAndPredecessor}
Any v-node $v(B)$ has a clockwise successor (resp., predecessor) and their common point is the other endpoint of the last (resp., first) edge of $B$.
\end{observation}

\begin{observation}[\cite{DGRSS17a}]
\label{obs:sumCounts}
Consider any of the two (oriented) rings corresponding to the outer boundary (resp., an inner boundary). Then the sum of counts of the ring's v-nodes is 6 (resp., -6).
\end{observation}

\paragraph*{Erodable Points} 
A \emph{redundant point} $v \in S$ is a point whose removal does not disconnect its 1-hop neighborhood (in $S$). If $v$ is also on the outer boundary of $S$, then $v$ is an \emph{erodable point} w.r.t. $S$ -- see Figure \ref{fig:exampleCounts}. In which case, $v$ has a single local boundary $B$ in $S$, by Proposition \ref{prop:EquivalenceErodableSingleLocalOuterBoundary}. If, in addition, $v$ is strictly convex w.r.t. $B$, then $v$ is said to be \emph{strictly convex and erodable} (\emph{SCE}) w.r.t. $S$. We show below that if $S$ is simply-connected then it is possible to remove SCE points iteratively (call this an ``erosion process'') until a single point remains (Observation \ref{obs:noDisconnection} and Proposition \ref{lem:AllConvexOuterBoundaryNodesErodable} below).

\begin{observation}
\label{obs:noDisconnection}
For an arbitrary simply-connected shape $S$ (with at least two points) and an erodable point $v \in S$, the shape $S \setminus \{v\}$ is simply-connected.
\end{observation}

\begin{proposition}
\label{prop:EquivalenceErodableSingleLocalOuterBoundary}
A point $v \in S$ is erodable if and only if $v$ has a single local boundary $B$, and $B$ is a local outer boundary.
\end{proposition}

\begin{proof}
Assume, by contradiction, that some redundant point $v\in S$ has at least two local boundaries $B_1,B_2$. For $i \in \{1,2\}$, consider an arbitrary edge in $B_i$ and denote its other endpoint by $u_i$ (where $u_i \notin S$). Since $u_1$ and $u_2$ are not adjacent, the removal of $v$ would separate the 1-hop neighborhood of $v$, contradicting the fact that $v$ is redundant. Because a point that has a single local boundary is (trivially) redundant, a point is redundant if and only if it has a single local boundary. The statement follows from the definition of an erodable point.
\end{proof}

\begin{proposition}
\label{lem:AllConvexOuterBoundaryNodesErodable}
If $S$ is simply-connected and has at least two points then it has at least one SCE point (w.r.t. $S$). 
\end{proposition}

\begin{proof}
Consider the clockwise oriented ring of v-nodes on the outer boundary of $S$. It is easy to show that there exists a point $v$ on the outer boundary of $S$ and a path $\mathcal{P} = v(B),u_1(B_1), \ldots, u_k(B_k),v(B')$ on the ring such that $\mathcal{P}$ is of length at least 3 (i.e., $k \geq 1$) and such that for any strictly positive integer $i$, $u_i \neq v$ and in addition, for any strictly positive integer $j \neq i$, $u_i \neq u_j$.  (Intuitively, $v$ is the only point this path encounters twice.) The set of points $v,u_1,\ldots,v$ form a simple polygon (i.e., non-intersecting) in the Euclidean plane. (Note that not all points are vertices of the polygon, since some of the points may be in a straight line.)  Hence, the sum of this polygon's exterior angles is $360 \degree$. Assume $v$ is a vertex of the polygon. Since the polygon is non-intersecting, $v$ has an exterior angle of at most $180 \degree$. After removing $v$'s exterior angle, the remaining sum of exterior angles is at least $180 \degree$ (if $v$ is not a vertex, the sum is $360 \degree$). For every point $u_i$ that is also a vertex (for $i \geq 1$), the count of $u_i(B_i)$ is the exterior angle of $u_i$ divided by $60 \degree$. Thus, there exists $k \geq 1$ such that $u_k(B_k)$ has strictly positive count, or in other words, $u_k$ is strictly convex w.r.t. one of its local (outer) boundaries. In addition, since the path only encounters $u$ once, $u$ has exactly one local outer boundary. Finally, since $S$ is simply-connected, $u$ does not have any other local boundaries and thus is erodable by Proposition \ref{prop:EquivalenceErodableSingleLocalOuterBoundary}.
\end{proof}

\subsection{System Definitions}
\label{subsec:system}

In the amoebot model, particles occupy (at most two) points of the triangular grid. 
A particle that occupies two points is \emph{expanded}, whereas one which occupies a single point is \emph{contracted} -- see Figure \ref{fig:particleMovement}.
Two particles that occupy adjacent points are said to be neighboring particles. For any particle $p$, the set of its neighboring particles is denoted by $\mathcal{N}(p)$.
To communicate, particles write to their own local memory, and read from their neighboring particles' memory.
Importantly, a particle stores in its memory whether it is contracted or expanded. 
Due to the memory constraint, particles do not have unique identifiers.  
Instead, each particle $p$ relies on a cyclical ordering of (an occupied point's) incident edges (i.e., port numbers in $\{0,\ldots,5\}$) to distinguish between neighboring points. That ordering's (clockwise or counter-clockwise) direction is referred to as the \emph{chirality} of that particle. This work makes the common assumption~\cite{DGRSS16,DDGPRSS18,PR18,DGSBRS15,DGRSS17a,BB19,GAMT18} that particles have common chirality -- taken to be, without loss of generality, the clockwise direction.
Moreover, for any particle $p$ occupying some point $u$ and any point $v$ adjacent to $u$, let $port(p,u,v)$ denote the port $p$ assigns to $v$ from $u$.
This work makes the common assumption~\cite{EKLM19,BB19,DGRSS16,DFPSV18} that, for any two neighboring particles $p$ and $q$ occupying two adjacent points, respectively $u$ and $v$, $p$ knows $port(q,v,u)$.

\paragraph*{Movements and Connectivity Assumptions} Particles can move by expanding or contracting along the edges of the triangular grid. 
More concretely, a contracted particle, occupying some point $v$, can expand into any empty adjacent point $u$. 
The particle now occupies both point $v$ (said to be its \emph{tail}) and point $u$ (said to be its \emph{head}) -- see Figure \ref{fig:particleMovement}.
(If a particle occupies a single point $v$, then $v$ is both the tail point and the head point of the particle.) As for an expanded particle occupying two adjacent points $u$ and $v$, it can contract into either $u$ or $v$.
Finally, two neighboring particles $p$ and $q$, where $p$ is contracted and $q$ is expanded, can participate in a \emph{handover} (performed by either $p$ or $q$) where $p$ expands into a point previously occupied by $q$ and $q$ becomes contracted. 

The set of occupied points (or also, by abuse of notation, the corresponding induced subgraph) is said to be the \emph{shape of the particle system}. It is commonly required \cite{EKLM19,DFSVY20shapeformation,CDGRR19,CDRR16} that the particle system maintains system-wide connectivity: that is, an algorithm must not move a particle such that the particle system's shape becomes disconnected. However, in this work, temporary disconnection is allowed: we only require that the particle system's shape is connected at the beginning and at the end of an algorithm.

\paragraph*{Additional Definitions} The \emph{state} of a particle $p$ consists of $p$'s local memory and whether $p$ is contracted or not. A state from which $p$ does nothing, when activated, is said to be a \emph{final state}. The \emph{configuration} of the particle system consists of the particle system's shape and of all particles' states and occupied points.
A configuration is said to be \emph{connected} if the particle system's shape is connected, \emph{non-empty} if the particle system is non-empty, and \emph{contracted} if all of the particles are contracted.
A \emph{problem} is defined by a set of permitted initial configurations, particle outputs (i.e., problem-specific variables in particles' memories), and a problem predicate (i.e., a predicate on the system's configuration). As is common in the literature~\cite{EKLM19,DFSVY20shapeformation,DGRSS16,DGSBRS15},  the set of permitted initial configurations here consists exactly of connected, non-empty contracted configurations. 

The particle system progresses through a sequence of atomic particle \emph{activations}. An activated particle executes the following 3 actions in order: (i) it reads the memories of its neighbors, (ii) it performs some arbitrarily bounded computations, and updates its local memory as well as its neighbors' memories, and (iii) it may execute a single movement operation (described above).\footnote{Note that if the activated particle is in a final state, then it performs none of the 3 steps.} An \emph{execution fragment} is a sequence $C_0,a_1,C_1,\ldots$ of configurations alternating with activations such that configuration $C_i$ is obtained by applying activation $a_i$ to $C_{i-1}$. If additionally, $C_0$ is a permitted initial configuration, then the execution fragment is called an \emph{execution}. An execution is said to be \emph{fair} if each particle is activated infinitely often. An \emph{asynchronous round} is a minimal execution fragment in which each particle is activated at least once. An algorithm \emph{terminates} if, for any fair execution, all particles reach a final state. An algorithm \emph{solves} problem $\mathcal{P}$ if it terminates and any fair execution reaches a configuration that satisfies $\mathcal{P}$'s predicate. The \emph{round complexity} of an algorithm (solving some problem $\mathcal{P}$) is the number of rounds needed until it terminates, in the worst-case.

\paragraph*{Notations.} For an arbitrary execution and configuration $C_t$ (i.e., the configuration obtained after $t$ activations in the execution), the particle system's shape in $C_t$ is denoted by $S_{P}(C_t)$ and its area by $S_{A}(C_t)$.
When evident, we omit $C_t$ and use the simpler notations $S_P$ and $S_A$. 
The eccentricity of point $v \in S_P$ w.r.t. $G$ in the initial configuration $C_0$ is denoted by $\epsilon_G(v)$. 
The diameters of $S_P$ w.r.t. $dist_{S_P}$, $dist_{S_A}$, and $dist_G$ in $C_0$ are respectively denoted by $D$, $D_A$ and $D_G$ (where $D_A = O(D)$, see (1) of Observation \ref{obs:parameterObservations}). The number of points of $S_P$ and $S_A$ in $C_0$ are denoted by $n$ and $n_A$ (where $n_A = O(D_A\!{}^2)$, see (2) of Observation \ref{obs:parameterObservations}).

\subsection{Problem Definitions}
\label{subsec:prob-defs}



\paragraph*{Leader Election and Disconnecting Leader Election} The output of a particle $p$ is the variable $p.status$, which can take values in $\{undecided, leader, follower\}$. The predicate of the \emph{disconnecting leader election} problem (DLE) is satisfied if there exists a unique particle with the leader state, and all other particles are in the follower state. The predicate of the \emph{leader election} problem is satisfied if the particle system is connected and the predicate of DLE is satisfied.

\section{Some of the Technical Ideas}

The main results are described in the very next section in Subsections~\ref{subsec:DisconnectingLeaderElection} and~\ref{subsec:AnalysisDLE}. The current section is intended for an informal description of the technical ideas behind the {\em additional} results. That is, an informal description of the process by which the particle system is reconnected (if this is desired) efficiently after algorithm \DLE~ terminates is given in Subsection \ref{subsec:inf-rec}. Intuition about the removal of the assumption that the outer boundary is known is given in Subsection \ref{subsec:inf-boundary}. 
The much more detailed description of the assumption's removal is discussed in length later in Section \ref{sec:boundaryDetection}.

\subsection{Informal Description of the Reconnection Algorithm}
\label{subsec:inf-rec}

After executing Algorithm DLE, the particle system may be disconnected. As opposed to a general disconnected particle system, we prove in Lemma \ref{lem:betterParticleDistribution} that Algorithm DLE leaves particles at some points at every grid distance from the eventual leader (up to the furthest distance).
This allows Algorithm $\Collect$ to collect the particles in phases.
Intuitively, in the first phase, the leader particle collects its neighboring particles; by the lemma it collects at least one (besides itself). Then, intuitively, $i$ particles are enough to collect every particle at distance $2i$ from the leader, 
by rotating around the leader, not unlike a blade of a fan. By the lemma, they collect $2i$ particles (including themselves) unless they have reached the furthest particle already. Various technicalities are still encountered.

\subsection{Intuition Regarding the Removal of the Known Outer Boundary Assumption}
\label{subsec:inf-boundary}

The algorithm to detect the outer boundary is given in this paper for completeness, to show that the known boundary assumption can be removed. This primitive is a speed-up-by-pipelining version of the similar primitives of \cite{BB19,EKLM19}. Like them, it uses mostly classical distributed computing methods in the sense that no particle moves during its execution.

Following \cite{BB19}, the primitive uses the sum of the boundary counts (of the points on each boundary) to decide whether this is an outer boundary. This sum is positive iff this is the outer boundary. The difficulties arise from the anonymity and from the fact that the memory of each particle is constant. For example, it is not easy to collect the sum to one particle,
since some  partial sums may be larger than the memory of the particle. 
As in \cite{BB19,EKLM19}, up to 6 leaders are elected on each boundary, so that each can perform the sum starting from itself. For simplicity, assume for now that a global boundary visits each particle at most once (in other words, each particle has at most one local boundary that is part of that global boundary). 
During the election, sets of consecutive particles along a boundary form segments. The particles of one segment $s_1$ may have a different collection of boundary counts than the particles of another segment $s_2$. The counts are compared, and one of these segment loses.
Each such segment has its own leader (initially, everybody is the leader of a segment including only itself). Eventually, at most 6 segments remain. 

 We manage to save in runtime compared to \cite{BB19,EKLM19} by using pipelining. Each particle produces a token (mobile agent) carrying its sum, and the collection of mobile agents of one segment is pipelined into the other segment. A complication arises from the fact that segments may change during the comparison. Previous algorithms ``froze'' the segments -- that is, blocked any changes -- during comparison, creating a runtime bottleneck. Here, the segment initiating the comparison cannot grow during the comparison, while the other one can. 
 Hence, if the initiating segment is smaller, it remains smaller even if the other segment grew in the meantime. We thus have the smaller segment win. Moreover, we show that the time for that win is proportional to the size of the smaller segment. (Additional time is used later for the winning segment to ``absorb'' the particles of the losing one.)  After the election of leaders, careful cancellation of negative partial sums with positive ones allows the boundary leader to decide whether the boundary is an outer one. In this case, the termination is announced using flooding, to save time.

\section{Leader Election}
\label{sec:LeaderElection}

In this section, an $O(D_A)$ round leader election algorithm  is presented in two parts. The first part -- Sections \ref{subsec:DisconnectingLeaderElection} and \ref{subsec:AnalysisDLE} -- presents and analyzes Algorithm DLE, an $O(D_A)$ round leader election algorithm. However, when Algorithm DLE terminates, the particle system may be disconnected. 
The second part -- Section \ref{subsec:ReconnectingLeaderElection} -- reconnects the particles system for the sake of later algorithms that may require an initial connected configuration. The runtime of the reconnecting algorithm is  $O(D_G)$. When executed after the leader election algorithm, it ensures that upon termination, the particle system is connected.

\subsection{Disconnecting Leader Election (DLE)}
\label{subsec:DisconnectingLeaderElection}

Algorithm DLE starts in a permitted initial configuration $C_0$ and carries out an erosion process starting from the area of the particle system in $C_0$, denoted by $S_A(C_0)$.
Define the set $S_e^t$ of \emph{eligible} points to contain initially all the points in the area $S_A(C_0)$. 
The algorithm sometimes marks a point $v$ ineligible. Define $S_e^t$ in the resulting configuration $C_t$ to be $S_e^{t-1}$ (i.e., $S_e$ before executing this operation) minus $\{v\}$. Neither the algorithm nor the definition of $S_e^t$ include any operation of adding a point into $S_e$.  Henceforth, we do not mention $t$ when it is clear from the context.  

\begin{observation}
\label{obs:nonEligibleRemainsNonEligible}
During the execution of Algorithm DLE, a non-eligible point does not become eligible. In particular, once a point becomes non-eligible, it remains non-eligible.
\end{observation}

The pseudo code is given below. Informally, the algorithm proceeds by an activated particle $p$ who occupies an SCE (w.r.t. $S_e$) eligible point $v$, 
making $v$ ineligible. For the sake of the analysis, recall that this removes $v$ from $S_e$; $v$ is said to \emph{leave} $S_e$. 
(If possible, $p$ expands out of $v$ into another eligible point, otherwise, $p$ remain in $v$, but marks $v$ as ineligible (using the variables of neighboring particles);
Moreover, a particle keeps track of its ineligible unoccupied neighboring points.)
Eventually, a single point remains in $S_e$, occupied by some particle $p_l$ who becomes the leader.

\paragraph*{Inputs and Variables.} 
The known outer boundary assumption (removed later) is defined using an  input (read-only) variable $p.outer[0..5]$ at every particle $p$.  Let $u$ be the point reached via port $i$ (for $i \in \{0,\ldots,5\}$) of $p$ in the initial configuration.
Specifically, the (temporary) assumption is that $outer[i] = true$ iff point $u$ is in the outer face of the particle system's shape initially. 
The $p.status$ variable is the leader election output and takes value in $\{undecided,leader,follower\}$. 
The $p.eligible[i]$ variable,
for any $i \in \{0,\ldots,5\}$, indicates whether the point $u$ reached via port $i$ of $p$'s head is eligible or not. 
Initially, $eligible[i]$ is set to $true$ if the point reached via port $i$ is either occupied or $p.outer[i]= false$.

\begin{definition}
The $eligible$ variable of a particle $p$ is said to be \emph{consistent} if for any $i \in \{0,\ldots,5\}$, $p.eligible[i] = true$ if and only if the point $u$ reached via port $i$ of $p$'s head is in $S_e$.
\end{definition}

\begin{algorithm}
\caption*{\textbf{Algorithm DLE.} \enspace Solution for the disconnecting leader election problem}
\begin{algorithmic}[1]
\State \textbf{Input:}
\State $p.outer$: a boolean array of length 6  \Comment{Indicates which neighbors are initially outside the system's shape.} 

\State
\State \textbf{Initialization:}
\State $p.status := undecided$
\For{$i \in \{0,\ldots,5\}$} $p.eligible[i] := (p.outer[i] = false)$   \label{line:initializationDLE} \Comment{True for occupied or hole neighbors}
\EndFor

\State
\State \textbf{During the atomic activation of particle $p$:}
\If{$p$ is expanded} $p$ contracts into its head. \label{alg:contractionDLE}
\ElsIf {$p.status \neq undecided$ and $\forall q \in \mathcal{N}(p)$, $q.status \neq undecided$} 
	\State $p$ terminates  \Comment{If $p$ and all of its neighboring particles have decided, then $p$ terminates.} \label{alg:terminationDLE}
\ElsIf{$p.status = undecided$}  
	\State \emph{// $p$ is contracted and occupies some point $v$}
	\If{$v$ has no adjacent points in $S_e$}  \label{alg:singleLeaderCandidateCondition} \Comment{According to $p.eligible[0..5]$}
		\State $p.status := leader$ \label{alg:leaderElected}
	\ElsIf{$v$ is an SCE point w.r.t. $S_e$}   \Comment{And otherwise, do nothing.}  \label{alg:startOfErosionBlock}
		\State \emph{// $p$ removes $v$ from $S_e$: }    \label{alg:leaveSe}
		\For{every particle $q \in \mathcal{N}(p)$} \label{alg:notify1}
			\If{$q$'s head point $w$ is adjacent to $v$} $q.eligible[port(q,w,v)] := false$ \label{alg:notify2}
			\EndIf
		\EndFor
		\State \emph{// Afterward, $p$ keeps the outer boundary of $S_e$ occupied by moving if necessary.}
		\If{$v$ has an adjacent empty point $u$ in $S_e$}   \Comment{By Claim \ref{claim:SingleAdjacentEmptyNode}, there is exactly one such point.}
			\State \emph{// Set the neighborEligible flags in preparation for $u$: only that corresponding to $v$ is set to false.} \label{alg:startOfExpansionBlock}
			\State $i_v := port(p,v,u) + 3 \; (mod \; 6)$       \Comment{Once $p$ expands, $port(p,u,v) = i_v$.}
			\For{$i \in \{0,\ldots,5\} \setminus \{i_v\}$}  $\; p.eligible[i] := true$   \label{alg:modify3}
			\EndFor
			\State $p.eligible[i_v] := false$  \label{alg:modify4}
			\State $p$ expands into $u$ \label{alg:endOfExpansionBlock}
		\Else
			\State $p.status := follower$    \label{alg:endOfErosionBlock}
		\EndIf
	\EndIf
\EndIf
\end{algorithmic}
\end{algorithm}

\subsection{Analysis of Algorithm DLE}
\label{subsec:AnalysisDLE}

\subsubsection{Correctness}
\label{subsec:correctnessDLE}

First we prove a geometric claim and an invariant on Algorithm DLE (see Claim \ref{claim:SingleAdjacentEmptyNode} and Lemma \ref{lem:EligibleLemma}). 

\begin{claim}
\label{claim:SingleAdjacentEmptyNode}
Consider a configuration $C$ reached by Algorithm DLE such that the boundary points of $S_e$ are occupied.
In $C$, every SCE point $v$ w.r.t. $S_e$ has at most one adjacent, empty point in $S_e$. Moreover, if the SCE point $v$ has an adjacent empty point in $S_e$, then $v$ has a boundary count of 1 w.r.t. $S_e$.
\end{claim}

\begin{proof}
Consider such a point $v$, if it exists. Since $v$ is erodable w.r.t. $S_e$, $v$ has a single local boundary $B$ in $S_e$, and $B$ is a local outer boundary (by Proposition \ref{prop:EquivalenceErodableSingleLocalOuterBoundary}). In addition, $v$ is strictly-convex w.r.t. $B$, and thus, $v$ has at most 3 neighbors in $S_e$. If $v$ has exactly one neighbor $z$ in $S_e$, $z$ is trivially a boundary point of $S_e$. Thus, $z$ is occupied and the statement follows. Otherwise, there exists two neighbors $u,w \in S_e$ of $v$, such that $u,w$ are respectively adjacent to some other neighbors $y,y' \notin S_e$ of $v$. Since $u$ and $w$ are boundary points of $S_e$, they are occupied and the statement follows.
\end{proof}

\begin{lemma}
\label{lem:EligibleLemma}
In every configuration reached during a fair execution of Algorithm DLE, the following holds:
\begin{enumerate}
	\item For any expanded particle, its head (resp., tail) is in $S_e$ (resp., not in $S_e$).
	\item $S_e$ is simply-connected and non-empty.
	\item The boundary points of $S_e$ are occupied.
	\item The $eligible$ variables of all particles are consistent.
\end{enumerate}
\end{lemma}

\begin{proof}
Let us prove this statement by induction. The initial configuration $C_0$ is contracted so (1) holds vacuously. Since $S_P(C_0)$ is non-empty and connected, and $S_e^0 = S_A(C_0)$, (2) also holds. Part (2) implies that all boundary points of $S_e^0$ are in the outer boundary of $S_e^0$. These points are occupied, by definition of $S_A(C_0)$, so (3) holds. Now, note that all particles are contracted in $C_0$ and that in the initialization of Algorithm DLE (see line \ref{line:initializationDLE}), a (contracted) particle occupying an interior (resp., boundary) point of $S_e^0$ sets all $eligible$ flags to $true$ (resp., except for those that lead to points not in $S_e^0$, using the input). Since the boundary points of $S_e^0$ are exactly the points that are adjacent to non-eligible points (i.e., not in $S_e^0$), (4) holds.

Now, consider a configuration $C_t$ of the execution that satisfies the induction hypothesis and an activated particle $p$, and let us show that the statement holds for the resulting configuration $C_{t+1}$. \\
First, assume that $p$ does not enter the code block from line \ref{alg:startOfErosionBlock} to line \ref{alg:endOfErosionBlock}. Then, $p$ either terminates, becomes the leader, does nothing, or contracts to its head (when $p$ is expanded). Since (1),(2),(3), and (4) hold for $C_t$ (and for the last case, the consistency of $p.eligible$ depends only on $p$'s head), (1),(2),(3), and (4) also hold for $C_{t+1}$.
 Otherwise, in $C_t$, $p$ is contracted and occupies some point $v$. Since (4) holds in $C_t$, $p.eligible$ is consistent. Therefore, $v$ is SCE w.r.t. $S_e^t$. 
 (Note that $p$ has enough information to decide the condition in line \ref{alg:startOfErosionBlock}.)
 During $p$'s activation, $v$ leaves $S_e^t$ (see line \ref{alg:leaveSe}): $S_e^{t+1} = S_e^{t} \setminus \{v\}$. Moreover, if $p$ expands, it expands into some eligible point (see line \ref{alg:endOfExpansionBlock}). Thus, (1) holds in $C_{t+1}$. 
Since $v$ is SCE w.r.t. $S_e^t$, $S_e^{t+1}$ is simply-connected and non-empty, and (2) holds in $C_{t+1}$. Let us next show that (3) holds in $C_{t+1}$. Note that since $v$ leaves $S_e^t$, all the neighbors of $v$ in $S_e^t$ are boundary points in $S_e^{t+1}$. 
Now, from the consistency of $p.eligible$ in $C_t$, particle $p$ executes line \ref{alg:endOfErosionBlock} if $v$'s neighbors in $S_e^t$ are all occupied, and the code block from line \ref{alg:startOfExpansionBlock} to line \ref{alg:endOfExpansionBlock} otherwise. In the first case, (3) holds trivially in $C_{t+1}$. In the second case, (3) holds in $C_t$, and thus, by Claim \ref{claim:SingleAdjacentEmptyNode}, $v$ has exactly one empty adjacent neighbor in $S_e^t$, denoted by $u$. Since $p$ expands into $u$ (see line \ref{alg:endOfExpansionBlock}), (3) holds in $C_{t+1}$. Finally, let us show that (4) holds in $C_{t+1}$. First, the particles in $\mathcal{N}(p)$ whose head is adjacent to $v$ have their $eligible$ variables modified to be consistent in $C_{t+1}$ (see lines \ref{alg:notify1}-\ref{alg:notify2}). For all other particles, excluding $p$, their $eligible$ variables, without any change, are consistent in $C_{t+1}$. As for $p$ itself, if it does not expand, then $p.eligible$ does not change and is consistent in $C_{t+1}$. Otherwise, $p$ expands into some empty adjacent neighbor $u$ in $S_e^t$ (defined previously). Since (2) and (3) hold in $C_t$, $u$ is an interior point of $S_e^t$: that is, all of $u$'s neighbors are in $S_e^t$. Since $v$ is the only point to become non-eligible during the activation, $p.eligible$ is consistent in $C_{t+1}$ (see lines \ref{alg:modify3}-\ref{alg:modify4}). In summary, (4) holds in $C_{t+1}$.
\end{proof}

\begin{theorem}
\label{thm:correctness}
Algorithm DLE solves leader election.
\end{theorem}

\begin{proof}
Since $S_e$ remains connected during the execution (see Lemma \ref{lem:EligibleLemma}), when $|S_e| > 1$, no leader is elected (see lines \ref{alg:singleLeaderCandidateCondition}-\ref{alg:leaderElected} of the algorithm). Also, note that by the algorithm definition, no point joins $S_e$, so $|S_e|$ is non-increasing. Therefore, consider an arbitrary configuration reached during Algorithm DLE with $|S_e|>1$, and let us show that $|S_e|$ decreases eventually. By the algorithm definition (see lines \ref{alg:leaveSe}-\ref{alg:endOfErosionBlock}), particles decrease $|S_e|$ when they expand. Moreover, an expanded particle contracts upon activation (see line \ref{alg:contractionDLE}). Therefore, eventually either $|S_e|$ decreases, or all particles are contracted, starting from some configuration $C^*$. In the latter case, the set of SCE (w.r.t. $S_e$) points in $C^*$, denoted by $X$, is non-empty, by Proposition \ref{lem:AllConvexOuterBoundaryNodesErodable}. A contracted particle that occupies a point in $S_e \setminus X$ does nothing when activated. Therefore, when a (contracted) particle occupying some point $v \in X$ is eventually activated, $v$ is SCE w.r.t. $S_e$. During that activation, $v$ leaves $S_e$ so $|S_e|$ decreases.
Finally, from the above and Lemma \ref{lem:EligibleLemma}, a single point $l$ remains in $S_e$ eventually; the particle occupying $l$ at that point  becomes the leader (see line \ref{alg:leaderElected}). 
\end{proof}

\begin{remark*}
Since $S_e$ initially contains $n_A = O(D_A\!{}^2)$ points, the above proof also suggests a naive round complexity of $O(D_A\!{}^2)$ for Algorithm DLE. In the next section, a more involved analysis leads to an $O(D_A)$ round complexity. 
\end{remark*}

\begin{remark*} An $O(D_A\!{}^2)$ round algorithm for leader election -- that maintains the connectivity of the particle system's shape at all times -- can be obtained from Algorithm DLE, by translating the ideas of \cite{DDDNP20ieee} into our work. In Algorithm DLE, particles may end up disconnected from the leader particle. Very informally, a disconnection occurs in Algorithm DLE when a particle $p$ makes a point ineligible and simultaneously marches inwards away from the boundary of $S_e$; to avoid disconnection, $p$ can then also ``pull'' some neighboring particle $q$ who occupies an ineligible point, if such a particle $q$ exists; we do not follow this approach here.
\end{remark*}

\subsubsection{Time Complexity}
\label{subsec:runtimeDLE}

The following definitions and invariants are used in the analysis. For any given execution of Algorithm DLE, consider $l$, the last eligible point occupied by the leader when the algorithm terminates (see Theorem \ref{thm:correctness}) and some configuration $C_t$, reached during the execution after $t$ activations. Then, the \emph{level sets} and \emph{closed neighborhoods} of $l$ in $S_e^t$ are respectively defined as $L_i^t = \{u \in S_e^t \; | \; dist_{S_e^t}(l,u) = i\}$ and $N_i^t = \{u \in S_e^t \; | \; dist_{S_e^t}(l,u) \leq i\}$, for all integers $i \geq 1$. (Note that if $i > D_A$, $L_i^t = \emptyset$ and $N_i^t = N_{D_A}^t$.) Moreover, some point $v \in L_i^t$ is said to be \emph{$(i,k)$-SCE-related} if there exists some SCE (w.r.t. $N_i^t$) point $u \in L_i^t$ such that $dist_{L_i^t}(v,u) \leq k$.
As a finite induced subgraph of the infinite triangular grid, $S_e^t$ is $K_4$-free. Since $S_e^t$ is, in addition, simply-connected (see Lemma \ref{lem:EligibleLemma}), it is a bridged graph \cite{CDV02}. Known results on $K_4$-free bridged graphs give Lemma \ref{lem:convexityResults}. Building upon it, the following three lemmas (Lemmas \ref{lem:invariant1}, \ref{lem:SCErodability}, and \ref{lem:distanceToSCErodable}) provide invariants on the execution of Algorithm DLE. These invariants allow us to analyze how many rounds it takes for points in $S_e^0$, excluding $l$, to become non-eligible (see Lemma \ref{lem:fastNonEligibility} and Theorem \ref{thm:LEImprovedRuntime}).  

\begin{lemma}[\cite{CDV02}]
\label{lem:convexityResults}
For any integer $i \geq 1$, the following holds:
\begin{enumerate}
	\item $N_i^t$ is \emph{convex} w.r.t. $S_e^t$: that is, for any two vertices $u,v \in N_i^t$ and for any vertex $z$ on any of the shortest paths in $S_e^t$ between $u$ and $v$, $z$ is also in $N_i^t$.
	\item $L_i^t$ does not contain three pairwise adjacent vertices.
\end{enumerate}
\end{lemma}

\begin{lemma}
\label{lem:invariant1}
For any integer $i \geq 1$, the following holds:
\begin{enumerate}
	\item $L_i^{t+1} = L_i^t \cap S_e^{t+1}$ and $N_i^{t+1} = N_i^t \cap S_e^{t+1}$, and thus, $L_i^{t'} = L_i^t \cap S_e^{t'}$ and $N_i^{t'} = N_i^t \cap S_e^{t'}$ for any integer $t' > t$.
	\item $N_i^t$ is simply-connected.
	\item For any point $v$ in $L_i^t$ and two neighbors $u, w \in L_{i-1}^t$ of $v$, $u$ and $w$ are adjacent.
	\item Any point $v$ in $L_i^t$ has at most two neighbors in $L_{i-1}^t$ and at most two neighbors in $L_{i}^t$.
	\item Points in $L_i^t$ are (outer) boundary points of $N_i^t$.
\end{enumerate}
\end{lemma}

\begin{proof}
(1) If $S_e^{t+1} = S_e^t$, (1) holds trivially. Otherwise, by the algorithm definition, $S_e^t = S_e^{t+1} \cup \{v\}$ for some point $v \in S_e^t$. For any two points $u_1,u_2 \in S_e^{t+1}$ and some shortest path between $u_1$ and $u_2$, since $v$ is SCE w.r.t. $S_e^t$, if the shortest path goes through $v$, there exists a shortest path between $u_1$ and $u_2$ that goes through the neighbors in $S_e^{t+1}$ of $v$. Thus, for any integer $i \geq 1$, $L_i^{t+1} = L_i^t$ if $v \notin L_i^t$, and $L_i^{t+1} = L_i^t \setminus \{v\}$ otherwise. As a result, (1) holds.
(2) Assume, by contradiction, that $N_i^t$ is not simply-connected. In other words, the set of hole points $H$ of $N_i^t$ is non-empty. Since $S_e^t$ is simply-connected (by (2) of Lemma \ref{lem:EligibleLemma}), all points of $H$ are in $S_e^t$. However, for any point $h \in H$, $h$ is on the shortest path between two points $v_1,v_2 \in N_i^t$ by Proposition \ref{prop:holesNodesAreOnShortestPath}.
Since $N_i^t$ is \emph{convex} w.r.t. $S_e^t$ (by (1) of Lemma \ref{lem:convexityResults}), this leads to a contradiction and (2) holds. 

(3) Points $u$ and $w$ are at distance at most 2 in $S_e^t$. Assume, by contradiction, that they are at distance exactly 2. Then, $v$ is on one of the shortest paths between $u$ and $w$. However, by (1) of Lemma \ref{lem:convexityResults}, $N_{i-1}^t$ is convex w.r.t. $S_e^t$, which leads to a contradiction. As a result, (3) holds. (4) Assume, by contradiction, that $v$ has at least three neighbors $u_1,u_2,u_3$ in $L_{i-1}^t$. By (3), $u_1,u_2,u_3$ are pairwise adjacent. As a result, $v, u_1, u_2, u_3$ form a 4-clique, which leads to a contradiction. Now, assume, by contradiction, that $v$ has at least three neighbors $w_1,w_2,w_3$ in $L_{i}^t$. Let us first show, by contradiction, that at least two points in $w_1,w_2,w_3$ are adjacent. Assume that $v$ has exactly three neighbors in $L_{i}^t$ and that no two of these neighbors are adjacent (otherwise the contradiction is trivially obtained). Since $N_i^t$ is convex w.r.t. $S_e^t$ (by (1) of Lemma \ref{lem:convexityResults}), the other three points $y_1,y_2,y_3$ adjacent to $v$ are not in $L_{i+1}^t$. Hence, $y_1,y_2,y_3$ are either in $L_{i-1}^t$ or not in $S_e^t$. Since $v \in L_i^t$ and $N_{i-1}^t$ is convex w.r.t. $S_e^t$ (by (1) of Lemma \ref{lem:convexityResults}), at least two points, without loss of generality $y_2$ and $y_3$, are not in $S_e^t$. However, $N_i^t$ is simply-connected by (2). Hence, $y_2$ and $y_3$ are points in the outer face and removing $v$ from $S_e^t$ would have disconnected $l$ from one of $w_1,w_2,w_3$, w.l.o.g. $w_3$. This contradicts with $w_3 \in N_i^t$. 
Then w.l.o.g., $v, w_1,w_2$ are pairwise adjacent, which contradicts with (2) of Lemma \ref{lem:convexityResults}, and (4) holds.
(5) By the definition of level sets, the neighbors of some point $v \in L_i^t$ are either in $L_{i-1}^t$ or $L_i^t$. Thus, by (2) and (4), (5) holds. 
\end{proof}

\begin{lemma}
\label{lem:SCErodability}
For any integer $i \geq 1$ and point $v \in L_i^t$, the following holds:
\begin{enumerate}
\item Point $v$ is convex and erodable w.r.t. $N_i^t$ (but not necessarily SCE w.r.t. $N_i^t$).
\item If $v$ becomes SCE, it remains SCE while eligible, that is, if $v$ is SCE w.r.t. $N_i^t$ then for any integer $t' > t$ such that $v \in S_e^{t'}$, $v$ is SCE w.r.t. $N_i^{t'}$.
\item Point $v$ is $(i,2i)$-SCE-related.
\end{enumerate}
\end{lemma}

\begin{proof}
(1) If $v$ had two (resp., three) local outer boundaries in $N_i^t$, then removing $v$ would have disconnected $N_i^t$. As a result, $N_i^t$ would have been separated into two (resp., three) components (with $l$ in one of them). Hence, $v$ would have a neighbor in $L_{i+1}^t$, leading to a contradiction.
Thus, by (5) of Lemma \ref{lem:invariant1}, $v$ has one local outer boundary in $N_i^t$. 
Moreover, $N_i^t$ is simply-connected (by (2) of Lemma \ref{lem:invariant1}), so $v$ does not have a local inner boundary. Thus, by Proposition \ref{prop:EquivalenceErodableSingleLocalOuterBoundary}, $v$ is erodable w.r.t. $N_i^t$.
It remains to show that $v$ is convex w.r.t. $N_i^t$. If $i=1$, this is trivial. Assume that $i > 1$. Since $v$ has a single local (outer) boundary in $N_i^t$ and at most four neighbors in $N_i^t$ (two neighbors in $L_{i-1}^t$ and two neighbors in $L_i^t$, by (4) of Lemma \ref{lem:invariant1}), $v$ is convex w.r.t. $N_i^t$ and (1) holds.
	
(2) Let $v \in L_i^t$ be some SCE point w.r.t. $N_i^t$. Since $v \in S_e^{t'}$, $v \in L_i^{t'}$ and $v$'s set of neighbors in $N_i^{t'}$ is included in $v$'s set of neighbors in $N_i^{t}$, by (1) of Lemma \ref{lem:invariant1}. Since $v$ is, in addition, convex and erodable w.r.t. $N_i^{t'}$ by (1), $v$ is SCE w.r.t. $N_i^{t'}$ and (2) holds.
(3) Intuitively, part (1) implies that any level set consists of straight line segments (each consisting of points with a boundary count of 0 w.r.t. $N_i^t$) whose extremities are strictly convex points (w.r.t. $N_i^t$). Since $L_i^t$ is contained within a hexagon of radius $i$ centered in $l$ (since $L_i^t \subseteq N_i^t$), each straight line segments can consist of at most $2i$ points (e.g., a diagonal of the hexagon), and (3) holds. 
\end{proof}

\begin{lemma}
\label{lem:distanceToSCErodable}
For any integer $i \geq 1$ and $k \geq 0$, and $(i,k)$-SCE-related point $v \in L_i^t$, all neighbors of $v$ in $L_{i+1}^t$ are $(i+1,k+1)$-SCE-related points.
\end{lemma}

\begin{proof} Consider some integer $i \geq 1$ and point $v \in L_i^t$. Let us show the statement by induction on $k$.
In the base case, $v$ is erodable w.r.t. $N_i^t$ and thus has a single local boundary $B$ in $N_i^t$ by Proposition \ref{prop:EquivalenceErodableSingleLocalOuterBoundary}. Recall that $B$ is a cyclic interval of incident edges. Consider any edge of $B$, excluding the first and last edges, such that the other endpoint $w \notin N_i^t$ is in $S_e^t$. Let us show that $w$ is SCE w.r.t. $N_{i+1}^t$.
Note that $w$ is in $L_{i+1}^t$. By definition of $w$, and since any two neighbors of $w$ in $L_i^t$ neighbor each other (by (3) of Lemma \ref{lem:invariant1}), $v$ is the only neighbor of $w$ in $L_{i}^t$. Moreover, by (4) of Lemma \ref{lem:invariant1}, $w$ has at most two neighbors in $L_{i+1}^t$.
Therefore, $w$ has at most three neighbors in $N_{i+1}^t$. Also, $w$ is erodable w.r.t. $N_{i+1}^t$ by (1) of Lemma \ref{lem:SCErodability}. Therefore, $w$ is SCE w.r.t. $N_{i+1}^t$.
It remains to consider the endpoints $w_1,w_2$ of the first and last edges of $B$. If the successor (resp., predecessor) edge's endpoint is in $S_e^t$, the above claim implies that $w_1$ (resp., $w_2$) is $(i+1,1)$-SCE-related or not in $S_e^t$. Otherwise, one can show that $w_1$ is SCE w.r.t. $N_{i+1}^t$ or not in $S_e^t$ (using (1) of Lemma \ref{lem:SCErodability} and (2) of Lemma \ref{lem:convexityResults}). Hence, the base case holds. 

Now, assume that the induction hypothesis holds for some $k \geq 0$. Let point $v \in L_i^t$ be an $(i,k+1)$-SCE-related point. Assume that $v$ is not $(i,a)$-SCE-related, for any $0 \leq a \leq k$ (otherwise, the induction hypothesis trivially implies the induction step). Since $v$ is convex and erodable w.r.t. $N_i^t$ by (1) of Lemma \ref{lem:SCErodability}, $v$ has a single local boundary (by Proposition \ref{prop:EquivalenceErodableSingleLocalOuterBoundary}). Moreover, $v$ is not $(i,0)$-SCE-related, so by (4) of Lemma \ref{lem:invariant1} and the definition of level sets, $v$ has exactly two neighbors in $L_{i-1}^t$, two neighbors $v_1,v_2 \in L_i^t$ and two neighbors $y_1,y_2$ that are either in $L_{i+1}^t$ or not in $S_e^t$.
Also, w.l.o.g., $y_1$ is adjacent to $v_1$ and $y_2$ is adjacent to $v_2$. If exactly one of $y_1,y_2$ is not in $S_e^t$, then one can show (using (1) of Lemma \ref{lem:SCErodability} and (2) of Lemma \ref{lem:convexityResults}) that the other point is SCE w.r.t. $N_{i+1}^t$, and the induction step follows. Hence, consider that both $y_1,y_2 \in L_{i+1}^t$. Since $v$ is $(i,k+1)$-SCE-related, one neighbor of $v$, w.l.o.g. $v_1$, is $(i,k)$-SCE-related. So, by the induction hypothesis, $y_1$ is $(i+1,k+1)$-SCE-related. Since $y_1$ and $y_2$ are adjacent, $y_2$ is $(i+1,k+2)$-SCE-related, and the induction step follows.
\end{proof}

\begin{lemma}
\label{lem:fastNonEligibility}
For any two integers $0 \leq j \leq D_A-1$ and $k \geq 0$, any $(D_A-j,k)$-SCE-related point $v \in L_{D_A-j}^0$ becomes non-eligible within the first $6j+2(k+1)$ rounds. 
\end{lemma}

\begin{proof} 
Let us prove the lemma statement by induction on $j$, and for each $j$, by induction on $k$. 

For the base case of $j = 0$, if $L_{D_A}^0 = \emptyset$, then the inner induction on $k$ is vacuously correct. Otherwise, let us first show that for any point $u \in L_{D_A}^0$, there exists a contracted particle $p_u$, initially occupying $u$, such that $p_u$ remains contracted and occupies $u$ while $u$ is eligible. Since $u$ is a boundary point of $N_{D_A}^0 = S_e^0$ (by (5) of Lemma \ref{lem:invariant1}), $u$ is occupied, respectively, by some particles $p_u$ in the initial configuration $C_0$ by (3) of Lemma \ref{lem:EligibleLemma}. By the definition of permitted initial configurations, $p_u$ is contracted in $C_0$. Assume by contradiction that $v$ is eligible in some configuration $C_t$ for $t > 0$ and occupied by some different particle $p_u'$. Then, $p_u$ must have expanded once. However, once a particle expands, its occupied point becomes ineligible (see lines \ref{alg:startOfErosionBlock}-\ref{alg:endOfExpansionBlock}). Since ineligible points remain so (by Observation \ref{obs:nonEligibleRemainsNonEligible}), a contradiction is reached and the claim follows. 
For $k=0$, let $v \in L_{D_A}^0$ be some $(D_A,0)$-SCE-related point. By definition, $v$ is SCE w.r.t. $N_{D_A}^0 = S_e^0$. 
Therefore, for any configuration $C_t$ ($t>0$) such that $v \in S_e^t$, $v$ is SCE also w.r.t. $N_{D_A}^t = S_e^t$, by (2) of Lemma \ref{lem:SCErodability}. Moreover, by the claim shown above, for any configuration $C_t$ ($t \geq 0$) such that $v \in S_e^t$, $v$ is occupied by a contracted particle $p_v$. Hence, when $p_v$ is first activated (within the first round of the execution), $v$ leaves $S_e$ (see lines \ref{alg:startOfErosionBlock}-\ref{alg:leaveSe}). 
Now, assume that the induction hypothesis (IH) holds for $j=0$ and some $k \geq 0$. Let $v' \in L_{D_A}^0$ be some $(D_A,k+1)$-SCE-related point (if no such point exists, the inner induction step is satisfied). By definition, there exists some SCE (w.r.t. $S_e^0$) point $u \in L_{D_A}^0$ such that $dist_{L_{D_A}^0}(v',u) \leq k+1$. If $dist_{L_{D_A}^0}(v',u) < k+1$, then the inner induction step follows from the IH on $k$. 
Otherwise, $v'$ has a neighbor $w \in L_{D_A}^0$ such that $dist_{L_{D_A}^0}(w,u) = k$. By the inner IH (for $k$), $w$ leaves $S_e$ by some configuration $C_{t'}$ within the first $2(k+1)$ rounds. If $v' \notin S_e^{t'}$, the inner induction step is satisfied. Otherwise, note that for any configuration $C_t$ ($t>0$) such that $v' \in S_e^t$, $v' \in L_{D_A}^t$ (by (1) of Lemma \ref{lem:invariant1}), and thus, $v'$ is convex and erodable w.r.t. $N_{D_A}^t = S_e^t$ by (1) of Lemma \ref{lem:SCErodability}. Because $w \notin S_e^{t'}$, for any configuration $C_t$ -- for $t \geq t'$ -- such that $v' \in S_e^t$, $v'$ is also strictly convex w.r.t. $N_{D_A}^t = S_e^t$.
Moreover, by the claim shown above, for any configuration $C_t$ ($t \geq 0$) such that $v' \in S_e^t$, $v'$ is occupied by a contracted particle $p_{v'}$. Hence, when $p_{v'}$ is activated (within one round of $C_{t'}$), $v'$ leaves $S_e$ (see lines \ref{alg:startOfErosionBlock}-\ref{alg:leaveSe}).

Next, assume that the induction hypothesis holds for some integer $0 \leq j < D_A-1$, for every integer $k \geq 0$. If $L_{D_A-(j+1)}^0 = \emptyset$, the induction step is vacuously correct. Otherwise, let us prove the induction for $j+1$ by induction on $k$. For $k = 0$, let $v \in L_{D_A-(j+1)}^0$ be some $(D_A-(j+1),0)$-SCE-related point. By definition, $v$ is SCE w.r.t. $N_{D_A-(j+1)}^0$.  By Lemma \ref{lem:distanceToSCErodable}, every neighbor of $v$ in $L_{D_A-j}^0$ is $(D_A-j, 1)$-SCE-related.
Therefore, by the IH (for $k=1$ and $j$), every neighbor of $v$ in $L_{D_A-j}^0$, if there is any, leave $S_e$ by some configuration $C_{t'}$ within the first $6j+4$ rounds. If $v \not\in S_e^{t'}$, then the base case is satisfied. Otherwise, since for any integer $t \geq t'$ such that $v \in S_e^t$, $v$ is SCE w.r.t. $N_{D_A-(j+1)}^{t}$ (by (2) of Lemma \ref{lem:SCErodability}) and all of $v$'s neighbors in $L_{D_A-j}^0$ (if there are any) are not in $S_e^{t}$. Hence, $v$ is also SCE w.r.t. $S_e^{t}$. 
Thus, $v$ is an outer boundary point in $C_{t'}$ by Proposition \ref{prop:EquivalenceErodableSingleLocalOuterBoundary}, and so $v$ is occupied by some particle $p$ in $C_{t'}$ (by (3) of Lemma \ref{lem:EligibleLemma}). If $p$ is contracted in $C_{t'}$, then when $p$ is activated (within one round of $C_{t'}$), $v$ leaves $S_e$ (recall that $v$ is SCE w.r.t. $S_e^{t}$, and see lines \ref{alg:startOfErosionBlock}-\ref{alg:leaveSe}). Otherwise ($p$ is expanded in $C_{t'}$), $v$ is the head of $p$ (by (1) of Lemma \ref{lem:EligibleLemma}). Within two rounds of $C_{t'}$, $p$ is activated twice. Thus, $p$ contracts into $v$, and $v$ leaves $S_e$ (recall that $v$ is SCE w.r.t. $S_e^{t}$, and see lines \ref{alg:contractionDLE} and \ref{alg:startOfErosionBlock}-\ref{alg:leaveSe}). In both cases, $v$ leaves $S_e$ within the first $6j+6 \leq 6(j+1)+2$ rounds.

Finally, assume that the IH holds for $j+1$ for some $k$, and also holds for $j$. Let $v' \in L_{D_A-(j+1)}^0$ be some $(D_A-(j+1),k+1)$-SCE-related point (if no such point exists, the induction step is satisfied). By definition, there exists some SCE (w.r.t. $N_{D_A-(j+1)}^0$) point $u \in L_{D_A-(j+1)}^0$ such that $dist_{L_{D_A-(j+1)}^0}(v',u) = k+1$. If $dist_{L_{D_A-(j+1)}^0}(v',u) < k+1$, then the induction step follows from the IH on $j+1$ and $k$. 
Else, by Lemma \ref{lem:distanceToSCErodable}, every neighbor of $v'$ in $L_{D_A-j}^0$ is $(D_A-j, k+2)$-SCE-related.
By the IH (for $j$ and $k+2$), every neighbor of $v'$ in $L_{D_A-j}^0$, if there is any, leave $S_e$ within the first $r_1 = 6j+2(k+3)$ rounds. Additionally, since $dist_{L_{D_A-(j+1)}^0}(v',u) = k+1$, $v'$ has a neighbor $w \in L_{D_A-(j+1)}^0$ such that $dist_{L_{D_A-(j+1)}^0}(w,u) = k$. By the IH (for $j+1$ and $k$), $w$ leaves $S_e$ within the first $r_2 = 6(j+1)+2(k+1)$ rounds. (Note that $r_2 \geq r_1$.) Consequently, there exists a configuration $C_{t'}$ within the first $r_2$ rounds such that $w$ and all neighbors of $v'$ in $L_{D_A-j}^0$, if there are any, are not in $S_e^{t'}$. 
If $v' \not\in S_e^{t'}$, then the induction step is satisfied. Otherwise, it is now just left to show that for any $t \geq t'$ such that $v \in S_e^t$, $v'$ is SCE w.r.t. $S_e^{t}$. Because this implies, similarly to the base case (for $k=0$ and $j$), that $v$ leaves $S_e$ within the first $6(j+1)+2(k+2)$ rounds.
To prove that, note that $v \in L_{D_A-(j+1)}^t$ (by (1) of Lemma \ref{lem:invariant1}). Thus, $v$ is convex and erodable w.r.t. $N_{D_A-(j+1)}^{t}$ (by (1) of Lemma \ref{lem:SCErodability}). Note also that $w$ and all neighbors of $v$ in $L_{D_A-j}^0$, if there are any, are not in $S_e^{t}$ (by the definition of $t$ and the IH). Hence $v$ is SCE w.r.t. $S_e^{t}$. 
\end{proof}

\begin{theorem}
\label{thm:LEImprovedRuntime}
Algorithm DLE terminates in $O(D_A)$ rounds.
\end{theorem}

\begin{proof} 
By (3) of Lemma \ref{lem:SCErodability}, every point is $(D_A,2 D_A)$-SCE-related. Hence, all points initially in $S_e$, excluding $l$, leave $S_e$ in the first $O(D_A)$ rounds by Lemma \ref{lem:fastNonEligibility}. Let $C^*$ be the first configuration reached in which $l$ is the last eligible point. Hence, $l$ is an outer boundary point of $S_e$. Thus, $v$ is the head of some occupying particle $p_l$ by (1) and (3) of Lemma \ref{lem:EligibleLemma}. If $p_l$ is contracted in $C^*$, then when $p_l$ is activated (within one round of $C^*$), $p_l$ becomes the leader (see line \ref{alg:leaderElected}). Otherwise, $p_l$ is activated twice within two rounds of $C^*$. During these activations, $p_l$ contracts and becomes the leader (see lines \ref{alg:contractionDLE} and \ref{alg:leaderElected}).
\end{proof}


\subsection{Reconnecting Leader Election}
\label{subsec:ReconnectingLeaderElection}

During Algorithm DLE, the particle system may become disconnected. However, the particle system, even if disconnected, satisfies an important property w.r.t. the grid distance (i.e., distance w.r.t. $dist_G$) -- see Lemma \ref{lem:betterParticleDistribution} below.
Leveraging this property, Algorithm $\Collect$ below collects all of the particles gradually such that eventually, the system is connected. 

On the way to a connected system, let us define the notion of \emph{collected} particles inductively. Intuitively, the set of collected particles may not be connected, but this set is already arranged such that it will be easy for the algorithm to connect it later.
  Initially, the leader $p_l$ (who occupies the last eligible point $l$ in Algorithm DLE) is the only \emph{collected} particle. The algorithm executes in phases. In each phase $i \geq 1$, the particles within grid distance $2^{i-1}-1$ from $l$ are already \emph{collected}, and they cooperate in order to collect particles up to grid distance $2^{i}-1$. In steps (1) and (2) of the more detailed algorithm description (Section \ref{subsec:algorithmCollectDescription}), we describe the  \emph{collect} action that turns uncollected particles into \emph{collected} ones. Intuitively, such an action is applied in some of the cases that a collected particle moves and touches an uncollected one. Moreover, the algorithm now adds the newly collected particle into the structure of collected ones such that it will be easy to reconnect.  
Eventually, some phase is executed with no additional particles collected.

When that happens, the algorithm performs the, by now, easy operation of reconnecting the collected particles. Then the leader terminates and the particle system is connected.

Very informally, the intuition for a phase is as follows: in phase $i$, we would have liked to arrange the collected (contracted) particles into a straight line of length $2^{i}$ leading from the leader outward. Then, having the line simply rotate around $l$ -- just like the blade of a fan -- would have collected every particle up to grid distance $2^{i}-1$. After which, we would want to use Lemma \ref{lem:betterParticleDistribution} to show that enough particles are collected during phase $i$ and (earlier phases) to perform phase $i+1$. The actual implementation of Algorithm $\Collect$ is somewhat more involved than that. Three complications that come up in that implementation are described next. The first arises from the fact that there may not be enough collected particles to form a blade of length $2^i$ in phase $i$. (Luckily, Lemma \ref{lem:betterParticleDistribution} ensures that we have collected at least one half of that number, since we have already collected all the particles up to grid distance $2^{i-1}-1$ from $l$; hence, we overcome the above mentioned shortage by having a blade of only length $2^{i-1}$ -- one half of the desired length, but letting that shorter blade rotate only over grid distances $2^{i-1}$ till $2^{i}-1$; unfortunately, this make the description of the algorithm more complex.) The second complication consists in dealing with the collected particles that are not used to form the blade. Where can we put the additional particles -- collected in earlier phases -- beyond the $2^{i}$ particles needed for the blade? (Intuitively, the algorithm organizes these ``extra'' particles at points within distance $2^{i-1}-1$ from $l$, such that they do not impact the blade's rotation yet remain easily accessible.) Similarly, how does the algorithm handle the particles collected during phase $i$ until the end of the phase? (Intuitively, those ``extra'' particles are ``hung'' behind the blade -- that is, counter-clockwise from the blade -- and rotate with it.) The last complication appears during the rotation. Indeed, the blade particles try to move in parallel; however, an uncollected particle delays the blade particle collecting it. (This is avoided, intuitively, by switching roles between the collecting particle and the collected one: the first does not proceed as a part of the blade, instead, it is hung behind the blade; the newly collected particle becomes a part of the blade.)  The ``hanging behind the blade'' procedure hinted at above is the reason that in the description below we abandon the blade metaphor and speak, instead, of a stem of a tree, where the ``hung'' particles are referred to as branches.

\subsubsection{Properties of the Disconnected Particle System upon Termination of Algorithm DLE}
\label{subsec:disconnectedSystem}

Algorithm DLE allows the particle system to disconnect. However, since $S_e$ remains connected throughout Algorithm DLE, particles are left behind not arbitrarily, but in an incremental manner. Intuitively, the way particles are left behind resemble a trail of ``breadcrumbs'' -- see Lemma \ref{lem:betterParticleDistribution} below for the precise statement. 

\begin{lemma}
\label{lem:betterParticleDistribution}
For any $i \in \{0,\ldots,\epsilon_G(l)\}$, there exists a contracted particle at grid distance $i$ of $l$ (when algorithm DLE terminates). Moreover, there are no particles at grid distance $i > \epsilon_G(l)$ of $l$.
\end{lemma}

\begin{proof}
During Algorithm DLE, all eligible points are at grid distance less than or equal to $\epsilon_G(l)$ from $l$ (by definition of $\epsilon_G(l)$ and Observation \ref{obs:nonEligibleRemainsNonEligible}). Since particles occupy eligible points initially (by definition of $S_e^0$) and only expand into eligible points (see lines \ref{alg:startOfExpansionBlock}-\ref{alg:endOfExpansionBlock}), no particle occupies a point at grid distance $i > \epsilon_G(l)$ of $l$ when Algorithm DLE terminates. 

For $i=0$, the statement holds trivially. Now, assume $0 < i \leq \epsilon_G(l)$. By definition of $\epsilon_G(l)$, there exists a point $v$ at grid distance $\epsilon_G(l)x    $ of $l$ in the initial configuration $C_0$. Since $S_e^0$ is connected (by (2) of Lemma \ref{lem:EligibleLemma}) and $l \in S_e^0$ (by Observation \ref{obs:nonEligibleRemainsNonEligible}), there exists a point $v' \in S_e^0$ on an eligible path (consisting of eligible points) between $v$ and $l$, such that $dist_G(v',l) = i$. Let $C_t$ be the last configuration reached during the execution of Algorithm DLE in which there exists an eligible point at grid distance $i' \geq i$ of $l$. Since each activation removes at most one eligible point from $S_e$ (see line \ref{alg:leaveSe} of Algorithm DLE), there exists exactly one such point, denoted by $u$, in $C_t$. By the definition of $C_t$, $u$ becomes ineligible between $C_t$ and $C_{t+1}$. In addition, since $S_e^t$ is connected, $i' = i$. Since the triangular grid $G$ is a bridged graph, one can obtain similar results to Lemmas \ref{lem:convexityResults} and \ref{lem:invariant1} for the sets $\{z \in G \; | \; dist_{G}(l,z) = i\}$ and $\{z \in G \; | \; dist_{G}(l,z) \leq i\}$. Thus, one can easily show, in particular, that $u$ has at most 2 neighbors at grid distance $i-1$ from $l$. Moreover, $u$ has no eligible neighbors at grid distance $i'' \geq i$, by the definition of $C_t$. Hence, $u$ is a boundary point of $S_e^t$, with at most two neighbors in $S_e^t$. 

Since $u$ is a boundary point of $S_e^t$, it is occupied by some particle $p$ in $C_t$ by (3) of Lemma \ref{lem:EligibleLemma}. If $p$ was expanded in $C_t$, then $u$ wouldn't leave $S_e$ in $C_t$, which is a contradiction. Hence, $p$ is contracted in $C_t$. Since $S_e^t$ is simply-connected (by (2) of Lemma \ref{lem:EligibleLemma}), $u$ is erodable w.r.t. $S_e^t$. Moreover, recall that $u$ has at most two neighbors in $S_e^t$. Thus, $u$ has a boundary count of 2 or 3 w.r.t. $S_e^t$. Hence, during the activation, $p$ stays in point $u$ by (the contrapositive of) Claim \ref{claim:SingleAdjacentEmptyNode} (see line \ref{alg:endOfErosionBlock} of Algorithm DLE). Consequently, $p$ never moves again until Algorithm DLE terminates, and the lemma holds.
\end{proof}

\subsubsection{Algorithm \Collect}
\label{subsec:algorithmCollectDescription}

At the start (and end) of each phase, all of the collected particles are contracted, and the set of collected particles $S_c$ -- initially the set $\{p_l\}$ -- is organized into a collection of disjoint line segments. A part of the collected particles -- starting at $l$ and stretching along increasing grid distances from $l$ -- forms the main segment, called the \emph{stem}. 
All of the other particles form secondary segments, called \emph{branches}. Recall that the stem has a single particle $p_j$ at distance $j$ from $l$. The branch of distance $j$ contains all of the particles whose grid distance from $l$ is $j$, except for $p_j$. The branch is located counter-clockwise from the stem and one of the particles, $p_j'$, of the branch of distance $j$ is a neighbor of $p_j$. Sometimes, it is convenient to view the stem together with the branches as a tree directed towards its root $p_l$. Particle $p_j'$ is then the root of the branch.
 The number of particles in a segment is said to be the segment's \emph{size}. A segment is \emph{contracted} if all of its particles are contracted. A segment \emph{occupies} points $u_1,\ldots,u_k$ if its particles occupy these points.

\begin{figure}[ht]
     \centering
     \begin{subfigure}{0.45\textwidth}
        \centering
        \includegraphics[width=0.9\linewidth]{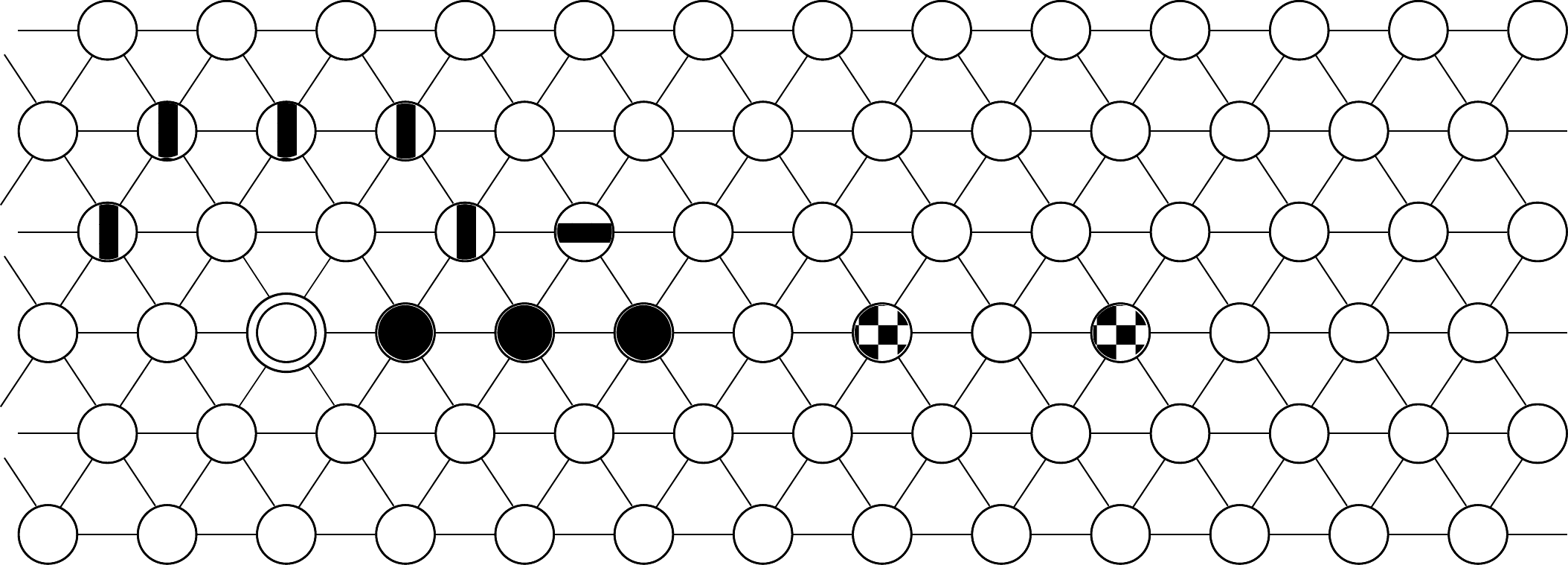}
        \caption{At the start of the phase, the stem of size $k=4$ initially occupies point $l$ (with a ``ring'') and the black points $v_1,\ldots,v_{k-1}$. The points with a ``stripe'' pattern form two branches of already collected particles (one branch is formed by the vertically striped points and the other by the horizontally striped point). The points with a ``checkered'' pattern is occupied by an uncollected particle.}
        \label{fig:collectionFirstMovementa}
     \end{subfigure} \hfill
     \begin{subfigure}{0.45\textwidth}
        \centering
        \includegraphics[width=0.9\linewidth]{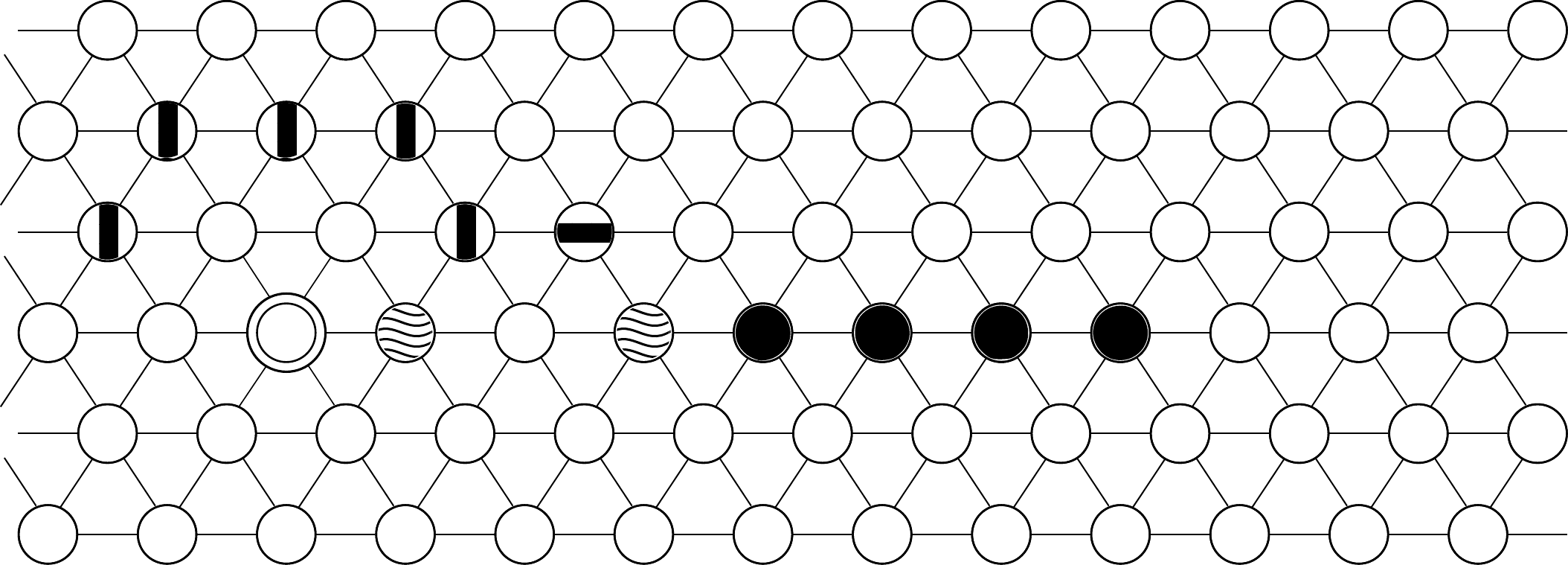}
        \caption{The stem moves $k$ points away from $l$, thus occupying the black points $v_k,\ldots,v_{2k-1}$. Because two of these points are occupied at the start of the first step, two particle are left behind -- in the points with a ``wave'' pattern.}
        \label{fig:collectionFirstMovementb}
     \end{subfigure} \vspace{0.2cm}

     \begin{subfigure}{0.45\textwidth}
        \centering
        \includegraphics[width=0.8\linewidth]{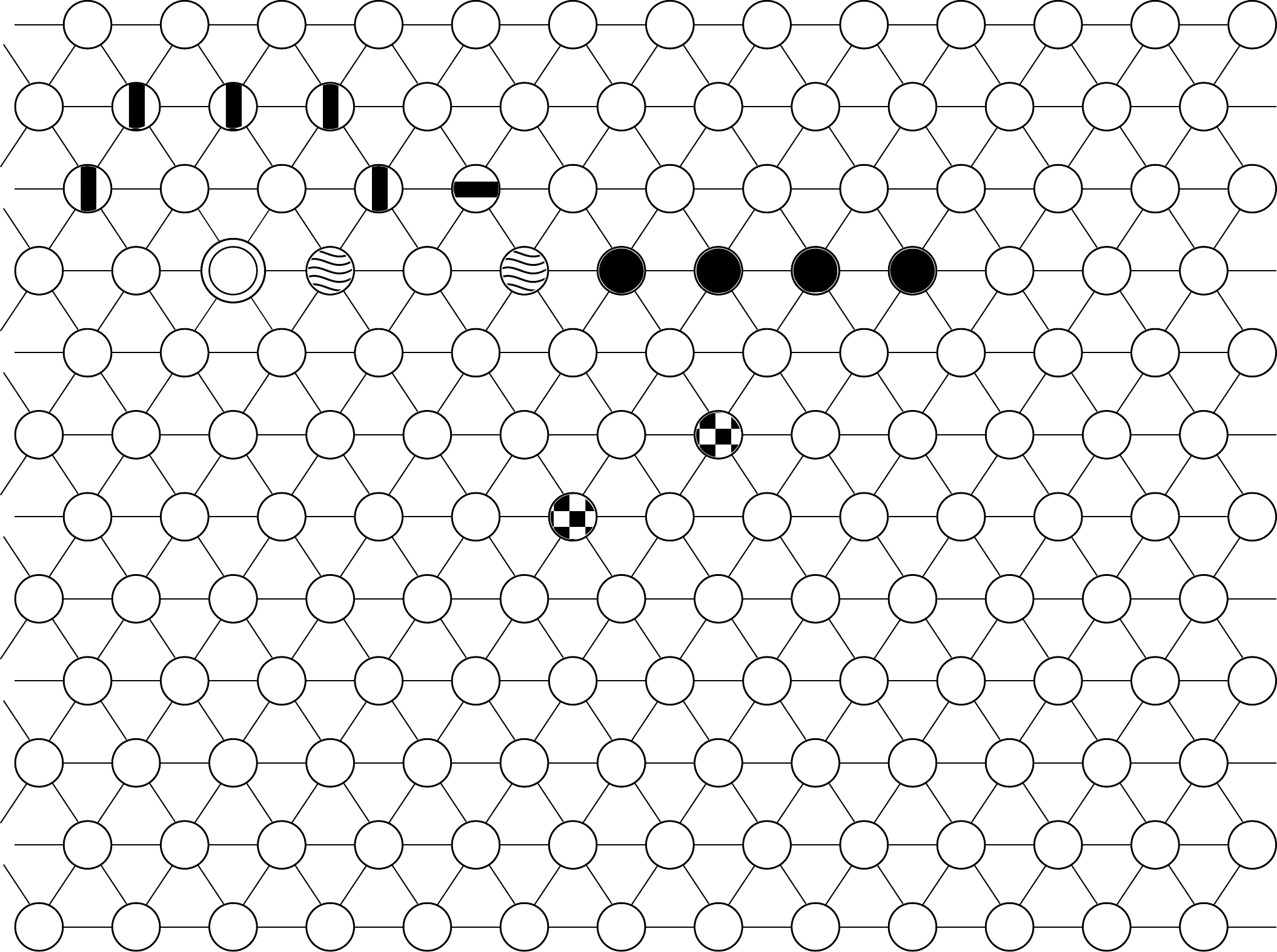}
        \caption{The stem occupies the black points $v_k,\ldots,v_{2k-1}$ at the start of the second step. The points with a ``checkered'' pattern are occupied by uncollected particles.}
        \label{fig:collectionSecondMovementa}
     \end{subfigure} \hfill
     \begin{subfigure}{0.45\textwidth}
        \centering
        \includegraphics[width=0.8\linewidth]{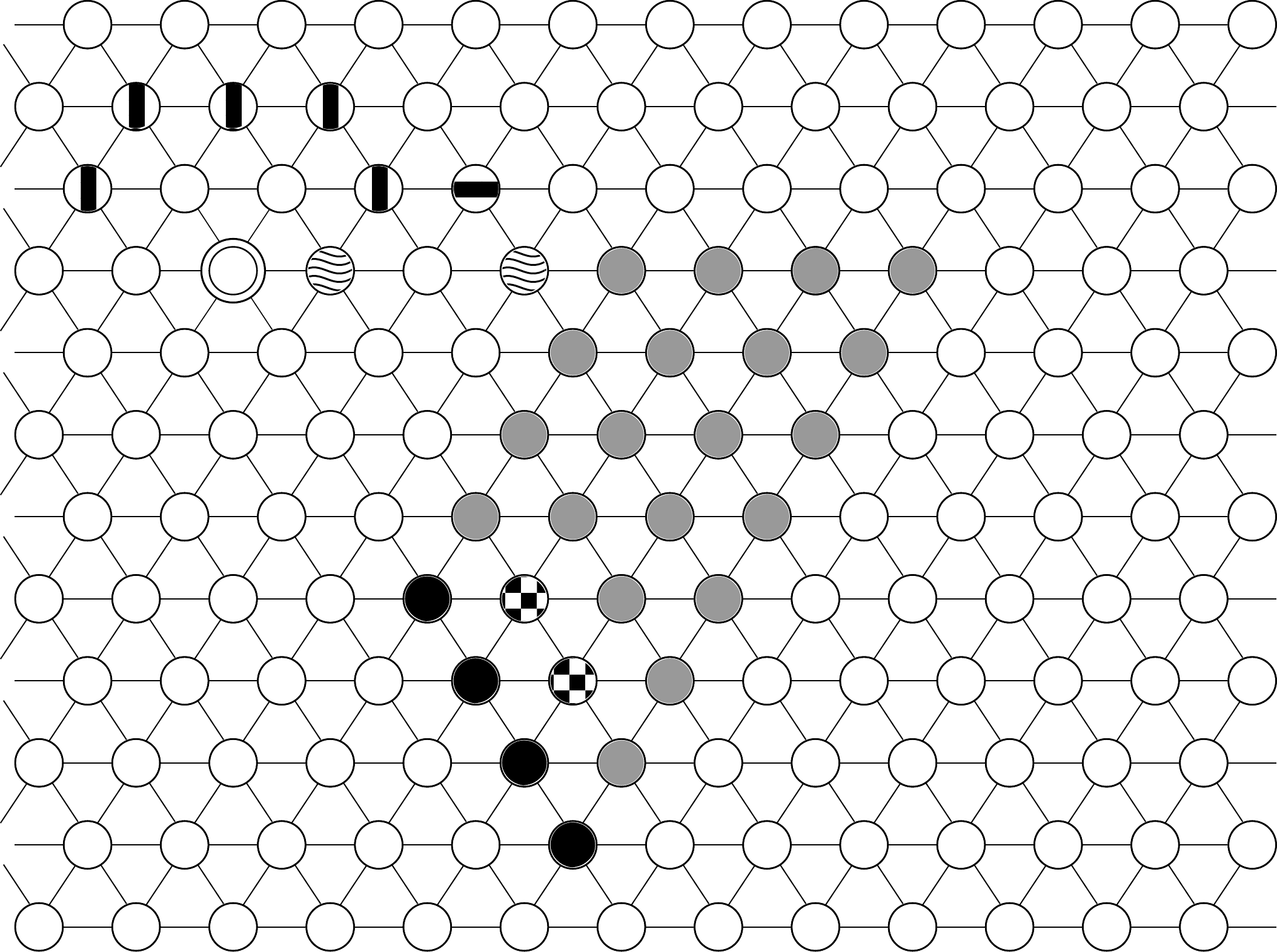}
        \caption{The stem rotates by $60 \degree$ clockwise around $l$. By doing so, it sweeps all of the grey points for uncollected particles. Here, two particles are collected, forming two new branches (see the checkered points). }
        \label{fig:collectionSecondMovementb}
     \end{subfigure} \vspace{0.2cm}

     \begin{subfigure}{0.45\textwidth}
        \centering
        \includegraphics[width=0.8\linewidth]{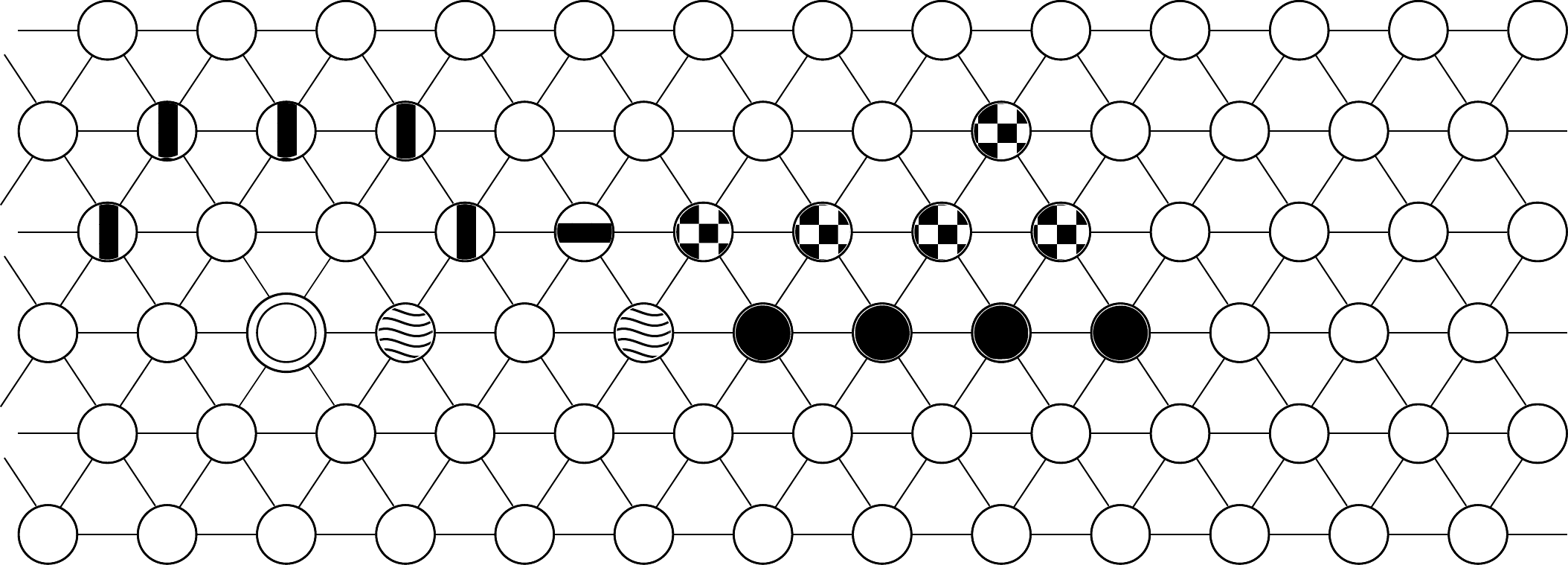}
        \caption{After a full rotation around $l$, the stem occupies the black points $v_k,\ldots,v_{2k-1}$ and the newly-collected particles form new branches (see the checkered points).}
        \label{fig:collectionThirdMovementa}
     \end{subfigure} \hfill
     \begin{subfigure}{0.45\textwidth}
        \centering
        \includegraphics[width=0.8\linewidth]{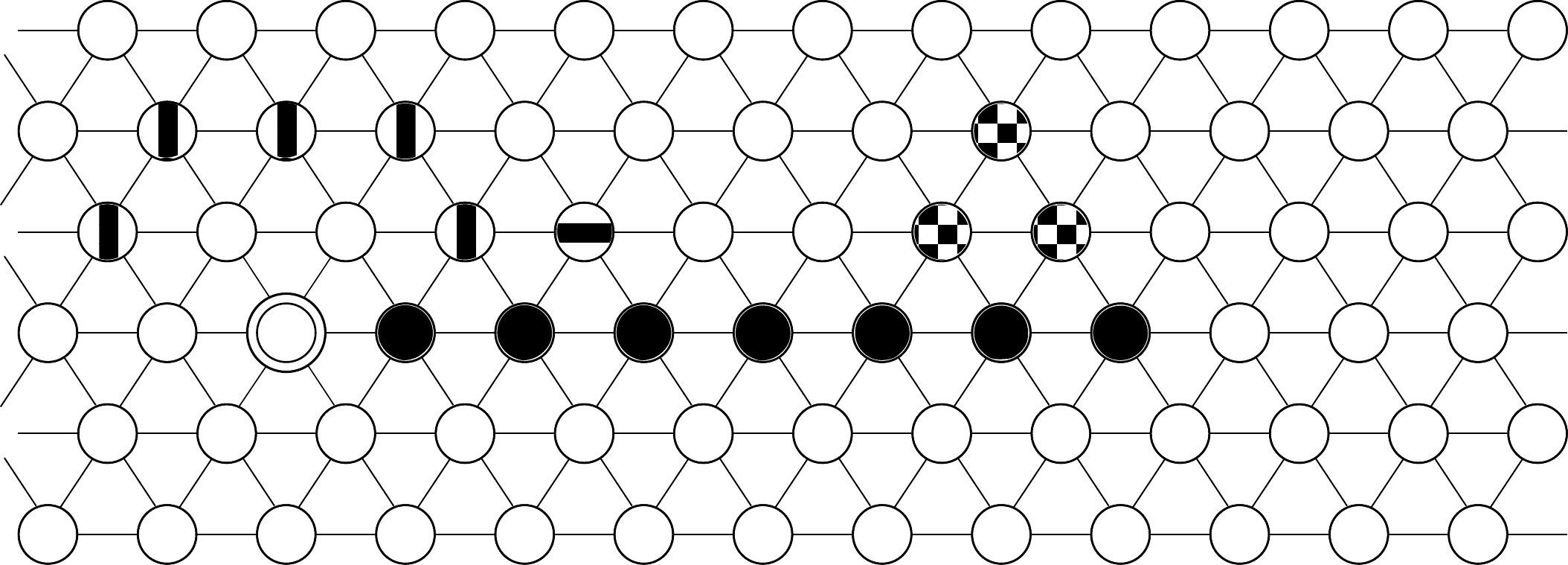}
        \caption{Using the newly-collected particles, the stem moves back to $l$ and doubles in size, thus occupying $l$ and the black points $v_1,\ldots,v_{2k-1}$.}
        \label{fig:collectionThirdMovementb}
     \end{subfigure}
\caption{Description of a phase's three steps in Algorithm $\Collect$}
\label{fig:explanationsAlgorithmCollect}
\end{figure}

\paragraph*{Phase} Initially, the set of collected particles, as well as the stem, consist only of $p_l$. The branches are empty. 
In phase $i=1,2, ...$ (except, possibly, the last two phases), the stem starts the phase with $k=2^{i-1}$ particles and collects at least as many (as shown later, using  Lemma \ref{lem:betterParticleDistribution}). 
Those particles are thus enough for the stem to double its size before the next phase starts. 
The rest of the particles (if any exist, beyond those needed for the doubling) remain in the branches. 
Let us now provide a more precise description of one phase (accompanied by Figure \ref{fig:explanationsAlgorithmCollect}) using movement primitives whose descriptions appear in Section \ref{subsubsec:movementPrimitives}. 
Let $v_0,\ldots,v_{k-1}$ (where $v_0 = l$) be the points occupied by the $k$ particles of the stem initially (recall that the particles are initially contracted).

\begin{enumerate}
\item First, the stem disconnects from its branches and moves $k$ points away from $l$, to occupy all of the points $v_{k},\ldots,v_{2k-1}$ (where $v_0,\ldots,v_{2k-1}$ is a straight line of points). If any (up-to-now uncollected) particles occupy these points, a \emph{collect} action is applied on each such particle and it becomes a part of the new stem. (This is described in more detail as primitive {\OMP} in Section \ref{subsubsec:movementPrimitives}.) 
In exchange, some of the particles initially in the stem are left behind -- in some of the points $v_0,\ldots,v_{k-1}$ -- during the stem's movement. When the first step ends, the stem is disconnected from all branches of distance $j \leq k-1$.  The particles left behind are not considered parts of the stem whether they are connected to it or not. 
See Figures \ref{fig:collectionFirstMovementa} and \ref{fig:collectionFirstMovementb}.
\item Afterward, the stem performs one complete clockwise rotation around $l$ and returns to its original location of points $v_{k},\ldots,v_{2k-1}$. The rotation is composed of 6 partial (clockwise) rotations each of $60\degree$. (For details see the description of primitive {\PRP} in Section \ref{subsubsec:movementPrimitives}.) During these rotations, a stem particle at distance $j$ from $l$ remains at distance $j$. While executing a rotation, consider the case that an uncollected particle $q$ at distance $j$ from $l$ becomes a neighbor of one of the stem's particles, call it $p_j$, who too is at distance $j$ from $l$. Particle $q$ is an ``obstacle'' in the sense that $p_j$ cannot continue performing the rotation in parallel to the rest of the stem. Instead, a \emph{collect} action is applied on $q$. Particle $q$ takes the place of $p_j$ in the stem, and $p$ becomes the first particle in the branch of distance $j$. If $p_j$ has already been a neighbor of a branch of distance $j$ whose root was, say, $p_j'$, then $p_j'$ becomes the child of $p_j$ in the branch. Branches remain connected to the stem during the rotations and perform the rotation with the stem. See Figures \ref{fig:collectionSecondMovementa}, \ref{fig:collectionSecondMovementb} and \ref{fig:collectionThirdMovementa}.
\item Finally, if no particle was collected in the first two steps, then Algorithm $\Collect$ terminates. 
Otherwise, the algorithm starts enlarging the stem.
(If the size of the stem is already at least $\epsilon_G(l)/2$ then the following enlargement process may not end up enlarging beyond $\epsilon_G(l)$; 
we show later that by Lemma \ref{lem:betterParticleDistribution}, this means that all the particles are already collected; hence, no particle is collected in the next phase and the algorithm will terminate; henceforth, for the sake of clarity, we ignore this special case in the current paragraph.)
We show that enough particles were collected in the current phase for the stem to double in size. So it does double, moving to occupy points  $\{l, v_1, \ldots, v_{2k-1}\}$. (For details see the description of primitive {\SDP} in Section \ref{subsubsec:movementPrimitives}.)
Note that we must be careful not to grow the stem beyond doubling even if there are enough particles for that. (Otherwise, the stem, together with the runtime, could have grown far beyond the eccentricity $\epsilon_G(l)$ and thus the diameter $D_G$; avoiding extra growth is done, intuitively, by ``charging'' every addition of a new particle to the stem to one expansion and extraction of an ``old'' stem particle.) 

In the process, the stem absorbs any particle left behind in $\{l, v_1, \ldots, v_{k-1}\}$ and reconnects with all branches of distance $j \leq k-1$.
After that, a new phase starts. See Figures \ref{fig:collectionThirdMovementa} and \ref{fig:collectionThirdMovementb}.
\end{enumerate}

\paragraph*{Analysis of Algorithm $\Collect$} 
The proofs of Lemmas \ref{lem:collectedIsConnected} and \ref{lem:collectionPhaseProperties} appear in Section \ref{subsubsec:movementPrimitives}. Corollary \ref{cor:collectionPhaseCorollary} follows straightforwardly from Lemma \ref{lem:collectionPhaseProperties}. 

\begin{lemma}
\label{lem:collectedIsConnected}
The set of collected particles $S_c$ is connected at the beginning and end of each phase.
\end{lemma}

\begin{lemma}
\label{lem:collectionPhaseProperties}
Consider a phase in which the stem is initially of size $k$. The phase takes $O(k)$ rounds and all the particles at grid distance $j \in \{k,\ldots,2k-1\}$ from $l$ are collected. Moreover, the following holds:
\begin{itemize}
	\item If $k \leq \epsilon_G(l)$, then the stem grows to a size of $k' \in \{\min\{2k,\allowbreak \epsilon_G(l)\}, \ldots,2k\}$ at the end of the phase. 
	\item If $k > \epsilon_G(l)$, then no particle is collected during the phase.
\end{itemize}
\end{lemma}

\begin{corollary}
\label{cor:collectionPhaseCorollary}
For $i \in \{1,\ldots, \lfloor \log{\epsilon_G(l)}  \rfloor + 1 \}$, the stem is of size $2^{i-1}$ at the start of phase $i$.
\end{corollary}

\begin{theorem}
\label{thm:collectionCorrectnessAndRuntime}
Algorithm $\Collect$ terminates in $O(D_G)$ rounds. Upon termination, the particle system is connected.
\end{theorem}

\begin{proof}
By Lemma \ref{lem:betterParticleDistribution}, all particles are at grid distance $j \leq \epsilon_G(l)$ from $l$. In phase $i$, the algorithm collects all the particles at grid distance $j \in \{2^{i-1},\ldots,2^i-1\}$ by Lemma \ref{lem:collectionPhaseProperties} and Corollary \ref{cor:collectionPhaseCorollary}. Thus, at least one particle is collected in every phase $i \leq \lfloor \log{\epsilon_G(l)}  \rfloor + 1$ by Lemma \ref{lem:betterParticleDistribution}. Furthermore, after $\lfloor \log{\epsilon_G(l)} \rfloor + 1 \leq \lfloor \log{D_G}  \rfloor + 1$ phases, all particles have been collected. Hence, the algorithm executes one last phase, in which no particle is collected, and terminates. When that happens, the particle system is connected by Lemma \ref{lem:collectedIsConnected}. The runtime follows directly from the number of phases and Lemma \ref{lem:collectionPhaseProperties}.
\end{proof}

\subsubsection{Movement Primitives} 
\label{subsubsec:movementPrimitives}

Let us first give some preliminary definitions. 
Similarly to the above definition of the branch root,  the closest stem particle to $l$ (in grid distance) is called the stem's \emph{root} and the furthest particle the \emph{leaf}. Naturally, this defines for each other particle in the stem, a single parent and child neighboring particle. (Note that during most of the movement primitives described below, it is simple to maintain this tree structure while particles are moving.)
The root (either that of the stem, or of a branch) is considered as the segment's first particle, and the leaf is the segment's last.
Consider a contracted stem of size $k$ and denote its particles by $p_1,\ldots,p_k$ starting from the root.

When Algorithm {\Collect} starts, the leader particle $p_l$ assumes its six incident edges (in order of increasing port number) to lead, respectively, W, NW, NE, E, SE and SW. Furthermore, $p_l$ chooses two opposite directions (e.g., E and W) to indicate the outwards and inwards direction for the stem. These two directions are denoted, respectively, by $\vout$ and $\vin$. Then, whenever a particle, say $q$, is collected, $q$ ``learns'' all the above mentioned directions of the leader (including $\vout$ and $\vin$). Indeed, recall that $q$ is being touched by some already collected particle $p$. In general, this would not have been enough for $q$ to learn directions from $p$. However, $p$ and $q$ also share common chirality. See primitive {\Communicate} below.

\paragraph*{Virtual Movement Operations and Virtual Particles.} 
Consider the case that a contracted particle $p$ is instructed by the algorithm to expand from a point $u$ along some edge $\{u,v\}$, but the other endpoint $v$ of that edge is already occupied by some contracted particle $q$. 
Particle $p$ \emph{virtually expands} into $q$. In particular, $p$ writes into both $q$'s and $p$'s memories that $q$ is the head of a virtual particle denoted $(p,q)$, together with the port numbers of $q$'s port leading to $p$ and $p$'s port leading to $q$. The virtual particle plays the role that $p$ was playing up to that point in time (e.g., as a stem particle). Particles $p$ and $q$ now starts cooperating in order to simulate the actions of $(p,q)$.

 After virtual expansion, let us now speak of virtual contraction. 
Consider some virtual particle $(p,q)$ (playing the role of some particle $p$ who virtually expanded and now occupies points $u$ and $v$ correspondingly).  First, consider the case that no handover is involved. That is, particle $(p,q)$ decides to contract to point $v$, the head of $(p,q)$. Particle $p$ leaves the simulation and the virtual particle is now simulated by particle $q$ alone. At first glance, this may create a strange situation, since the virtual particle used to play the role of $p$, and $p$ is the one leaving the simulation.
However, recall that particles do not have identities, and the names $p$, $q$, etc. we used for our convenience only. As a part of the above virtual contraction, we say that the actual particle occupying points $u$ and $v$ (who used to be named $p$ and $q$ correspondingly) have lost their old names. Instead, the actual particle occupying point $v$ is now called $p$, and it plays every role $p$ used to play, in particular, as a stem particle. The new name of the particle occupying $u$ is not important. (For the sake of convenience, we can call it $q$.)
We also say that $p$ \emph{virtually moved} from $u$ to $v$ by way of the following two virtual movement operations: first, a virtual expansion (to $v$) and then, a virtual contraction (from $u$).

 Similar to the non-virtual case, a virtual contraction can take place also as a part of a handover. Let us show how to simulate the contraction action where a \emph{non}-virtual particle $p$ is \emph{non}-virtually expanded from some point
 $u$ to a neighboring point $v$. There, some neighboring contracted particle $r$ could expand into $u$, thus forcing $p$ to contract to $v$. 
 
 If $p$ is, instead, the virtual particle $(p,q)$ (occupying points $u$ and $v$
 correspondingly), then still some neighboring contracted particle $r$ may wish to expand into (w.l.o.g) $u$ (occupied by $p$). Note that since $r$ is contracted, $r$ is \emph{not} virtual.  
 Reading $p$, particle $r$ learns that $p$ participates in the simulation of a virtual particle $(p,q)$. 
 Then $r$ \emph{virtually expands} into the tail $u$ of $(p,q)$ by having 
 $r$ and $p$ simulate (the tail and head of) a virtually expanded particle $(r,p)$, playing the role of $r$. Note that $p$ needs to act too in order to let $q$ know that the simulation of $(p,q)$ is over. This leaves $q$ contracted. 
 As in the previous case of a virtual contraction from the tail, the particle occupying $v$ is thus renamed $p$. (The new name of the particle occupying $u$ is not important.) 
 Note that we have shown here the case of a contraction from the tail, otherwise, no renaming of the particles would have been necessary.

\paragraph*{Auxiliary Implementations.} 
Auxiliary primitive {\Communicate} involves an edge $e$ between some points $u,v$ occupied by some particles $p,q$ (correspondingly). 
Let $i_p$ and $i_q$ be the port numbers of $p$ and $q$ for $e$. (Recall that both port numbers are known to $p$ and $q$, see Section \ref{subsec:system}.)
Particle $p$ writes in $q$ the instruction that the port $i_q$ should, instead, be numbered $i_p + 3 \; mod \; 6$.
Particle $q$ changes its other port numbers accordingly.
It is easy to see that, since $p$ and $q$ have common chirality, the two particles now have the same port numbering. Moreover, particle $p$ also writes in $q$ the numbers of the ports corresponding to directions $\vin$ and $\vout$.  Since the port numbering of $p$ and $q$ are now the same, it is easy for $p$ to point at other ports of $q$ as well. Thus, we also use this primitive (with the obvious additional instruction) to convey an additional direction (the stem rotation direction) in the second movement primitive.

The movement primitives sometimes rely on either the root or leaf to detect when the stem particles (and possibly the connected branch particles) are all contracted or all expanded (including being virtually expanded). This can be done using the following auxiliary primitive, called {\Detect}. Let us show how to detect the global predicate that all stem particles are contracted. (Detecting that all stem particles are expanded can be done similarly.) The root (resp., the leaf) first performs a ``local validity check'', that is, it verifies that both itself and its neighbor in the stem are contracted.
If so, it sends a token along the stem. Whenever a stem particle $p$ receives that token, $p$ similarly does a local validity check. If this check fails, $p$ (destroys the token and) returns an ``answer'' token towards the root (resp., leaf) with value $false$ -- thus informing the root (resp., the leaf) that not all particles are contracted. If $p$ is the leaf (resp., root) and its local validity check succeeds, it sends back an ``answer'' token with a $true$ value to the root (resp., leaf). 

To also detect whether the branch particles are all contracted, one can modify the above algorithm as follows. A stem particle $p$ that holds the token sent by the root (resp., leaf) also requests the corresponding branch root(s) (at the same distance(s) from $l$ as $p$) to perform a similar check along its branch. When an ``answer'' token (either with a $true$ or with a $false$ value) is sent back towards the root (resp., leaf), the stem particle $p$ holding that answer token also waits for the answer from the branch root. The answer token forwarded by $p$ towards the root (resp. leaf) carries the logical ``and'' of both values (arriving from the stem and from the branch).

\begin{figure}[ht]
     \centering
     \begin{subfigure}[t]{0.31\textwidth}
        \centering
        \includegraphics[width=0.95\linewidth]{CollectionFirstMovement}
        \label{fig:collectionFirstMovement1}
        \caption{Initially, the stem occupies point $l$ (with a ``ring'') and the black points $v_1,\ldots,v_{k-1}$. The striped points form two branches of already collected particles. The checkered point is occupied by an uncollected particle.}
     \end{subfigure} \hfill
     \begin{subfigure}[t]{0.31\textwidth}
        \centering
        \includegraphics[width=0.95\linewidth]{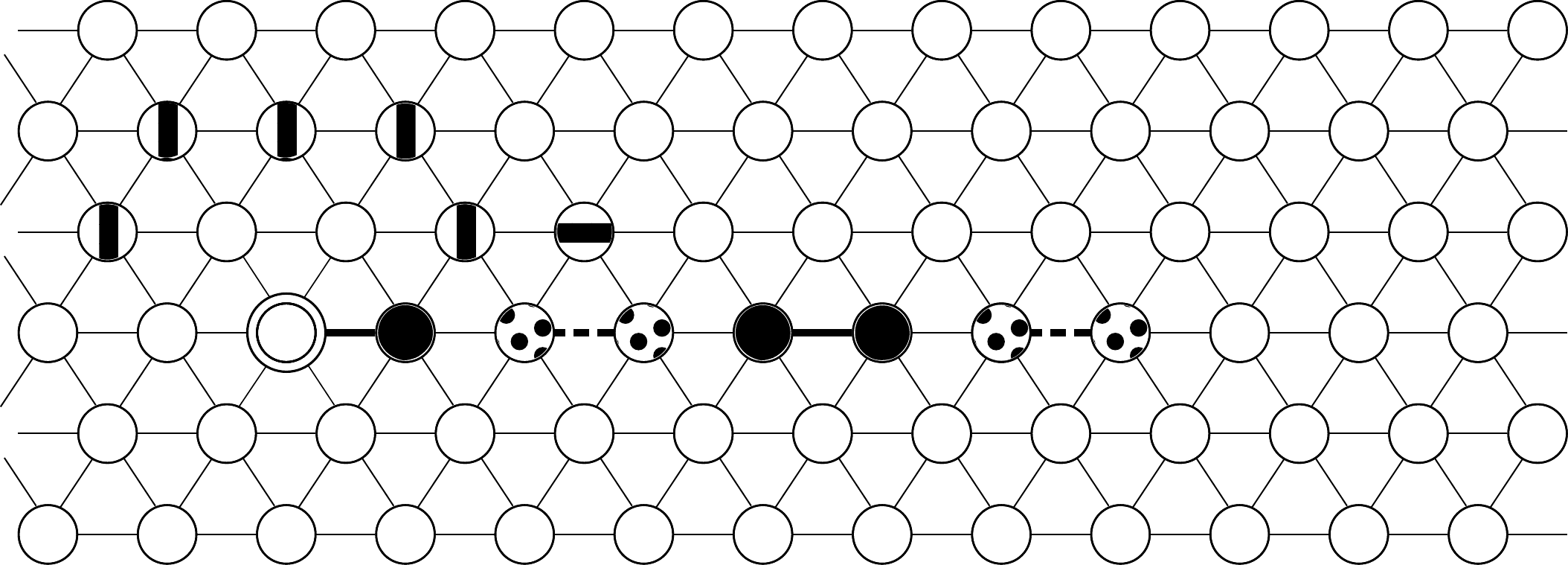}
        \caption{After the first part, all of the points $l,v_1,\ldots,v_{2k-1}$ are occupied by either an expanded particle or a virtually expanded particle (i.e., two contracted particles -- see the points with ``spots''). Note that the uncollected particles from the left figure are collected during the first part.}
        \label{fig:collectionFirstMovement3}
     \end{subfigure} \hfill
     \begin{subfigure}[t]{0.31\textwidth}
        \centering
        \includegraphics[width=0.95\linewidth]{CollectionFirstMovement2}
        \caption{After the second part, the stem occupies the black points $v_k,\ldots,v_{2k-1}$. Note that because two of these points were initially occupied, two particle are left behind -- in the points with a ``wave'' pattern. Furthermore, the stem is now disconnected from its branches -- see the striped points.}
        \label{fig:collectionFirstMovement2}
     \end{subfigure}
\caption{First Step: Moving a stem of size $k=4$ away from $l$ by $k$ points}
\label{fig:explanationsFirstStep}
\end{figure}

\paragraph*{Moving away from $l$} The first primitive we now describe (Outerwards Movement Primitive, or \OMP), is applied to the stem initially occupying $v_0,\ldots,v_{k-1}$. \OMP~ moves  (all the particles of) the stem by $k$ points in the direction of $\vout$ (see Algorithm \ref{alg:firstMovementPrimitive} and Figure \ref{fig:explanationsFirstStep}). 
From a high level, the first part of OMP expands all the stem particles (such that the tail of the root does not move). This doubles (to $2k$) the number of points occupied by the stem. (Intuitively, if an expansion ordered by OMP is blocked by an uncollected particle, this is handled using virtual expansions).
Then the second part contracts all the stem particles, such that the head of the leaf does not move. This performs the moving of the stem.
 Once  \OMP~ terminates, the stem occupies the destination points $v_k,\ldots,v_{2k-1}$, where $v_{k+ i} = v_i + k \cdot \vout$ for $0 \leq i \leq k-1$. 
 If no particle occupies these destination points initially, then when the primitive terminates, particles $p_1,\ldots,p_k$ occupy $v_k,\ldots,v_{2k-1}$ respectively. Otherwise, when the primitive terminates, some of the original particles have been replaced; that is, such a particle $p_i$ moved virtually, and its name $p_i$ was assigned to some other particle.

In more details, during the first part of the primitive (see procedure $Expansion$ in Algorithm \ref{alg:firstMovementPrimitive}), all of the  stem's particles expand in the direction of $\vout$ whenever possible. Initially, all the particles are contracted. Whenever the leaf particle is activated and contracted, it expands in the direction of $\vout$ (expands virtually if the grid point in question is occupied). Whenever a contracted non-leaf particle $p$ is activated and its child $p_+$ is expanded, $p$ expands in the direction of $\vout$ (that is, towards $p_+$). This is a virtual expansion if $p_+$ is virtually expanded, otherwise, $p$ expands by handover with $p_+$.
As a result, the stem occupies points $v_0,\ldots,v_{2k-1}$.
At this point, it may be composed of two sets of particles. One set consists of the particles that were in the stem when procedure $Expansion$ started. The other set (possibly empty) consists of particles that were uncollected before the procedure was invoked, but became \emph{collected} (and parts of the stem) when a stem particle expanded into them virtually.

The root detects when every stem particle is expanded (virtually or otherwise) (using {\Detect}) and informs all particles to start the second part (see procedure $Contraction$ in Algorithm \ref{alg:firstMovementPrimitive}). 
At this point, all the stem particles are expanded (possibly virtually expanded). Whenever an expanded non-root stem particle $p_+$ is activated and its parent $p$ is contracted, $p_+$ contracts to its head (in the direction of $\vout$) through a handover. Note that if $p_+$ is virtually expanded, then this contraction is virtual. Whenever the root particle $p_1$ is activated, it contracts into its head if it is expanded. Note that if $p_1$ is a virtually expanded particle occupying some points $u$ and $v$, then the particle at the tail $u$ leaves the stem and the role of $p_1$ is now played by the particle at point $v$. Once the root detects that all of the stem particles are contracted (using {\Detect}), the primitive is done.

\begin{algorithm}
\caption{\enspace Moving by $k$ points in the direction of $\vout$}
\label{alg:firstMovementPrimitive}
\begin{algorithmic}[1]
\Procedure{Expansion}{}
\State \textbf{During the atomic activation of a stem particle $p$:}
\If{$p$ is contracted and not virtually expanded}
	\State Let $u$ be the other endpoint of $\vout$ (for $p$)
	\If{$u$ is not occupied}  \label{line:contractStartStep1}  
		\State $p$ expands into $u$ \label{line:contract1} 
	\ElsIf{$u$ is occupied by an expanded particle $q$}   
		\State $p$ expands (through a handover with $q$) into $u$ \label{line:contractHandoverStep1}
		\State // See the definition of the virtual expansion (through handover) operation for more details.
	\ElsIf{$u$ is occupied by a virtually expanded particle or an uncollected (contracted) particle}
		\State $p$ virtually expands into $u$  \label{line:contractEndStep1}
		\State // See the virtual expansion operation's definition for more details.
	\EndIf
\EndIf
\EndProcedure
\State

\Procedure{Contraction}{}
\State \textbf{During the atomic activation of a stem particle $p$:}
\If{$p$ is the root}
	\If{$p$ is expanded}
		\State $p$ contracts into its head   \label{line:contractStep1b}
	\ElsIf{$p$ is contracted and the tail of a virtually expanded particle $(p,q)$}
	    \State $p$ virtually contracts into its head \label{line:changeStep1b}
	    \State // See the virtual contraction operation's definition for more details, in particular regarding $p$'s renaming. 
	\EndIf
\ElsIf{$p$'s parent $q'$ is not expanded nor in a virtually expanded particle}
	\If{$p$ is expanded}
		\State $p$ contracts into its head (through a handover with $q'$)  
	\ElsIf{$p$ is the tail of a virtually expanded particle $(p,q)$}
	    \State $p$ virtually contracts into its head through a handover with $q'$ 
	    \State // See the definition of the virtual contraction (through handover) operation for more details.
	\EndIf
\EndIf
\EndProcedure

\end{algorithmic}
\end{algorithm}

\begin{lemma}
\label{lem:firstMovementPrimitive}
When executed by a stem of size $k$, primitive \OMP~ terminates in $O(k)$ rounds. Once  \OMP~ terminates, the stem is contracted and occupies the destination points $v_k,\ldots,v_{2k-1}$.
\end{lemma}

\begin{proof}
Let us first show that procedure Expansion takes $O(k)$ rounds. A similar proof provides the same runtime for procedure Contraction.
At the start of procedure Expansion, name the stem particles $p_1,\ldots,p_k$ starting from the root to the leaf. We assume that at this point, each of those is a non-virtual particle, and is contracted. This is clearly true the first time OMP is used. We shall show that all the primitives end up that way too.

During the execution, particles can be renamed as defined above for virtual movement operations. We shall show that at any time during the execution, the stem consists of exactly $k$ particles named $p_1,\ldots,p_k$, but some of them may be virtual. (Intuitively, this may result from the stem expanding into points occupied by uncollected particles.)

Let us say that the leaf particle $p_k$ holds an expansion permit if it is contracted. A non-leaf particle $p_j$ holds an expansion permit if $p_j$ is contracted and $p_j$'s child is expanded. Since the leaf is initially contracted, the leaf $p_k$ initially holds an expansion permit. It expands in direction $\vout$ in $O(1)$ rounds (see lines \ref{line:contract1} and \ref{line:contractEndStep1} in Algorithm \ref{alg:firstMovementPrimitive}) and thus the leaf parent's now has a permit. We say that the leaf forwarded the very same expansion permit to its parent. The parent $p_{k-1}$ now expands (see lines \ref{line:contractHandoverStep1} and \ref{line:contractEndStep1} in Algorithm \ref{alg:firstMovementPrimitive}) too, forwarding the very same permit to its own parent $p_{k-2}$. More generally, if some particle $p_i$ having a permit ${\mathring{p}}$ expands when $p_{i-1}$ is contracted, we say that $p_i$ forwarded ${\mathring{p}}$ to $p_{i-1}$. Similarly, if $p_i$ having a permit ${\mathring{p}}$ expands  when $p_{i-1}$ is \emph{expanded}, but $p_{i-1}$ later contracts, this still is a forwarding of permit ${\mathring{p}}$ to  $p_{i-1}$.
It is easy to see by induction (from the leaf to the root) that this expansion permit is thus forwarded all the way towards towards the root. Moreover, each particle holding a permit expands in $O(1)$ rounds, so the permit proceeds from each stem particle $p_i$ toward the root in $O(1)$ rounds.

Notice that the expansion operation of $p_{k-1}$ forced the leaf to contract. Thus the leaf now has a second permit. 
We now claim that the total number of permits is bounded by $k$. (In fact, it is exactly $k$).
Once the root receives the first permit, it expands in $O(1)$ rounds. Since the root $p_1$ cannot contract in procedure Expansion, $p_1$ remains expanded during this procedure, so no permit can be forwarded to it. Similarly, it is easy to see by induction that no more than $i$ permits can be forwarded to $p_i$. Note that the leaf is the only one who can have new permits (not forwarded to it), however, the leaf cannot have a new permit before forwarding any permit it possesses. The claim follows. 

We have established that there are up to $k$ different permits. Moreover, the forwarding of a permit, given no previous permit blocks its forwarding, takes $O(1)$ time. These are the conditions for which pipelining is known to exist. That, is, using \cite{cidon1995greedy,mansour1993greedy}, the total time until no permit can be forwarded is bounded by the length of the longest route of a permit plus the number of permits which is $O(k)$.

The first part -- Procedure Expansion -- does not terminate while there exists a contracted particle $p$ that is not a part of a virtual expanded particle. (See lines \ref{line:contractStartStep1}-\ref{line:contractEndStep1}.)
Furthermore, every contraction is the result of an expansion, while the expansion of the leaf does not cause any contraction. Hence, eventually, all the stem particles are expanded.
Moreover, virtual expansions and virtual contractions (as used here) do not change the number of a stem particles. Hence, upon termination of procedure Expansion, there exists exactly one virtually expanded particle for every (initially uncollected) particle that occupies a point in $\{v_k,\ldots,v_{2k-1}\}$. 
By induction on time (and using the definition of virtual movement operations), 
if there are $0 \leq j \leq k$ such uncollected particles, then the stem eventually consists of $j$ virtually expanded particles and $k-j$ expanded particles when procedure Expansion terminates. Hence, $2k$ points are occupied. Since the stem's root stays in $l$ and expands into $v_1$, and the stem remains connected, the points $v_0,\ldots,v_{2k-1}$ -- and only these points -- are occupied by the stem particles.  

The second part, Procedure Contraction, does not terminate while there exists an expanded particle or a virtually expanded particle. In the beginning, the stem's root is expanded -- it is either an expanded particle $p$ or a virtually expanded particle $(p,q)$. In either case, the root eventually contracts. 
In the first case, $p$ contracts into its head (see line \ref{line:contractStep1b}); in the second, $p$ is removed from the stem and $q$ replaces $p$ as the stem's root (see line \ref{line:changeStep1b}).
After which, the root's child eventually contracts through a handover with the root. Once again, the root eventually contracts. Eventually, all of the stem particles are contracted. Note that every virtually expanded particle at the start of Procedure Contraction results in a particle being removed from the stem. Hence, there remain exactly $k$ stem particles. Since, in addition, the stem's leaf stays in $v_{2k-1}$ and the stem remains connected at all times, the stem particles occupy points $v_{k},\ldots,v_{2k-1}$ (and only these).
\end{proof}

\begin{figure}[ht]
     \centering
     \begin{subfigure}[t]{0.31\textwidth}
        \centering
        \includegraphics[width=0.95\linewidth]{CollectionSecondMovement}
        \label{fig:collectionSecondMovement1}
        \caption{Initially, the stem occupies the black points $v_k,\ldots,v_{2k-1}$. The points with a ``checkered'' pattern are occupied by uncollected particles.}
     \end{subfigure} \hfill
     \begin{subfigure}[t]{0.31\textwidth}
        \centering
        \includegraphics[width=0.95\linewidth]{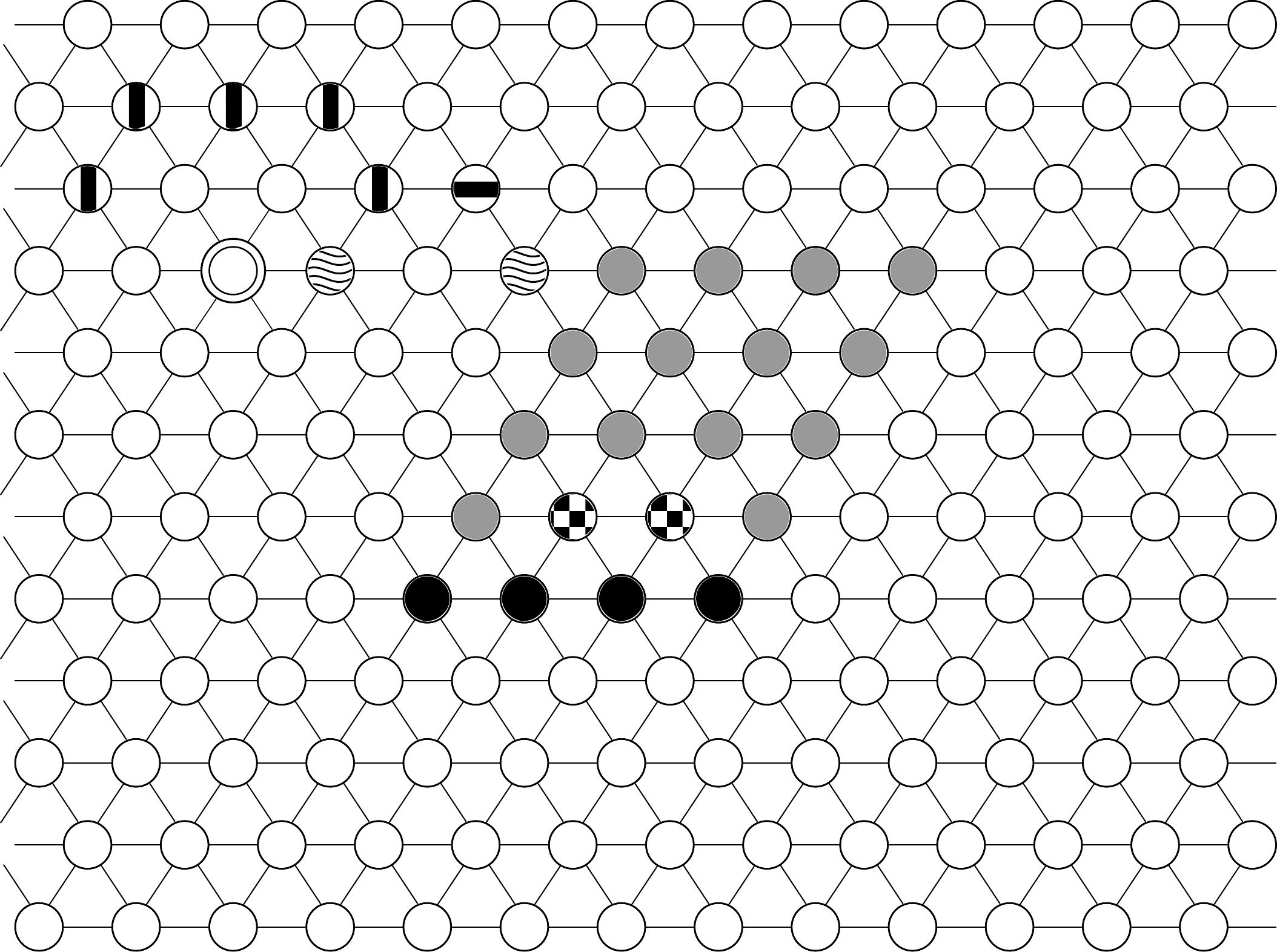}
        \caption{After the first part, the stem has moved by $k$ points in the direction of $\vrot$. At the same time, it sweeps all of the grey points for uncollected particles. Here, two particles are collected, forming two new branches (see the checkered points). }
        \label{fig:collectionSecondMovement3}
     \end{subfigure}  \hfill
     \begin{subfigure}[t]{0.31\textwidth}
        \centering
        \includegraphics[width=0.95\linewidth]{CollectionSecondMovement2}
        \caption{During the second part, the stem rotates around the stem's root -- while ``pulling'' the branches of collected particles. Hence, after the second part, the stem has rotated by $60 \degree$ clockwise around $l$. }
        \label{fig:collectionSecondMovement2}
     \end{subfigure} 
\caption{(Partial) Second Step: Rotating a stem of size $k=4$ around $l$ by $60 \degree$}
\label{fig:explanationsSecondStep}
\end{figure}

\paragraph*{Rotating around $l$} The second step of a phase, the Partial Rotation Primitive (or \PRP) rotates a segment of size $k > 1$ around $l$ by $60\degree$ clockwise -- see Figure \ref{fig:explanationsSecondStep}. For the first rotation, the stem's root sets $\vrot$ to the clockwise predecessor of $\vin$ (e.g., if $\vin$ is W, then $\vrot$ is SW) and communicates the direction to all the stem particles. Then, to rotate around $l$, the stem (1) moves $k$ points in the direction of $\vrot$ (see Algorithm \ref{alg:secondMovementPrimitive}). After which, the stem (2) rotates clockwise around its root by $60\degree$ (described as a modified version of Algorithm \ref{alg:secondMovementPrimitive}). Finally, the stem's root (3) sets $\vrot$ to $\vrot$'s clockwise successor (e.g., from SW to W) and communicates the new direction to all the stem's particles, to prepare for the next rotation. For the next 5 rotations, the stem similarly executes (1), (2) and (3). 

Let us first describe part (1) of primitive {\PRP}. Some challenge is posed by the fact that a particle cannot count (beyond a constant number). However, $k$ particles can count (even in unary) to $k$. Each particle just floods the step with a ``move'' message.  To  move, each particle uses a move message originated by a different particle (and moves that message away such that it never  returns; recall that the stem is a directed tree). Hence, eventually, each particle moves $k$ steps. In the description below, we actually use two kind of move messages -- one for the expansion and one for the contraction. A second minor challenge is posed by the requirement that the stem remains connected. Note, however, that under the scheme just described, the parent and the child of a particle $p$ move once between each of $p$'s move, ensuring the continued connection -- giving the first half of Observation \ref{obs:stemNoDisconnect}. Finally, the role of generating new ``move'' messages (as opposed to forwarding them) starts with the root and is passed from point to point until the leaf. 
Note that every move is in the direction of $\vrot$. The runtime follows easily from the fact that all the messages are pipelined through the stem (see again \cite{cidon1995greedy,mansour1993greedy}).

In Algorithm \ref{alg:secondMovementPrimitive}, each (stem) particle can hold at most one (``expand'' or ``contract'') message and initially has a $start$ flag set to $true$ if and only if it is the root. A (stem) particle $p$ who holds a message that has been received from $p$'s parent (resp., child), can only forward that message to its child (resp., parent) $p_+$, if it has any, and can delete the message if it has none. (Moreover, to forward or delete  a ``contract'' message, $p$ must be able to contract through a handover with its branch child, if it has one; Algorithm \ref{alg:secondMovementPrimitive} ensures this is eventually possible by having non-leaf branch particles contract through a handover with their child whenever possible, and the leaf branch particle contract into its head when expanded.) Note that to be able to receive this message, $p_+$ must not hold a message (if $p_+$ does hold a message, then the message at $p$ waits). Whenever $p$ either forwards or deletes a message, $p$ also executes the message's action (i.e., $p$ either expands or contracts), as described in Algorithm \ref{alg:secondMovementPrimitive}. (Note that the stem particles do not break connectivity while moving in this manner by Observation \ref{obs:stemNoDisconnect}.)
Every message is created by a particle $p$ with a $start$ flag set to $true$.
(This is not described in Algorithm \ref{alg:secondMovementPrimitive}). Actually, $p$ creates a ``double'' expand message, in the sense that when possible, $p$ forwards one expand message to its parent and one to its child (if they exist) simultaneously. After forwarding these expand messages, $p$ creates and sends, similarly, ``contract'' messages to both its parent and child (if they exist) simultaneously once this is possible. If $p$ is not the leaf, then after its child $p_+$ no longer holds a ``contract'' message, $p$ sets $p_+.start$ to $true$ and $p.start$ to $false$.  Otherwise (p is now the leaf), $p$ waits until all of the particles in the stem and branches (of the connected component) are contracted -- using {\Detect} -- before it orders all stem (and connected branches) particles to terminate Algorithm \ref{alg:secondMovementPrimitive}.

\begin{algorithm}
\caption{\enspace Moving by $k$ points in the direction of $\vrot$}
\label{alg:secondMovementPrimitive}
\begin{algorithmic}[1]
\State \textbf{During the atomic activation of a stem particle $p$:}


	\If{$p$ has a ``contract'' message and can forward it} \label{line:contractStartStep2}
		\If{$p$ is expanded}
			\If{$p$ has a branch child $q'$ and $q'$ is contracted} 
				\State $p$ contracts through a handover with $q'$, and forwards the ``contract'' message  \label{line:contractHandoverStep2}
			\ElsIf{$p$ has no branch child}  
				\State $p$ contracts and forwards the ``contract'' message \label{line:contractStep2}
			\EndIf
		\ElsIf{$p$ is the head of a virtual particle $(q,p)$}
		    \State the virtual particle $(q,p)$ virtually contracts into its head
		    \State // $p$ and $q$ are renamed: the details are given in the virtual contraction operation's definition.
			\State $p$ forwards the ``contract'' message \label{line:contractEndStep2}
		\EndIf
	\ElsIf{$p$ has an ``expand'' message and can forward it}  \Comment{One easily sees that $p$ must be contracted} \label{line:expandStartStep2}
		\If{the other endpoint $u$ of $\vrot$ is occupied by some particle $p'$}
			\State $p$ virtually expands into $p'$
		\Else \enspace $p$ expands into $u$  \label{line:expansionStep2} 
		\EndIf
		\State $p$ forwards the ``expand'' message	 \label{line:expandEndStep2}
	\EndIf
\State
\State \textbf{During the atomic activation of a branch particle $p$:}
\If{$p$ is expanded}
	\If{$p$ has no child} $p$ contracts into its head  \label{line:branchPull1}
	\ElsIf{$p$'s child is contracted} $p$ contracts through a handover with its child \label{line:branchPull2}
	\EndIf
\EndIf
\end{algorithmic}
\end{algorithm}

Now, let us describe part (2) of primitive {\PRP} which rotates the stem by $60 \degree$ clockwise.
Intuitively, the rotation of the stem is not accomplished by any rotation-like behavior from particles. Instead, they continue walking in the same direction as in part (1). However, the second particle of the stem (the child of the root) walks one point forward, the third particle walks 2 points forward, etc. To do so, particles do not count (beyond some constant). Instead, for each point walked by particle $p_j$ on the stem, particle $p_{j+1}$ walks one point. Then, when particle $p_j$ is done, its child $p_{j+1}$ walks yet another point.

Specifically, this is done just by making three simple modifications to Algorithm \ref{alg:secondMovementPrimitive}.  First, the $start$ flag is set to $true$ for the root's child instead of the root. (If the root has no child, then the rotation is done.) Second, the particle that creates an ``expand'' or ``contract'' message (i.e., the particle with the $start$ flag set to $true$) only forwards the messages to its child (and not to its parent). 
The last modification concerns the stem's tree structure. Once a particle sets its start flag to $false$, it modifies its parent and child edges to their clockwise successors (to maintain the stem's tree structure despite the staggered movement in direction $\vrot$).

It is easy to observe that these three modifications are enough to ensure the particles $p_1,\ldots,p_k$ move, respectively, by $0,\ldots,k-1$ points in the direction of $\vrot$. 
In the triangular grid, this corresponds to a rotation around the stem's root. Observe also that, although the particles do not move the same amount of times, the stem remains connected throughout the rotation because (1) each particle $p$ makes one move in between each of its parent's moves and (2) the additional move $p$ does when the parent is done moving induces no disconnection due to the choice of $\vrot$ -- giving the second half of Observation \ref{obs:stemNoDisconnect}. (The first point is simple to show; before particle $p$ moves, it does not forward its move message, and thus cannot accept another move message; hence, this delays the parent from forwarding its next move message to $p$ and moving again.) 

\begin{observation}
\label{obs:stemNoDisconnect}
The stem does not disconnect during part (1) nor part (2) of primitive \PRP~.
\end{observation}

\begin{lemma}
\label{lem:secondMovementPrimitive}
When executed by a stem of size $k$, primitive \PRP~ terminates in $O(k)$ rounds.
\end{lemma}

\begin{proof}
During the executions of parts (1) and (2) of primitive \PRP, the (contracting) behavior of the branch particles resembles procedure Contraction of primitive {\OMP}. Hence, any correctness and runtime result needed for the branch particles can be easily obtained by adapting the proof of Lemma \ref{lem:firstMovementPrimitive}. For part (1) of primitive {\PRP}, one can easily see that $2k$ messages are created. Specifically, every stem particle $p$, from the root to the leaf, creates an ``expand'' message followed by a ``contract'' message. These messages are forwarded away from $p$ (and never return because of the stem's tree structure). Hence, every stem particle receives all the $2k$ messages. Whenever a stem particle receives one of these two messages, it can execute the corresponding movement action in $O(1)$ rounds and does not disconnect the stem by Observation \ref{obs:stemNoDisconnect}. (Note that if $p$ has a branch child $q'$, but $q'$ is expanded, then $q'$ contracts in $O(1)$ rounds -- possibly through a handover.) Hence, every stem particles moves $k$ points in the direction of $\vrot$. Moreover, the messages are pipelined and thus, following \cite{cidon1995greedy,mansour1993greedy} here too, part (1) of primitive {\PRP} takes $O(k)$ rounds. The correctness and runtime of part (2) of primitive {\PRP} can be obtained similarly.
\end{proof}

\paragraph*{Primitive {\SDP} (Stem Doubling Primitive): Moving back to $l$ and doubling the stem's size} The third movement primitive \SDP~ resembles primitive \OMP~ (Algorithm \ref{alg:firstMovementPrimitive}). Initially, the stem is contracted and occupies points $v_k,\ldots,v_{2k-1}$. In the first part, the stem expands in the direction of $\vin$. This can be done by using the first part of Algorithm \ref{alg:firstMovementPrimitive} with the reverse direction $\vin$. 

Following the above first part of \SDP, there may be some  $j \in \{0,\ldots,k\}$ virtually expanded particles in the stem.
In a second part of \SDP, these virtually expanded particles break off into two normal contracted particles. After which, non-root expanded stem particles contract through handover with their parents, if possible.
When these movements are no longer possible, the last $j$ particles of the stem (including the leaf) are contracted.

The leaf detects when that happens -- using {\Detect} -- and starts the third part of \SDP~ -- the actual doubling of the stem.
Here, stem particles absorb newly-collected branch particles in the stem -- that is, stem particles contract through handovers with branch particles, thus moving them into the stem -- whenever possible. Let $p$ be some branch particle brought in that way into the stem. Note that $p$ is still expanded towards the branch. Particle $p$ then contracts (through handover with their branch child if such exists). Then, $p$ starts acting like the other stem particles.
A stem particle with no branch children contracts through handover with its child in the stem. The root detects -- using {\Detect} -- when the stem is contracted or when there are no longer any newly-collected branch particles. In the first case, the primitive terminates. 
In the latter case, the root first forces the stem to contract -- by allowing the leaf particle to simply contract if expanded -- after which the primitives terminates. Note that in the second case, the stem grows but does not double in size.

\begin{figure}[ht]
     \centering
     \begin{subfigure}[t]{0.47\textwidth}
        \centering
        \includegraphics[width=0.95\linewidth]{CollectionThirdMovement}
        \label{fig:collectionThirdMovement1}
        \caption{Initially, the stem occupies the black points $v_k,\ldots,v_{2k-1}$. The checkered points are occupied by (contracted) particles that were collected during the previous step (i.e., the rotation around $l$).}
     \end{subfigure} \hfill
     \begin{subfigure}[t]{0.47\textwidth}
        \centering
        \includegraphics[width=0.95\linewidth]{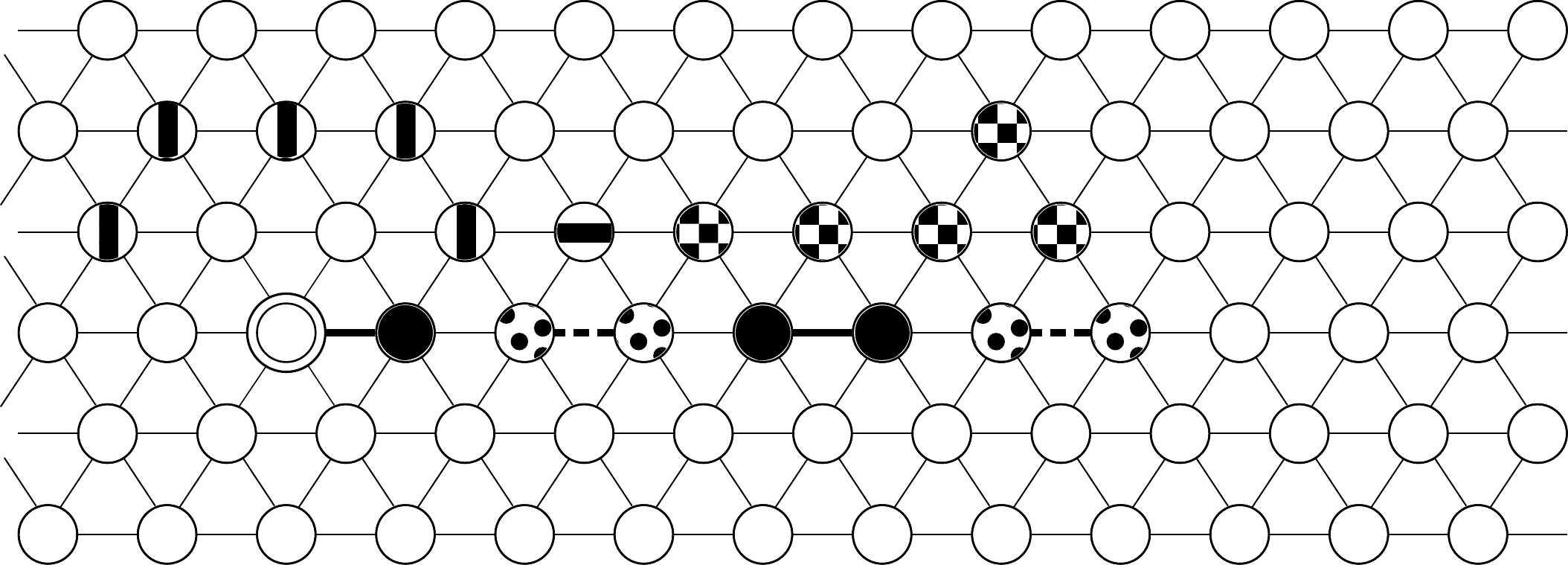}
        \caption{After the first part, all of the points $l,v_1,\ldots,v_{2k-1}$ are occupied by either an expanded particle or a virtually expanded particle (i.e., two contracted particles -- see the points with ``spots''). }
        \label{fig:collectionThirdMovement3}
     \end{subfigure}  \hfill
     \begin{subfigure}[t]{0.47\textwidth}
        \centering
        \includegraphics[width=0.95\linewidth]{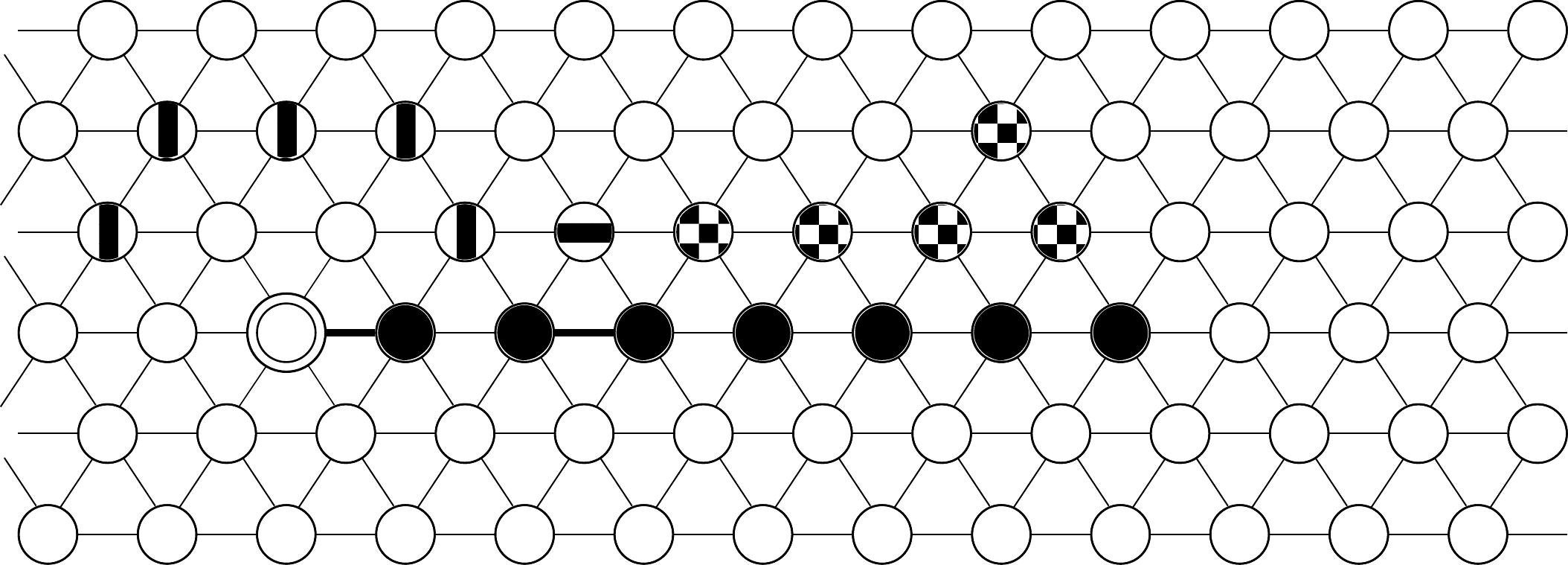}
        \caption{After the second part, the contracted particles occupy the last particles in the stem (see the black points with no incident bold edges). }
        \label{fig:collectionThirdMovement4}
     \end{subfigure}  \hfill
     \begin{subfigure}[t]{0.47\textwidth}
        \centering
        \includegraphics[width=0.95\linewidth]{CollectionThirdMovement2}
        \caption{During the third part, the stem moves backs to $l$ and doubles in size (if there are enough newly-collected particles to do so). }
        \label{fig:collectionThirdMovement2}
     \end{subfigure} 
\caption{Third Step: Moving a stem of size $k=4$ back towards $l$ while doubling its size}
\label{fig:explanationsThirdStep}
\end{figure}

\begin{lemma}
\label{lem:thirdMovementPrimitive}
When executed by a stem of size $k$, primitive {\SDP} terminates in $O(k)$ rounds.
\end{lemma}

\begin{proof}
The first part of primitive \SDP~ is essentially, the same as primitive \OMP, only with a different direction. Thus, correctness and runtime can be shown similarly to the proof of Lemma \ref{lem:firstMovementPrimitive}. The second and third parts highly resemble the second part of \OMP. Hence, their correctness and runtime is easily obtained by adapting the proof of Lemma \ref{lem:firstMovementPrimitive}.
\end{proof}

Now, we can prove Lemmas \ref{lem:collectedIsConnected} and \ref{lem:collectionPhaseProperties}.

\begin{proof}[Proof of Lemma \ref{lem:collectedIsConnected}]
Let us show by induction that the set of collected particles is connected at the start of phase $i \geq 1$. For $i > 1$, that statement is equivalent to the following: the set of collected particles is connected at the end of phase $i-1$. The base case holds trivially. Next, assume that the set of collected particles is connected at the start of some phase $i \geq 1$. Primitive {\OMP} disconnects the stem from its branches (and possibly leaves behind particles in $v_0,\ldots,v_{k-1}$). Note that the particles left behind do not move during primitive {\PRP}. After primitive {\PRP}, the set of stem particles and newly-collected branch particles is connected. Thus, after primitive {\SDP}'s first part (i.e., the expansion towards $l$), the set of collected particles is connected. It is easily shown that the set of collected particles remains connected throughout the remainder of primitive {\SDP}. Hence the induction step follows and the lemma statement holds.
\end{proof}

\begin{proof}[Proof of Lemma \ref{lem:collectionPhaseProperties}]
The phase's runtime follows from Lemmas \ref{lem:firstMovementPrimitive}, \ref{lem:secondMovementPrimitive} and \ref{lem:thirdMovementPrimitive}. Furthermore, primitives {\OMP} and {\PRP} sweep all the points at grid distance $j \in \{k,\ldots,2k\}$ from $l$ by Lemmas \ref{lem:firstMovementPrimitive} and \ref{lem:secondMovementPrimitive}. Since all particles are within distance $\epsilon_G(l)$ of $l$ (by Lemma \ref{lem:betterParticleDistribution}), no particle is collected if $k > \epsilon_G(l)$. Next, consider $k \leq \epsilon_G(l)$. Initially, there exists (at least) one uncollected particle at grid distance $j$ from $l$ -- for any $j \in \{k,\ldots,\min\{2k,\epsilon_G(l)\}\}$ -- by Lemma \ref{lem:betterParticleDistribution}. Primitives {\OMP} and {\PRP} collects all of these particles, such that they are either in the stem, in the branches of distance $k \leq j \leq 2k-1$ from $l$ or occupy some of the nodes $v_0,\ldots,v_{k-1}$ (since they were left behind during primitive {\OMP}) at the start of primitive {\SDP}. Let $k_1$ be the number of branches of distance $k \leq j \leq 2k-1$ from $l$ and let $k_2$ be the number of contracted particles in the stem after the second part of primitive {\SDP}. Note that $k_1 + k_2 \geq \min\{k,\epsilon_G(l)-k\}$, by Lemma \ref{lem:betterParticleDistribution} and the fact that (newly-collected) branch particles do not change their distance from $l$ throughout the rotation. Moreover, note that the $k_2$ contracted particles are the last $k_2$ particles of the stem. Hence, the stem can ``pull'' in at least one branch particle into the stem from each of the (above-mentioned) $k_1$ branches. Thus, the stem's size grows by at least $\min\{k,\epsilon_G(l)-k\}$ when primitive {\SDP} ends. Furthermore, primitive {\SDP} cannot grow the stem's size by more than double. Hence, the lemma statement holds.
\end{proof}

\newpage

\section{  The Outer-Boundary Detection Primitive (\OBD)}
\label{sec:boundaryDetection}

This section is included for the sake of completeness -- to demonstrate that the runtime cost of removing the boundary assumption is only linear in $L_{out}+D$. Recall that in the previous section, we assumed that a particle on the outer boundary in the initial configuration knows which of its ports lead to the outer face.\footnote{In this section, we still make the common assumption that particles have common chirality. Note that this assumption is made by most previous papers too 
(in fact, it is not too hard conceptually to avoid this assumption; in particular,  \cite{EKLM19} may have not bothered to avoid the additional runtime overhead of the chirality agreement
 since the runtime of the rest of their algorithm was cubic, so the runtime of the chirality agreement was not their bottleneck).}

To implement the detection procedure (\OBD), we used tools from traditional distributed computing in the sense that no particle movement is used.  
The detailed definition of the primitive is necessarily cumbersome, because the memory size of each particle is constant. Hence, coding the procedure resembles programming a Turing machine where the particles simulate both the tape, the head, and the finite control. 
The description becomes even more cumbersome since the algorithm uses pipeling to increase the efficiency. This makes its description resemble programming multiple Turing machines in parallel, working on the same tape. Hence, we first describe the primitive informally and then give a more formal definition. Similarly, it is pretty easy to understand the ideas of the proof informally (but a more detailed proof is also given for completeness).

Very informally, the particles over a global boundary simulate a virtual ring of v-nodes of a classical network type. Note that by abuse of notation, v-nodes -- defined in Section \ref{sec:prelim} -- are now considered to be constant-size computational agents whose communication is limited to the ring and with no movement capabilities. Moreover, they possess a ``sense of direction'' \cite{flocchini1998sense} allowing each v-node to know which neighbor is clockwise and which is counter-clockwise. The ring executes a classical style algorithm in which v-nodes sum up their boundary counts. Some grid geometry insight is used here to interpret the meaning of that sum: the sum is equal to 6 if the ring corresponds to the outer boundary, and to -6 otherwise -- see Observation \ref{obs:sumCounts}.
To perform this summing up process, the v-nodes first elect a leader (otherwise, over a virtual ring, it is not clear where to start summing up and when is this process terminated).\footnote{Up to 6 leaders may be elected because of the symmetry of the grid (another grid geometry insight taken from \cite{BB19}). } At any point of time during the election process, the ring is covered by non-overlapping sets of consecutive v-nodes. Some of those are called ``segments''. The others are single sets. Initially, every v-node is in a segment containing only itself (so a segment may be a singleton set too).
Let us first describe the election as if it is executed by segments (the non-segment singletons may just react) . We implement that view later. 

Associated with each segment of some $k$ v-nodes $v_1(B_1), ... v_k(B_k)$  is a vector (called ``label'') of length $k$, [$c(v_1(B_1)),\ldots,$ $ c(v_k(B_k))$] going clockwise. A segment $s$ needs to compare its vector with the vector of the next segment clockwise, denoted by $s_1$.
If $s$'s vector is smaller, segment $s$ ``kills'' the next segment $s_1$ and ``swallows'' its v-nodes one by one (unless all the v-nodes of $s$ are swallowed by some other segment before $s$ finishes swallowing all the v-nodes of $s_1$).

The speed up compared to the boundary detection algorithms in \cite{BB19,EKLM19} is obtained by removing three bottlenecks. First, two consecutive segments were compared v-node by v-node  in \cite{BB19,EKLM19} -- like string comparison on a Turing machine. 
We manage to pipeline the comparisons. Second, the comparison process is somewhat complicated by the fact that segments may change over time. For that reason, the previous algorithms slowed the comparison down even further to compare only when the segments were stable. We manage to overcome the segments' dynamic nature without slowing the competition down. Finally, in \cite{BB19,EKLM19}, each particle $p$ waited to hear a termination signal from the detection primitive over every boundary $p$ belonged to. We use flooding to have outer boundary particles announce their termination to all other particles, resulting in faster termination detection. The flooding saves the propagation time over possibly longer boundaries. Moreover, the announcement just by the outer boundary prevents the congestion created when the results of every boundary were announced. 

Following \cite{BB19,EKLM19}, a property that follows from the embedding on the grid is that eventually, some configuration $C$ is reached that is stable in the sense that the virtual ring is covered by up to 6 equal segments. Then, a segment cannot grow nor shrink. Moreover, a segment can know how many such eventual equal segments may exist in $C$ by considering only the sum of the counts over its own segment. 
Hence, a segment $s$ can detect that the ring has stabilized to $C$ by comparing with up to 6 segments clockwise. (Moreover, that comparison takes $O(|s|)$ in runtime, since the other segments are supposed to have the same size as that of $s$.) In addition, by multiplying its own sum at this point by the number of segments in the stable configuration, each segment can know the total sum of the counts over the ring. 

For this high level description, it remains to explain how the segment computation is implemented at the level of v-nodes. 
Implementing most of the above operations is trivial, since the head v-node (the one on the segment's clockwise side) can be the segment's leader and thus oversee operations. Moreover, implementing the communication of the segment with all the other v-nodes is trivial. That is, to communicate with the next segment clockwise (if such exists), the head v-node of a segment just communicates clockwise. A more interesting implementation is that of the comparison between two neighboring segments $s, s_1$. The difficulty comes from the v-nodes' constant memories (which implies, furthermore, that their communication has constant ``bandwidth''). Such more complex operations are detailed later, after presenting the implementation of the v-nodes. Still, as a very informal example, let us outline how the vectors of two consecutive segments of equal size are compared using pipelining. 
The algorithm (1) creates, conceptually, a ``train'' consisting of one mobile agent (called {\em ``token''} later) per v-node in the segment, each agent carrying the boundary count of its creating v-node, (2) reverses the direction of one of two neighboring trains such that the trains' head agents neighbor each other and the trains' tail agents are far away from each other, and then (3) compares the train head agent of $s$ with that of $s_1$, then delete those head mobile agents, advances the next token in each train until they neighbor each other and compares them, etc.  

The  implementation of a virtual ring algorithm over the particles on a global boundary is described in Section \ref{subsec:virtualParticlesPreliminaries} together with other basic definitions such as segments and their labels. Section \ref{subsec:lexicographicComparison} presents an efficient lexicographic comparison primitive to compare the length and the label of a segment $s$ to those of its neighboring (successor clockwise) segment. It turns out that the successor segment may still grow clockwise during the comparison, as well as between comparisons. However, $s$ cannot. Hence, if $s$ is found to be ``larger''  than (or equal to) its successor segment, $s$ compares again and again. However, if $s$ is found to be ``smaller'' then $s$ will remain smaller and the successor segment can be ``killed''. This is described in Section \ref{subsec:segmentCompetition}. If a segment $s$ and its successor are found to be equal, it {\em  may} be the case that the above mentioned ``stable'' configuration has been reached with up to 6 equal segments covering the virtual ring. The detection of the stable configuration, the summing up process and the termination announcement are described in Section \ref{subsec:boundaryDetection}.

Before delving into details about the algorithm, let us say a few informal words also about the analysis. The high level correctness arguments are basically those of \cite{BB19,EKLM19}. The correctness proofs of each individual primitive are rather standard (e.g. that there always remains at least one segment, since it cannot happen that every segment is smaller than the next one on a ring, and if segment $s$ is smaller than $s_1$, then $s$ remains smaller). To see the linear runtime complexity, note that a smaller segment $s$ ``beats'' a larger one $s_1$. Moreover, for $s$ to grow by $|s_1|$ v-nodes, $s$ spends $O(|s_1|)$ time on comparisons and ``swallowing''. In particular, the travel of a train of size $|s|$ into a segment of size $|s_1|$ is accomplished in runtime $O(|s|)$. The reversing of the order of a train of $|s_1|$ mobile agents ({\em tokens}) takes $O(|s_1|)$ time. Also, the travel to check whether there are 6 {\em equal} segments of the same size $|s|$ takes $O(|s|)$ time.

\subsection{Implementation and Definitions}
\label{subsec:virtualParticlesPreliminaries}

Recall that a single (contracted) particle $p$ may occupy a boundary point $v$ with multiple local boundaries and thus, may be associated with multiple v-nodes. In that case, $p$ simulates these multiple v-nodes simultaneously (and each v-node participates in an execution of primitive \OBD~ that is unrelated to the executions of ~\OBD~ involving other v-nodes of the same particle ). Particles simulate v-nodes by using designated memory variables and communicating (from a v-node to its clockwise successor or predecessor, which must always exist -- see Observation \ref{obs:existenceOfSuccessorAndPredecessor}) using the following implementation. (Note that this implementation leverages the particles' common chirality, which we have already assumed, without loss of generality, to be clockwise.) With this implementation, v-nodes form one virtual ring per global boundary. Although the ring's real orientation (from a global view outside the particle system) is clockwise if the global boundary is the outer one, and counter-clockwise otherwise, that ``real'' orientation has no impact in this section and thus the rings are all considered to be oriented clockwise. In particular, Observation \ref{obs:sumCounts} holds no matter what the ring's ``real'' orientation is.

Each particle $p$ occupying a point $v$ uses the following (two-dimensional) array to store incoming messages: $from[0,\ldots,5][0,\ldots,5]$. The first index indicates the message's sender particle -- more specifically, the port $p$ assigns to the edge leading to the sender particle -- and the second distinguishes the receiver v-node $v(B)$ -- or more specifically, the local boundary $B$. When $p$ simulates the v-node $v(B)$, it computes port numbers $\tilde{j_s}$ and $j_s$ (and, respectively, $\tilde{j_p}$ and $j_p$) to communicate with $v(B)$'s successor (respectively, predecessor) v-node $v_s(B_s)$ (respectively, $v_p(B_p)$). Ports $\tilde{j_s}$ and $\tilde{j_p}$ are used to indicate $p$ to the simulating particles of, respectively, $v_s(B_s)$ and $v_p(B_p)$. Ports $j_s$ and $j_p$ are used to specify, respectively, the receiver v-nodes $v_s(B_s)$ and $v_p(B_p)$ -- the crucial property being that $j_s$ (respectively, $j_p$) corresponds to one of the edges of $B_s$ (respectively, $B_p$). 

Specifically, let $i_f$ be the port corresponding to $B$'s first edge and $i_l$ to $B$'s last edge. Then, the ports $i_p = i_f - 1 \; mod \; 6$ and $i_s = i_l + 1 \; mod \; 6$ lead respectively to $v$'s clockwise predecessor and successor points w.r.t. $B$. Let us call the particles occupying these points (respectively) $p_p$ and $p_s$. Particle $p$ computes the port numbers $\tilde{j_p},\tilde{j_s}$ that $p_p$ and $p_s$ respectively assign to the edges $\{p_p,p\}$ and $\{p_s,p\}$. (Recall that $p$ has access to that information -- see Section \ref{subsec:system}.) From these, particle $p$ computes $j_p$ and $j_s$ as follows: $j_p = \tilde{j_p} - 1 \; mod \; 6$ and $j_s = \tilde{j_s} + 1 \; mod \; 6$.

When $v(B)$ wishes to write to its clockwise successor (respectively, predecessor) v-node, $p$ writes $v(B)$'s message in $p_s.from[\tilde{j_s}][j_s]$ (respectively, $p_p.from[\tilde{j_p}][j_p]$). When $v(B)$ wishes to read incoming messages from its clockwise predecessor (respectively, successor), $p$ reads $p.from[\tilde{j_s}][i]$ (respectively, $p.from[\tilde{j_p}][i]$) for any port $i$ that corresponds to an edge in $B$.

\begin{lemma}
The above implementation allows particles to correctly simulate communication between v-nodes. 
\end{lemma}

\begin{proof}
First, recall that particles have common chirality, which we consider without loss of generality to be clockwise. 
Consider that the v-node $v(B)$, simulated by a particle $p$, wishes to communicate a message to its clockwise successor v-node $v_s(B_s)$, simulated by some particle $p_s$ (necessarily in $\mathcal{N}(p)$). By definition of $v_s(B_s)$, there exist two unique edges $e \in B, e_s' \in B_s$ with a common unoccupied endpoint $u$. By Observation \ref{obs:existenceOfSuccessorAndPredecessor}, $e$ is the last (clockwise) edge of $B$. Since port $i_l$ corresponds to $e$, port $i_s = i_l + 1 \; mod \; 6$ corresponds to the clockwise successor of $e$ -- the edge $e_s=\{v,v_s\}$. By the model assumptions, $p$ correctly obtains the port $\tilde{j_s}$ that $p_s$ assigns to $e_s$ (and thus to $v$ and $p$). Then, $p$ writes in $p_s.from[\tilde{j_s}][j_s]$ where $j_s = \tilde{j_s} + 1 \; mod \; 6$.
Since $p$ and $p_s(B_s)$ have the same chirality (i.e., clockwise), $e_s$ is the clockwise predecessor of $e_s'$. Hence, port $j_s$ corresponds to $e_s'$ for $p_s$, and thus to a port in $B_s$. Thus, $p_s$ correctly decodes that the message is sent from the particle occupying $v$ (i.e., $p$) and is sent to $v_s(B_s)$ (since $j_s$ corresponds to an edge in $B_s$). Similarly, the communication between $v(B)$ and $v_p(B_p)$ can be shown to be correct. 
\end{proof}

\paragraph*{Segments.} The algorithm maintains disjoint {\em segments}; each is defined as a finite, contiguous sequence of v-nodes along a global boundary. (Initially, every v-node belongs to a segment containing only itself.) A segment is considered to be directed clockwise. The \emph{tail} of the segment is the first v-node of the segment encountered when walking on the ring clockwise and the \emph{head} the last one. A segment is dynamic in the sense that v-nodes can leave and join it. The only v-node who may leave a segment at some point in time is its tail at that time. 
The head is responsible for the segment's growth. That is, v-nodes may either belong to some segment -- those are \emph{pledged} -- or may not belong to a segment -- those are \emph{free}; if the head's successor is free, then the head requests its successor to join the segment. Then, that v-node is said to be \emph{absorbed} by the segment and becomes its new head. A segment's tail may be in a \emph{defector} state (initially, no v-node is in the defector state). Once activated, a defector v-node becomes free and exits the defector state; if this newly free v-node is not a segment head then it also forces its successor to become a defector. A segment with a defector tail is said to be \emph{disbanding}. Indeed, we show later, in Section \ref{subsec:segmentCompetition}, that a disbanding segment eventually disappears even if for some time its head still continues absorbing free v-nodes.

The \emph{label} of a segment $s$  --  denoted by $label(s)$  --  is the corresponding sequence of boundary counts, one per v-node; the \emph{length} of $s$  --  denoted $|s|$  --  is the number of v-nodes in the segment and the \emph{sum} of $s$  --  denoted $sum(s)$  --  is the sum of the boundary counts along the segment.
For some segment $s$, if the predecessor $v(B)$ of $s$'s tail (resp., successor of its head) is pledged, then $v(B)$'s segment is said to be the \emph{predecessor} (resp., \emph{successor}) of $s$.
Additionally, for any given (non-disbanding) segment $s$, we define $s$'s \emph{projection} on a given global boundary as the contiguous clockwise sequence of v-nodes starting from $s$'s tail $v_t(B_t)$ (and including all of $s$) to (but not including) the next non-defector tail $v_1(B_1)$. (Possibly, $v_t(B_t)=v_1(B_1)$.)
The projection may contain free v-nodes as well as pledged ones that belong to some other segment $\tilde{s}$. In the latter case, segment $\tilde{s}$ is necessarily disbanding. Projections play an important role in Section \ref{subsec:segmentCompetition}.

\subsection{Lexicographic Comparison Primitive ({\LCP})}
\label{subsec:lexicographicComparison}

A segment $s$ is {\em smaller} than its successor segment $s_1$ if either (a) the size of $s$ is smaller, or (b) the sizes are equal and the label of $s$ is smaller lexicographically. 
The comparison primitive is composed of two modules, one to check condition (a) and the other to check (b). 
Intuitively, with the separation, we can allow segments that change; we do not want to freeze the segments during comparison, since this would have resulted in the larger time complexity of the boundary detection of \cite{BB19,EKLM19}. Moreover, the separation allows the comparison initiated by a segment $s$ to have a runtime of $O(|s|)$.

The primitive is described using \emph{tokens}, that is, messages that (virtually) act as mobile computational agents. Very informally, in module (a) (length comparison), each v-node in $s$ generates a token with the tail token and the head token being distinguished. That is, a token in the train may be the {\em head token}, {\em the tail token} or even both (when the segment contains only one v-node). Let such a sequence consisting of one token per v-node be called {\em train}. This train of $s$'s tokens starts marching in a FIFO order into $s$'s successor segment $s_1$, such that only a constant number of tokens can reside in a single v-node. First, the head token moves into the tail of $s_1$. After that, for the head token to advance by one step to the next v-node in $s_1$, the head token ``consumes'' the token following it in the sequence. Thus, the number of tokens in the train shrinks as they advance. If the head token consumes the tail token but still has not reached the head v-node of $s_1$, then (and only then) $s$ is shorter than $s_1$. 

For module (b) (label comparison), both $s$ and $s_1$ generate sequences of tokens, one per v-node. A token generated by a v-node with boundary count $c$ carries $c$ with it. The module then reverses the order of the tokens in $s_1$. That is, let the train be $t_1$ (tail token), $t_2, ... t_h$ (head token). Token $h-i$ is moved to virtual token $i$, for $0\le i \le h$. (To reorder, the algorithm virtually completes $s_1$ to be a virtual ring $0, 1, ..., h-1, h, h-1, h-2, ... 0$ and the tokens march in a FIFO order over the ring to their designated places.)
Next, the head tokens of $s$ and $s_1$ compare themselves -- note that they now reside in neighboring v-nodes ($s$'s head v-node and $s_1$'s tail v-node). Then, the head tokens of both trains are deleted, the next tokens in those trains become the head tokens, march to the neighboring v-nodes and compare themselves, etc. This concludes the high level description. Let us now supply more details.

\paragraph{Multiplexing trains of tokens:} Note that at the same time segment $s$ utilizes a sequence of tokens to compare with its successor $s_1$, the predecessor $s_0$ of $s$ may initiate a comparison between $s_0$ and $s$. This implies additional trains traversing $s$, such as the train of $s_0$ and the train $s$ generates in order to facilitate the comparison with $s_0$. Luckily, it is easy to show that the number of trains that traverse the v-nodes of segment $s$ at any given time is constant -- see Observation \ref{obs:constantAmountOfTrains}.  
To prevent the case that one train blocks the other, we designate several memory variables in each v-node for each possible train in each direction (clockwise and counter-clockwise). Note that a constant number of trains can pass from a v-node $v(B)$ to its neighbor $u(B')$ simultaneously without increasing the runtime, since the amoebot model allows the writing of several memory variables at the same activation of a particle.

For the comparison primitive discussed now, two different train types are used: one for the length comparison and one for the label comparison.  Note that it is straightforward to translate the token based description into a corresponding algorithm for the amoebot model (with v-nodes): whenever a v-node holds a certain token, it simulates the token accordingly. If the v-node holds (necessarily a constant number of) multiple tokens, the v-node simulates them in their order of arrival, thus keeping the order also when forwarding them.

\paragraph*{Details for Length Comparison.} The initiating segment's head creates a \emph{length train creation} token, which moves along the segment counter-clockwise until it reaches the tail, instructing the v-nodes to generate \emph{length} tokens, the collection of which is a \emph{length} train encoding $|s|$ in unary (since the train is composed of $|s|$ tokens). The length train moves clockwise from $s$ into $s_1$. However, in $s_1$, forwarding the train head token \emph{consumes} a unit token. More precisely, a head token (of the length train of $s$) moves from a v-node $v_1(B_1)$ of $s_1$ to the next (clockwise) v-node $v_2(B_2)$ of $s_1$ if and only if (1) v-node $v_2(B_2)$ has a free memory location reserved to the length train of $s$ and (2) $v_1(B_1)$ holds at that time another token $t$ (beside the head token) of that train. Token $t$ is then deleted when the head token is forwarded. Note that every other token of any train marching clockwise (correspondingly, counter-clockwise) in a segment marches to the next v-node $v(B)$ of the segment clockwise (resp. counter-clockwise) provided $v(B)$ has a free slot (in those reserved for that train). No consuming takes place in the march of other tokens. 

If all the tokens of the length train are consumed before the tail token reaches the head of $s_1$ (i.e., a non-head v-node holds both the head token and tail token and no other token in that train), then $|s| < |s_1|$. If, on the other hand, the head of $s_1$ holds both the head token and tail token and no other token in that train, then $|s| = |s_1|$. In which case, the head v-node of $s_1$ is \emph{marked} and remains marked until primitive ~{\LCP} terminates. (In the case that $s_1$ absorbs additional v-nodes before or during the label comparison, the mark is used to prevent the runtime from growing beyond $O(|s|)$). 
Otherwise, $|s| > |s_1|$. (This is the case in which the head of $s_1$ holds the head token together with an additional one.)
In all three cases, when the head token computes the result of the computation, it becomes a \emph{result} token. This last token moves (counter-clockwise) to the head of segment $s$, deletes all the remaining tokens in the (length) train along the way and returns the length comparison's value.

\paragraph*{Label Comparison.} The initiating segment $s$ creates a {\em label token train} representing the segment's label. More precisely, it generates a \emph{count token} at each v-node $v(B)$ of segment $s$, holding $v(B)$'s boundary count $c(v(B))$. The token generated at the head is also a head token, and the one generated at the tail is also a tail token.

Segment $s_1$ is then instructed to create a {\em reverse count train} that is structured initially the same as the count train of segment $s$ above, with one important exception: for the reverse count train, tokens are only generated by v-nodes located between the tail of $s_1$ (including) and the marked v-node of $s_1$ (including). The head token of this train then marches counter-clockwise until it reaches the tail of $s_1$. Every other token of the reverse count train first marches clockwise until it reaches the marked v-node (of $s_1$) and then counter-clockwise. 
(If no additional actions are taken later, then eventually, this process leads to the reverse count train having its head token in the tail v-node of $s_1$ and its tail token in the marked v-node of $s_1$.) 

The initiating segment's head compares the count tokens of both trains in order to determine the result of the lexicographic comparison. More precisely, the head v-node of $s$ compares the first-to-arrive count tokens (of $s$) with the first-to-arrive token in the reverse train at the tail of $s_1$. If one of them is larger, then the corresponding segment is deemed larger; the head v-node of segment $s$ stores the result of the comparison and initiates a delete of the remaining tokens of both segments. Otherwise, the two compared equal tokens are deleted so the next tokens in both trains can be read by the head of $s$. (Recall that the count train marches clockwise while the reverse count token marches counter-clockwise.) Eventually, either one token in the compared pairs is larger (and the corresponding segment is deemed larger) or no count token remains. In the latter case, the segments are deemed equal. The result is stored at the head v-node of $s$. Finally, that head sends a \emph{clean} token to the marked v-node of $s_1$, which unmarks it, and then comes back to the head of $s$. At that point this primitive terminates. 

\begin{remark*}
\begin{itemize}
    \item The execution of this primitive may be interrupted by decision of the algorithm using it as a subroutine (namely, by the Outer-Boundary Detection primitive). Any v-node or token executing the primitive and being instructed to cancel, terminates its execution of the primitive. (This can happen if segment $s$ loses to its predecessor segment -- a process that takes place in parallel to the comparison between $s$ and $s_1$.)
    \item During the execution of the primitive, $s_1$ may grow. 
    \item This remark explains the conditions in the statement of lemma \ref{lem:correctnessLexicographicComparison} below. It turns out that even though conditional, the lemma is strong enough to prove later the correctness of the Outer-Boundary Detection primitive.
\end{itemize}
\end{remark*}

\begin{observation}
\label{obs:constantAmountOfTrains}
The number of trains that traverse the v-nodes of a segment $s$ at any given time is constant.
\end{observation}

\begin{observation}
\label{obs:tain-time}
Consider a train of length $x$. 
As long as the the token at the head of the train can move one step within $O(1)$ time (e.g., it has not arrived yet at its destination v-node) then each token moves $x$ steps within $O(x)$ time. 
If the head token can no longer move, then, within $O(i)$ time, a configuration is reached where the first $i$ tokens after the head cannot move (since the location each needs to move to is occupied by the previous token in the train). 
\end{observation}

\begin{proof}
See \cite{cidon1995greedy,mansour1993greedy}.
\end{proof}

\begin{lemma}
\label{lem:correctnessLexicographicComparison}
\label{lem:runtimeLengthScheme}
\begin{itemize}
    \item The primitive terminates in $O(|s|)$ time.
    \item The primitive returns the correct comparison result when $s_1$ (and $s$) does not grow and the execution is not cancelled. 
    \item if the execution is not cancelled, the following two (correctness) properties are guaranteed. First, if $label(s) < label(s_1)$ initially, than the result is <. Second, if the result is <, than $label(s) < label(s_1)$ at the end of the primitive.
\end{itemize}
\end{lemma}

\begin{proof}
First, note that $label(s_1)$ cannot become lexicographically smaller during the primitive. 
The termination and time complexity follow from Observation \ref{obs:tain-time} and from the observation that at any point, a token just waits for the token that precedes it in its own  train as well as for the other train.
\end{proof}

\subsection{Segment Competition}
\label{subsec:segmentCompetition}

Let us first describe this module as if it is performed by segments as computational agents, and explain the (rather straightforward) implementation of the segments' algorithm by v-nodes (and tokens and trains of tokens) later.
Every v-node starts as a segment of length 1: that is, as both the head and (non-defector) tail of a segment. Every segment performs \emph{expansion attempts} repeatedly until an outside signal (discussed in a later subsection) instructs it to stop. In an attempt, a segment $s$ whose successor v-node $v(B)$ is free (that is, unpledged) absorbs $v(B)$. Specifically, $s$'s head informs its unpledged successor who then becomes pledged and also the new head of segment $s$.
Alternatively, segment $s$ may have a successor segment $s_1$, that is, the successor v-node $v_1(B_1)$ of the head of $s$ is pledged. Hence, $v_1(B_1)$ is the tail of segment $s_1$. That tail $v_1(B_1)$ may be in a defector state (so $s_1$ is disbanding). This completes the attempt and another attempt starts. (Eventually, the defector v-node $v_1(B_1)$ becomes free, so $v_1(B_1)$ can be absorbed by a future attempt.)
 Otherwise, $s_1$ is not disbanding, and $s$ starts a \emph{competition and locking operation}. First, it starts a competition with $s_1$ by activating primitive ~{\LCP} (from the previous section). If $s$ wins (i.e., if $s$ has a strictly smaller label than $s_1$) then $s$ checks whether it is disbanding (that is, whether $s$'s tail v-node is in the defector state). If so, then no further action is taken by $s$ in the module described in this subsection, except for disbanding (v-nodes becoming defectors and then free). 
 Otherwise, $s$ locks its tail -- a {\em locked} v-node cannot be put in the defector state.
Next, $s$ forces $s_1$ to disband (once $s_1$ is unlocked if it is locked) and finally, $s$ unlocks itself. If this point is reached, the operation is said to be \emph{successful}. Otherwise the operation is \emph{unsuccessful}. \\
Whether successful or unsuccessful, the current expansion attempt is done and the segment starts a new one. Note that after an expansion attempt with a successful competition and a locking operation, each v-node of $s_1$ (one by one, starting from the tail of $s_1$) eventually becomes a defector, and then becomes free. Recall that these v-nodes are absorbed by $s$ in future expansion attempts (unless $s$ disbands too before absorbing all the v-nodes of $s_1$).

Importantly, the competition procedure described in this subsection never detects its own termination. An additional module is presented later to force the termination of the Outer-Boundary Detection primitive. Until the forced termination, each segment $s$ keeps competing with its successor $s_1$ if $s$'s label is greater or equal than $s_1$'s label. However, we show later that every global boundary is eventually \emph{stable}: that is, it is covered by non-disbanding segments with equal labels. Note that at that point, the segments of such boundaries no longer change. 

\paragraph*{Competition and Locking Operation} Although a simple description with segments as computational agents is given above, it is in fact, the segment's head who dictates its segment's expansion attempts. The following describes the competition and locking operation initiated by $s$. First, the head $v_h(B_h)$ of $s$ starts a lexicographic comparison with $s_1$. If (the result is that) $label(s) < label(s_1)$, then $v_h(B_h)$ sends a \emph{locking token} to $s$'s tail $v_t(B_t)$. Eventually, $v_t(B_t)$ receives that token. If $v_t(B_t)$ is a defector, then it sends a \emph{unsuccessful token} to $v_h(B_h)$ and the operation is done. Otherwise, $v_t(B_t)$ becomes locked and sends back a \emph{disband token} to $v_h(B_h)$. Once $v_h(B_h)$ receives this token, it waits until the tail $v_1(b_1)$ of $s_1$ is unlocked. When that happens, $v_h(B_h)$ instructs $v_1(b_1)$ to become a defector. Finally, $v_h(B_h)$ sends a \emph{unlocking token} to $v_t(B_t)$, which sends back an \emph{acknowledgement} token to $v_h(B_h)$. Upon $v_h(B_h)$'s reception of that token, the operation is over.

\begin{observation}
\label{obs:no-deadlock}
 A boundary always consists of at least one non-disbanding segment. 
\end{observation}

\begin{proof}
When the primitive starts, every v-node is a non-disbanding segment. Consider the competition between some segment $s$ and its successor $s_1$ as a result of which $s_1$ starts disbanding. (Note that this is the only case that $s_1$ can start disbanding.) From the description of the algorithm, at that time, $s$ is locked, and thus is not disbanding. 
Moreover, from the description of the algorithm and the correctness of the lexicographic comparison primitive (see Lemma \ref{lem:correctnessLexicographicComparison}),  segment $s$ locks itself only if its successor $s_1$ has a strictly greater label than $s$. Moreover, the label of $s$ cannot grow until its successor starts disbanding. At the same time, the label of $s_1$ cannot shrink until it starts disbanding, at which point, it cannot instruct any other segment (including $s$) to disband (because $s_1$ cannot lock). That is, if $s$ later disbands, it is because of another segment whose label is strictly smaller. The observation follows.
\end{proof}

\begin{observation}
\label{obs:stableConfiguration}
A stable boundary consists of either 1, 2, 3 or 6 segments. 
\end{observation}

\begin{proof}
A stable boundary consists of segments with equal labels only (and at least one, by the previous observation). Then, these segment's sums are equal integers. Since the sum of counts of all the v-nodes in a ring is either 6 or $-6$ by Observation \ref{obs:sumCounts}, the statement follows.
\end{proof}

\begin{lemma}
\label{lem:correctnessSegmentCompetition}
Every boundary is eventually stable. 
\end{lemma}

\begin{proof}
 By Observation \ref{obs:no-deadlock}, a boundary always has at least one (non-disbanding) segment. Also, note that a boundary (of length $L$) with a single non-disbanding segment of length $L$ is stable.
Thus, to show the lemma statement, it suffices to show that for a non-stable boundary, within finite time, either the number of non-disbanding segments strictly decrease or the sum $L^*$ of their lengths strictly increase. Note that the number of non-disbanding segments is non-increasing (since a disbanding segment remains disbanding until all its v-nodes are absorbed by its predecessor segment, and no segment ever splits otherwise). Also note that the length of a non-disbanding segment is non-decreasing. 

 First, consider the case that the boundary is covered by non-disbanding segments. If all of them are equal in their labels, then the boundary is stable. Otherwise, at least one of them starts disbanding
by Lemma \ref{lem:correctnessLexicographicComparison} and the action taken when a segment $s$ wins a comparison with its successor.

If the boundary is not covered by non-disbanding segments, then there either exists a disbanding segment or a free v-node. Recall that there must also exist a non-disbanding segment $s$ by Observation \ref{obs:no-deadlock}. Furthermore, without loss of generality, the clockwise successor $v_1(B_1)$ of the head of $s$ is either the tail of a disbanding segment $s_1$ or a free v-node. In the first case, $v_1(B_1)$ eventually becomes free. Hence, in both cases, eventually either $s$ absorbs $v_1(B_1)$ and $L^*$ grows by one, or $s$ becomes disbanding. Since there must always exist a non-disbanding segment, one can consider a maximal chain of segments $s_1, s_2=s, \ldots, s_{k}$ such that $s_k$ never disbands and each of $s_1, \ldots, s_{k-1}$ disbands before it absorbs an additional v-node. Then, again, the tail of $s_{k-1}$ eventually becomes free. Since $s_k$ never disbands, $s_k$ eventually absorbs that tail v-node and $L^*$ grows by one. Thus, the lemma holds.
\end{proof}

\paragraph{Round Complexity Analysis.} To analyze the round complexity of the segment competition algorithm, we focus not on segments but rather on their projections. 
Segment projections grow through successful competition and locking operations only, unlike segments which grow by absorbing individual free v-nodes. More precisely, consider some initiating segment $s$ and its successor segment $s_1$, such that $s$ performs an expansion attempt with a successful competition and locking operation. Then, $s$'s projection after the attempt is the concatenation of the projections of $s$ and its successor $s_1$ just prior to that competition.

\begin{lemma}
\label{lem:correctnessAndRuntimeSegmentCompetition}
A boundary of length $L$ becomes stable in $O(L)$ rounds.
\end{lemma}

\begin{proof}
Consider an arbitrary execution. By Lemma \ref{lem:correctnessSegmentCompetition}, the boundary is eventually stable and consists of segments of length $l^* \leq L$ each. By the definition of a projection, at that point, the projection of each segment equals the segment itself.
First, only the projections' growth is analysed. The core of the proof (see statement below) is to show that these projections grow to a size of $l^*$ in $O(l^*)$ rounds. (Note that segments do grow during these first $O(l^*)$ rounds, but not necessarily as much as their projections.) After that, the segments that are smaller than their projections grow to a size of $l^*$ in $O(l^*)$ rounds.

Let us show the following statement by strong induction on $1 \leq l \leq l^*$: any operation  (including initialization) in the execution, that results in a projection changing to be of length $l$, finishes within the first $T(l) = (2k_{c} + 5) l$ rounds (where $k_{c} = 10$). The base case is trivial, since all the projections of length 1 are obtained by initialization. Consider an operation (denoted $\mathcal{O}$) that results in a projection of length $1 < l' \leq l^*$ and assume that the induction hypothesis holds for all $l < l'$. Then, $\mathcal{O}$ must be a successful competition and locking operation, involving an initiating segment $s$ and its successor segment $s_1$. (Note that as a result, $s$ and $s_1$ cannot be disbanding before $\mathcal{O}$ ends.) Operation $\mathcal{O}$ ``merges'' $s$'s projection, of length $1 \leq a < l'$, and $s_1$'s projection, of length $l'-a \geq a$. We assume the ``label'' of the first $a$ counts of $s$'s projection is not lexicographically smaller than the ``label'' of the first $a$ counts of $s_1$'s projection (in the other case, a similar proof can be obtained). 

The prior operations (possibly initialization) that result in $s$'s projection changing to length $a$ and $s_1$'s projection changing to length $l'-a$ finish within the first $max\{T(a),T(l'-a)\}= T(l'-a)$ rounds. Consider a configuration $C$ -- within the first $T(l'-a)$ rounds -- that is after these two operations end, but before $\mathcal{O}$ ends. Note that in $C$, $|s|,|s_1|$ may be, respectively, much smaller than $a$, $l'-a$. Also note that $s_1$ cannot be locked in $C$ and until $\mathcal{O}$ ends; otherwise, $s_1$ would need to terminate its own expansion attempt before $\mathcal{O}$ ends and hence its projection would grow beyond $l'-a$.

If $|s| < a$ in $C$, then $s$ must grow to a length of $a$ before it can compare with $s_1$. Furthermore, if $|s_1| \leq a$ in $C$, then $s_1$ must grow to a length of (at least) $a+1$ before $s$ can compare successfully with $s_1$ (recall the assumption on the projections' ``labels''). This takes at most $a$ rounds for both segments. (Recall that $s$ and $s_1$ do not become non-disbanding before $\mathcal{O}$ ends and both segments take at most 1 round to absorb a free v-node.) 
However, $s$ may be executing an (unsuccessful) comparison -- started when $|s_1| \leq a$ -- when round $T(l'-1) + a$ starts.
Since $s$ does a single comparison with $s_1$ at a time, $s$ completes a successful comparison within $a + 2 k_{c} a$ rounds of $C$ by Lemma \ref{lem:runtimeLengthScheme}. Finally, recall that until $\mathcal{O}$ ends, $s$ and $s_1$ do not become disbanding, and $s_1$ is not locked. Hence, $s$ completes its expansion attempt within $a + 2 k_{c} a + 4a$ rounds of $C$ and the induction step follows.
\end{proof}

\subsection{Detecting the Termination of Primitive {\OBD}}
\label{subsec:boundaryDetection}

The segment competition algorithm from the previous subsection does not yet solve boundary detection. 
The remaining parts of that task are (1) to let the segment heads know when the boundary becomes stable, (2) to have the segment heads compute the sum of the v-nodes' boundary counts and (3) to start the termination announcement if the boundary is the outer boundary (i.e., if this sum is 6).

Whenever primitive {\LCP} -- invoked in the segment competition algorithm -- deems that the initiating segment $s$ has the same label as its successor and no stability check is currently being executed by $s$, segment $s$ checks if its boundary is stable. (On the other hand, whenever primitive {\LCP} deems that segment $s$ and its successor have different labels or $s$ becomes disbanding, $s$ halts any  stability check it is currently performing.)
To do so, $s$ first verifies that its sum is consistent with a stable boundary (i.e., $|sum(s)| \in \{1,2,3,6\}$). Then, $s$ compares its label (in a pairwise manner) with that of the previous $\frac{6}{|sum(s)|}$ segments. If all of these comparisons return that the segments are equal (and none of these segments change), then $s$ is a part of a stable boundary (see Theorem \ref{thm:stableBoundaryDetection} from \cite{BB19}). (Note that $s$ also does not change, since otherwise its stability check is cancelled.) Following the detection, a segment $s$ computes whether it is on the outer boundary -- if and only if the sum of (all the v-nodes of all the segments) counts is 6 -- and if so, informs the other segments on that boundary about the result. Only after all other segments of the outer boundary are notified does a particle-system-wide notification -- by flooding -- take place (thus allowing all other segments to terminate primitive {\OBD} fast). Indeed, the other segments of the outer boundary can (and may) compute that result themselves. However, it is important that they know the result before they receive the notification. (Since upon receiving the flooding message, the recipient terminates its participation in the primitive; hence, v-nodes on the outer boundary receiving the flooding too early could have concluded they were {\em not} on the outer boundary.)

\paragraph*{Segment Sum Verification.} 
For a segment $s$ to verify that the sum of the v-node counts is in $\{1,2,3,6\}$ is somewhat complicated by the fact that the sum can be too large (in its absolute value) to be held by one v-node. Luckily, recall that if the boundary is stable, the sum {\em is} in $\{1,2,3,6\}$.
Let us now describe a procedure that either detects that this sum is not in $\{1,2,3,6\}$ or computes the sum if it is. 
The segment creates two token trains -- a positive token train and a negative token train. Intuitively, the positive train (respectively, the negative train) initially encodes the segment's label restricted to positive (resp. negative) values. Specifically, $s$ generates two sequences of tokens, each consisting of one token per v-node. As before, the head and and the tail tokens of each train are distinguished. In the positive (resp., negative) token train, the token generated by a v-node carries its boundary count $c$ if $c > 0$ (resp., $c < 0$) and 0 otherwise. Although each token initially takes value in $\{-1,\ldots,3\}$, we allow the tokens to take values in $\{-6,\ldots,6\}$ in order to sum up the boundary counts. The two token trains march clockwise towards the head v-node of $s$. To ensure the trains do not block each other, v-nodes assign designated memory variables to each train.
 
 We say that two tokens are {\em mergeable} if their sum is in $\{-6,\ldots,6\}$. Whenever a v-node holds two mergeable tokens of the same train, the v-node forwards a single token carrying the sum. We say that the v-node merged these two tokens. If one of the merged tokens is the head (resp., tail) token, the resulting token becomes the new head (resp., tail) token (possibly both, if the head and tail tokens are merged). Whenever the head v-node $(v_h(B_h)$ of the segment cannot perform a merging operation as above, it attempts to perform a merging of the following kind: if $(v_h(B_h)$ holds two mergeable tokens -- one from each train -- $(v_h(B_h)$ replaces these two tokens by new tokens, one in each train. One of the new token carries the sum of the replaced tokens and the other carries 0. Note that a token carrying 0 is mergeable with any other token.

Eventually, one of the token trains is reduced to a single token (both the head and tail now) carrying a value of 0. Then, the head v-node waits until it either also holds the single token $t$ of the other train (when that token is both the head and the tail token), or two tokens of the other train such that their absolute values add up to at least 7. In the first case, if $t$ has an absolute value in $\{1,2,3,6\}$, then $sum(s)$ is equal to $t$'s value and is consistent with a stable boundary. Otherwise (or in the second case), $sum(s)$ is not computed but is known to be inconsistent with a stable boundary.

\paragraph*{Stable Boundary Check.} 
If a segment $s$ computes that $|sum(s)| \in \{1,2,3,6\}$, it remains for $s$ to verify that its label is equal to that of the previous $k = \frac{6}{|sum(s)|}$ segments (that includes $s$ itself, on a stable boundary). Segment $s$ can do this verification in $O(|s|)$ rounds.
This process is extremely similar to \LCP~ so we do not describe it in detail here. Similarly, the proof is too similar to be repeated. 
Let us just touch upon the minor additions on top of \LCP. First, of course, there are $k$ comparisons, not just one. Second, to compare with its predecessor, the token trains move in the reverse direction and the role of the head and tail tokens are exchanged. Third, when comparing to a far away segment rather than to the predecessor, the comparison train of $s$ marches not just to $s$'s predecessor segment $s_{-1}$, it needs to proceed until $s_{-1}$'s tail and then to $s_{-1}$'s predecessor $s_{-2}$ (when comparing to $s_{-2}$). To compare with such a ``far'' segment, the tail token needs to count how many segments it has passed so far on the way from $s$ (and later, back to $s$). For that, it counts segment heads encountered. Fourth, when comparing to a segment, the train of $s$ marks the tail of that segment. (Segment $s_{-1}$'s tail is marked by 1, segment $s_{-2}$'s tail by 2, etc; Note that in parallel, the tail of $s_{-2}$ may also be marked by the process initiated by $s_{-1}$, etc.) Fifth, whenever a marked segment $\tilde{s}$ becomes disbanding, its tail $v_t(B_t)$ sends two tokens. One token is sent counter-clockwise in order to cancel all (stability check) comparisons with $\tilde{s}$'s predecessor. (Note that this correctly cancels, for any stability check $\tilde{s}$ is participating in, the current comparison because $\tilde{s}$ must be the last already compared segment by Observation \ref{obs:noDisbanding}.) The other token is sent clockwise to inform all (stability check) initiators -- that have already compared with $\tilde{s}$ -- that $\tilde{s}$ is disbanding. Note that the segments between $\tilde{s}$ and $s$ cannot become disbanding before $\tilde{s}$ does by Observation \ref{obs:noDisbanding}, and hence the second token -- by leveraging $v_t(B_t)$'s marked number -- correctly reaches the intended initiators.

\paragraph*{Informing Other Segments on the Outer Boundary} Once a segment $s$ detects that the boundary is stable, it can deduce whether it is on the outer boundary or not. Specifically, segment $s$ is on the outer boundary if and only if the total sum of counts $k \cdot sum(s)$ equals 6 -- see Observation \ref{obs:sumCounts}.
 If $s$ is on the outer boundary, it informs all of the other segments on that boundary that they are on the outer boundary. To do so, the head v-node of $s$ sends an \emph{outer} token along the boundary. Once the token has gone around the boundary back to $s$'s head, the v-node requests its simulating particle to start the termination announcement, described in the next paragraph. 
However, multiple outer tokens may be travelling along the boundary (if multiple segments have detected that they are on the outer boundary). To ensure $s$'s head can detect when its token does a full walk of the boundary, an outer token is initially created with a value of 0. Whenever the outer token is forwarded into another segment's tail, the token's value is incremented. Hence, $s$ can detect whether a received token is its own token by checking if its value is $k$.

\paragraph*{Termination Announcement.}
To obtain fast termination detection, outer boundary particles announce the termination of primitive {\OBD} to all the other particles using flooding. Every particle receiving the flooding stops executing the primitive. Since flooding is a well-known algorithm, we omit the details.

\begin{theorem}[\cite{BB19}]
\label{thm:stableBoundaryDetection}
There exists a sequence of $1+\frac{6}{|\sigma|}$ adjacent equal label segments, whose common sum $\sigma$ has an absolute value in $\{1,2,3,6\}$, along a boundary if and only if that boundary is stable. Then, the first and last segments of the sequence are the same.
\end{theorem}

\begin{observation}
\label{obs:noDisbanding}
Consider a stable boundary check initiated by a segment $s$. As long as the last already compared segment $\tilde{s}$ remains non-disbanding, then so do all other already compared segments. Hence, the stable boundary check is properly cancelled if $\tilde{s}$ becomes disbanding. Moreover, either none of the already compared segments grow or $s$ cancels its stable boundary check.
\end{observation}

\begin{proof}
As long as segment $s$ is executing the stable boundary check, $s$ is non-disbanding and its label does not change (otherwise, $s$ cancels the stable boundary check). Given this, a simple induction leads to the statement.
\end{proof}

\begin{lemma}
\label{lem:correctnessTerminationDetection1}
While a boundary of length $L$ is not stable, its virtual particles do not detect that the boundary is stable.
\end{lemma}

\begin{proof}
Assume, by contradiction, that some segment $s$ executing the stable boundary check detects that the boundary is stable (while, in fact, it is not). First, note that $s$ cannot be disbanding nor does it grow (otherwise the stable boundary check fails). Moreover, none of the compared segments disband nor grow (otherwise the stable boundary check also fails -- see Observation \ref{obs:noDisbanding}). 
Thus, if the $k$ comparisons between segment $s$ and its $k$ previous segments returns that they have equal labels, then there is a sequence of $1+\frac{6}{|sum(s)|}$ adjacent equal label segments. However, this leads to a contradiction with Theorem \ref{thm:stableBoundaryDetection}, and thus the lemma statement holds.
\end{proof}

\begin{lemma}
\label{lem:correctnessTerminationDetection2}
Once a boundary of length $L$ is stable, its v-nodes detect that the boundary is stable in $O(L)$ rounds.
\end{lemma}

\begin{proof}
Once the boundary is stable, then from the algorithm description and Theorem \ref{thm:stableBoundaryDetection}, at least one segment detects that the boundary is stable. Moreover, the segment does so in $O(L)$ rounds. 
\end{proof}

\begin{corollary}
\label{cor:terminationAnnouncementInit}
There exists an outer boundary v-node that initiates the termination announcement within $O(L_{out})$ rounds. Moreover, it does so only after all outer boundary v-nodes are aware that they are on the outer boundary.
\end{corollary}

\begin{proof}
The first part of the statement follows from Lemma \ref{lem:correctnessTerminationDetection2} and the fact that the outer token traverses the outer boundary in $O(L_{out})$ rounds. The second part follows from the fact that a v-node waits for the outer token to traverse the outer boundary before initiating the termination announcement.
\end{proof}

\begin{theorem}
\label{thm:CorrectnessAndRuntimeOfEfficientBoundaryDetection}
Primitive {\OBD} terminates in $O(L_{out}+D)$ rounds.
\end{theorem}

\begin{proof}
The outer boundary becomes stable in $O(L_{out})$ rounds by Lemma \ref{lem:correctnessAndRuntimeSegmentCompetition}. Furthermore, there exists an outer boundary v-node that initiates the termination announcement in $O(L_{out})$ rounds by Corollary \ref{cor:terminationAnnouncementInit}. Since the termination announcement starts only after all outer boundary v-nodes know they are on the outer boundary (by Corollary \ref{cor:terminationAnnouncementInit}), all particles correctly deduce which of their local boundaries are part of the global outer boundary. Moreover, the flooding mechanism takes at most $O(D)$ rounds. Hence, the theorem statement holds.
\end{proof}

\newpage

\section*{Acknowledgments}
The research of Fabien Dufoulon was supported in part at the Technion by a fellowship from the Lady Davis Foundation. The research of Shay Kutten was partially funded by a grant from the Binational Science Foundation. The research of William~K.~Moses~Jr. was supported, in part, by NSF grants, CCF1540512, IIS-1633720, CCF-1717075, and BSF grant 2016419.

\bibliographystyle{plainurl}
\bibliography{main}

\newpage

\appendix
\appendix

\section{Figures Illustrating the Preliminary Definitions}

\begin{figure}[ht]
        \centering
        \includegraphics[width=0.8\linewidth]{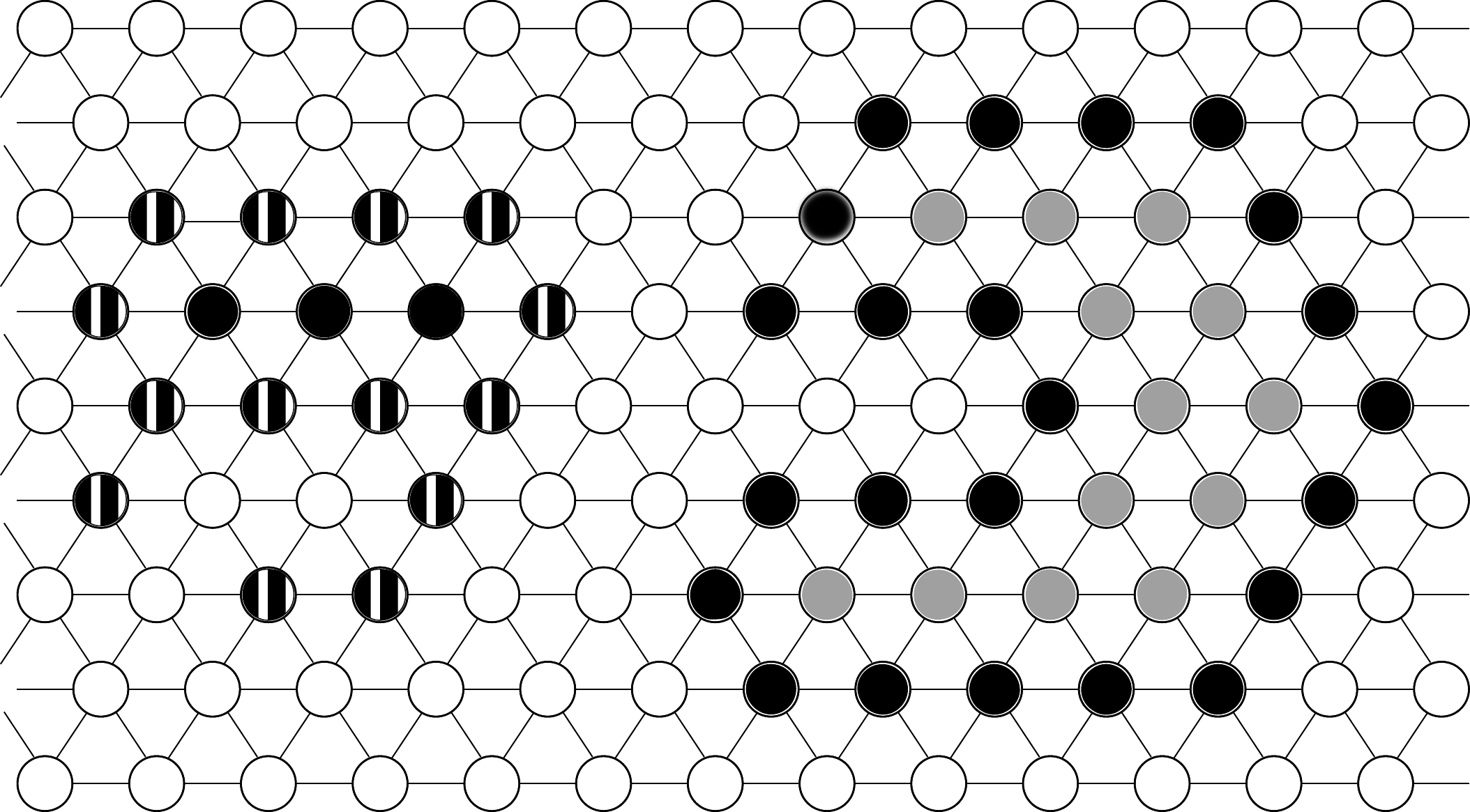}
        \caption{The black and striped grid points form two shapes in the triangular grid. All other grid points are colored in grey and white. The left shape is {\em simply-connected} (i.e., has no holes) and its outer boundary is the set of striped grid points. The right shape is not simply-connected and has one hole -- the set of grey grid points. The area of the right shape is the set of black and grey grid points. } 
        \label{fig:ShapesExample}
\end{figure}

\begin{figure}[ht]
        \centering
				\includegraphics[width=0.8\linewidth]{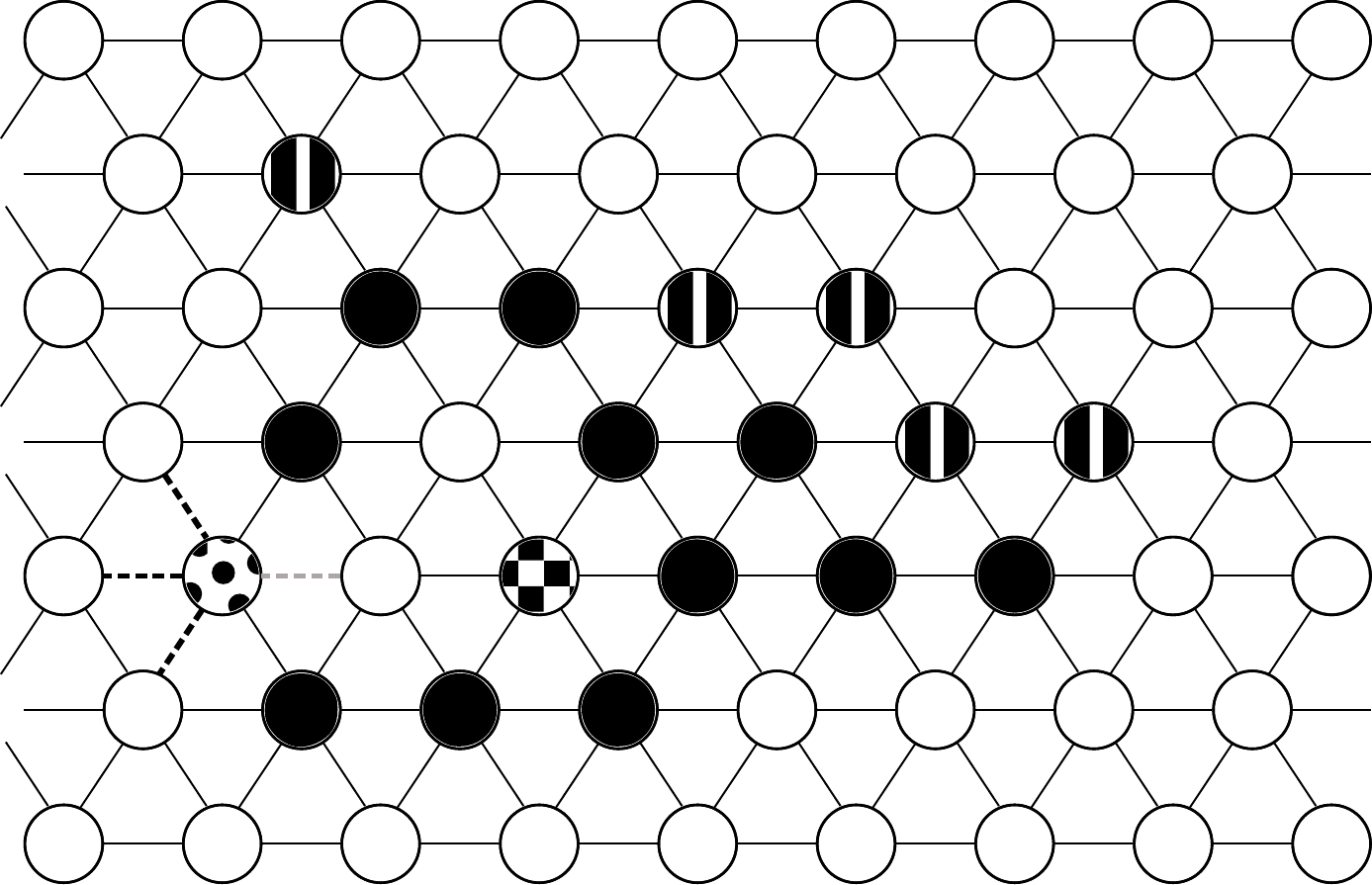}
				\caption{Black and patterned grid points form the shape. The striped grid points are erodable. Furthermore, from left to right, these grid points have a boundary count of, respectively, 3, 0, 1, -1 and 2. The checkered grid point is redundant but not erodable. The grid point with ``spots'' is not redundant. Note that it is strictly convex (with a count of 1) w.r.t. its first local boundary -- the black dashed edges -- and has a boundary count of -1 w.r.t. its other boundary -- the grey dashed edge.}
				\label{fig:exampleCounts}
\end{figure} 

\begin{figure}[ht]
\centering
		\includegraphics[width=0.8\linewidth]{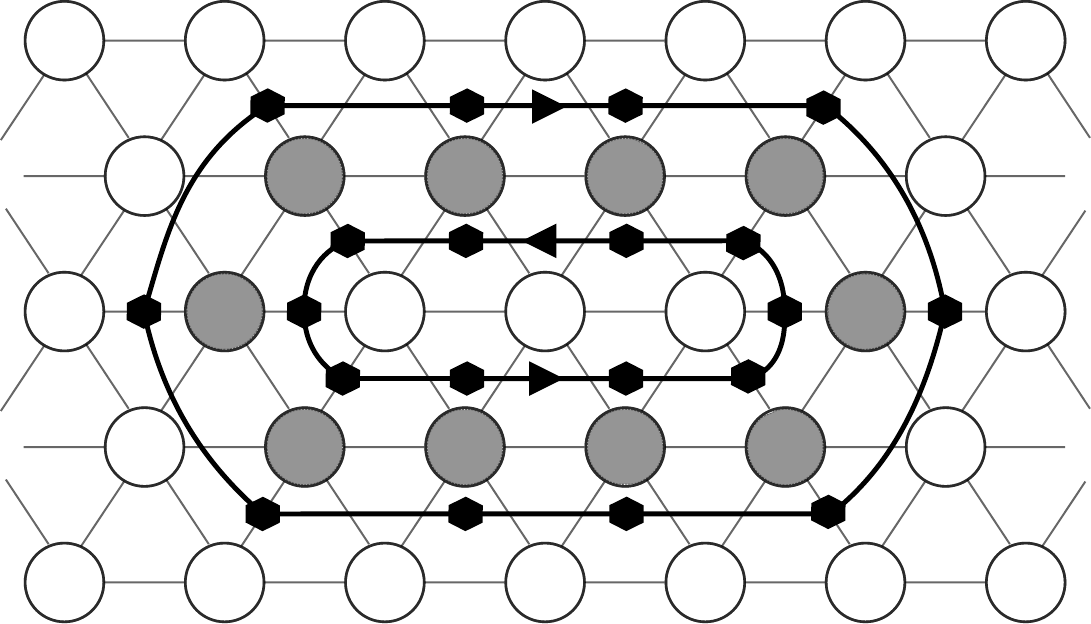}
		\caption{The shape's boundary grid points -- colored in grey -- are subdivided into v-nodes(s) -- one per local boundary -- represented here by black hexagons. Each global boundary is transformed into a virtual ring of v-nodes, where a v-node's successor in the ring is its clockwise successor. Note that the outer boundary's virtual ring is oriented clockwise and the other boundary's ring counter-clockwise.}
		\label{fig:exampleBoundaryTraversal}
\end{figure} 

\begin{figure}[ht]
     \centering
	\begin{subfigure}[t]{0.47\textwidth}
        \centering
        \includegraphics[width=0.9\linewidth]{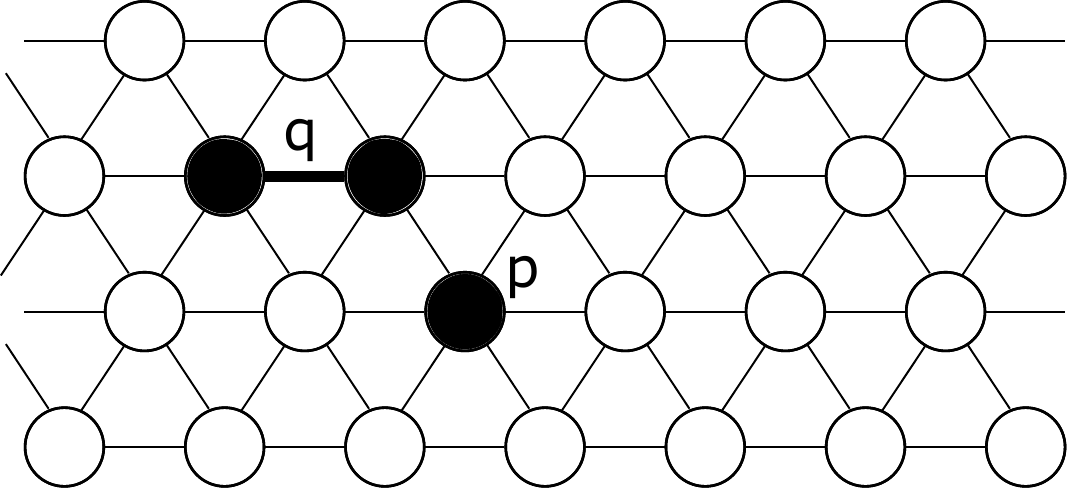}
        \subcaption{Triangular grid with white unoccupied grid points and black occupied grid points. Particle $p$ is contracted and particle $q$ is expanded.}
        \label{fig:triangularGridAndExample1} 
	\end{subfigure} \hfill
	\begin{subfigure}[t]{0.47\textwidth}
        \centering
        \includegraphics[width=0.9\linewidth]{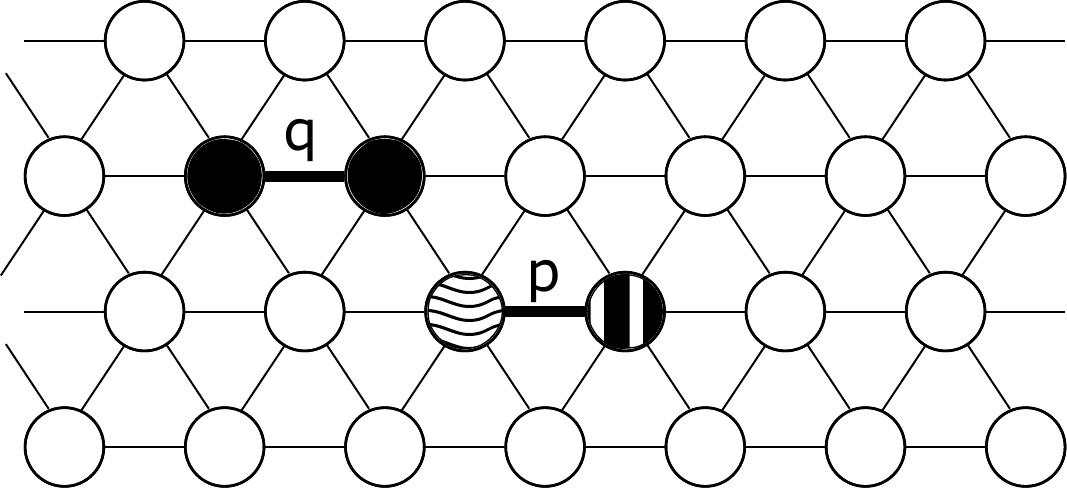}
        \subcaption{Particle $p$ expands from its original occupied grid point -- the one with a ``wave'' pattern, now $p$'s tail -- into the grid point with a ``stripe'' pattern, now $p$'s head.} 
        \label{fig:triangularGridAndExample2}
	\end{subfigure} 
\caption{Particle Movement}
\label{fig:particleMovement}
\end{figure}

\end{document}